\documentclass[a4paper,11pt]{article}
\RequirePackage{snapshot}  

\usepackage{jheppub} 

\usepackage{bm,amsmath,amssymb,slashed,graphicx,%
            enumerate,alltt,xspace,multirow,xcolor,mathrsfs}
\usepackage{fancyvrb}
\usepackage{booktabs,tabularx}
\usepackage{graphicx}
\usepackage{subcaption}
\usepackage{xspace}
\usepackage[utf8]{inputenc} 
\usepackage[compat=1.0.0]{tikz-feynman}
\usepackage{scalerel}
\usepackage{layouts}

\usepackage{lineno}
\usepackage{printlen}
\usepackage{wasysym}
\usepackage{grffile}
\usepackage{soul}

\makeatletter
\g@addto@macro\bfseries{\boldmath}
\makeatother

\definecolor{labelkey}{rgb}{0,0.5,0.0}

\usepackage{listings}
\definecolor{darkgreen}{rgb}{0,0.4,0}
\definecolor{lightblue}{rgb}{0.0,0.5,1.0}
\definecolor{grey}{rgb}{0.5,0.5,0.5}

\newcommand{\muR}{\mu_\text{\textsc{r}}}
\newcommand{\muF}{\mu_\text{\textsc{f}}}

\allowdisplaybreaks 

\definecolor{semiblue}{rgb}{0.3,0.3,0.8}
\newcommand{\logbook}[2]{}

\newcommand{\order}[1]{\mathcal{O}\left(#1\right)}
\newcommand{\as}{\alpha_s}

\newcommand{\dd}{\;\mathrm{d}}

\newcommand{\ktilde}{{\tilde k}}
\newcommand{\ptilde}{{\widetilde p}}

\newcommand{\nc}{N_\text{\textsc{c}}} 
\newcommand{\nf}{n_{\!\:\!f}} 
\newcommand{\betaps}{{\beta_\text{\textsc{ps}}}}

\newcommand{\itilde}{{\tilde \imath}}
\newcommand{\jtilde}{{\tilde \jmath}}

\tikzstyle{block} = [rectangle, minimum width=1.0cm, minimum height=0.75cm, thin, draw=black]
\tikzstyle{blob} = [circle, minimum width=0.5cm, thin, draw=black]
\tikzset{blackarrow/.style={-stealth, semithick, draw=black}}
\tikzset{connection/.style={inner sep=0,outer sep=0}}

\newcolumntype{C}{>{\centering\arraybackslash}X}

\title{PanScales parton showers for hadron collisions: formulation and
  fixed-order studies
}

\preprint{OUTP-22-06P}

\newcommand{\OXaff}{Rudolf Peierls Centre for Theoretical Physics, Clarendon Laboratory, Parks Road,
  University of Oxford, Oxford OX1 3PU, UK}
\newcommand{\ASCaff}{All Souls College, Oxford OX1 4AL, UK}

\author[a]{Melissa van Beekveld,}%
\author[a]{Silvia Ferrario Ravasio,}%
\author[a,b]{Gavin P.~Salam,}%
\author[c]{Alba~Soto-Ontoso,}%
\author[c]{Gregory Soyez,}%
\author[d]{Rob Verheyen}%

\affiliation[a]{\OXaff}
\affiliation[b]{\ASCaff}
\affiliation[c]{Universit\'e Paris-Saclay, CNRS, CEA, Institut de physique th\'eorique, 91191, Gif-sur-Yvette, France}
\affiliation[d]{Department of Physics and Astronomy, University College London, London, WC1E 6BT, UK}

\date{Received: date / Accepted: \today}

\abstract{
  We formulate PanScales parton showers for hadron collisions
  so as to achieve next-to-leading logarithmic (NLL) accuracy across a
  broad set of observables.
  We do so specifically for colour singlet production.
  Relative to the existing PanScales final-state showers, the main new
  question is that of how to redistribute momentum imbalances from
  initial-state branching across the remainder of the event.
  We present tests of the showers at fixed order, including the
  treatment of full colour for soft-collinear emissions and of
  spin correlations in both the soft and collinear domains.
  We also include comparisons to a formulation of a standard dipole
  shower, the current leading-logarithmic state of the art.
  A forthcoming companion paper~\cite{vanBeekveld:2022ukn}
  will explore all-order tests of the new showers.
}


\begin{document}

\maketitle

\section{Introduction}
\label{sec:intro}
One of the major achievements of CERN's Large Hadron Collider (LHC)
programme is the remarkable accuracy being achieved across a broad
range of measurements.
To maximally exploit the potential of the data, it is essential for
theoretical predictions to match that high accuracy, notably as
concerns the tools that can fully simulate the final state of hadron
collisions, i.e.\ general purpose Monte Carlo (GPMC) event generators.
These tools play a crucial role both in the interpretation of measurements
and in the extraction of those measurements from the raw experimental
data.

Two major components of GPMCs are under control in perturbative
Quantum Chromodynamics (QCD):
the hard scattering process, and the parton shower, which simulates
radiation from the hard scattering scale down to the hadronic scale.
Over the past two decades major advances have been made in improving
the accuracy of the hard scattering
description~\cite{Heinrich:2020ybq} and its matching to parton
showers, both for next-to-leading order~(NLO)
calculations~\cite{Frixione:2002ik,Nason:2004rx,Jadach:2015mza} and,
more recently, for (N)NNLO
computations~\cite{Hamilton:2012rf,Alioli:2013hqa,Hoche:2014dla,Monni:2019whf,Campbell:2021svd,Prestel:2021vww}.
Alternatively, multi-leg matrix elements can be merged with parton 
shower simulations up to NLO
accuracy~\cite{Catani:2001cc,Mangano:2001xp,Krauss:2002up,Lavesson:2008ah,Hoeche:2009rj,Hamilton:2009ne,Giele:2011cb,Platzer:2012bs,Lonnblad:2012ix,Frederix:2012ps,Lonnblad:2012ng,Bellm:2017ktr,Brooks:2020mab,Martinez:2021chk}.

Improvement of the accuracy of the hard scattering process is not the
only requirement for precision LHC phenomenology.
In particular parton showers are crucial for a correct physical
description of many of the features of events that are essential in
experimental measurements, such as the pattern of energy sharing
between particles, or the transverse momentum distribution of
colour-singlet objects such as a Drell-Yan (DY) pair or a Higgs boson and
correlations with final-state energy flow.
Since parton showers span disparate momentum scales, one natural way
of viewing their theoretical role is that they should account for
perturbative contributions that are enhanced by logarithms of the
ratios of those disparate scales.

All dipole-shower algorithms~\cite{Sjostrand:2006za,Giele:2007di,Schumann:2007mg,Platzer:2009jq,Hoche:2015sya,Cabouat:2017rzi} available in
the commonly-used general purpose Monte Carlo event
generators~\cite{Sjostrand:2006za,Sjostrand:2014zea,Bahr:2008pv,Bellm:2019zci,Gleisberg:2008ta,Sherpa:2019gpd}
reach leading-logarithmic (LL)
accuracy, i.e.~they are correct in the limit where emissions are
strongly ordered both in energy and angle.\footnote{Another class of shower is represented by the angular-ordered showers, as used in the \texttt{Herwig7} program~\cite{Bahr:2008pv,Bellm:2019zci}.
  They are based on the coherent branching
  formalism~\cite{Gieseke:2003rz}, that reproduces full-colour NLL accuracy
  for double logarithmic global observables~\cite{Catani:1990rr,Catani:1992ua}. Note that angular ordering is known~\cite{Banfi:2006gy} to lead to an incomplete treatment
  of the soft single (NLL) logarithms that affect
  non-global observables~\cite{Dasgupta:2001sh}, though the numerical impact is modest in many
  circumstances. }
The focus of this paper is to design a next-to-leading logarithmic
(NLL) accurate shower for hadron-hadron collisions, which represents a
key milestone in the PanScales
programme~\cite{Dasgupta:2018nvj,Dasgupta:2020fwr,Hamilton:2020rcu,Karlberg:2021kwr,Hamilton:2021dyz}
of developing parton showers with controlled logarithmic accuracy.
To achieve this, we use the guiding principles introduced in
Refs.~\cite{Dasgupta:2018nvj, Dasgupta:2020fwr} to assess the
logarithmic accuracy of a shower.
From the PanScales perspective, a shower can be said to be NLL if it
satisfies two types of condition:
\begin{itemize}
\item \label{item:panscales-1}
  Fixed-order: upon characterisation of the phase space of an
  emissions in terms of its transverse momentum, $k_t$, and rapidity
  $\eta$, the shower must reproduce the exact matrix element in the
  limit where every pair of emissions is well separated in at least one of
  the phase space variables, e.g.\ $k_t$ or $\eta$ are strongly
  ordered across emissions.
\item All-orders: the shower must reproduce analytic resummation
  results for a broad range of observables, including global event
  shapes, subjet multiplicity and non-global observables.
\end{itemize}
These principles were used in Ref.~\cite{Dasgupta:2020fwr} to design
the basic structure for several NLL-accurate dipole final-state parton
showers, including tests across a wide range of observables.
Full NLL accuracy for these showers (including full-colour (FC) at LL) was
subsequently achieved with the addition of soft-collinear subleading
colour corrections~\cite{Hamilton:2020rcu}, and of spin
correlations~\cite{Karlberg:2021kwr,Hamilton:2021dyz}.

Several other groups have also been investigating the logarithmic
accuracy of showers.
Refs.~\cite{Bewick:2019rbu,Bewick:2021nhc} showed that a careful
treatment of the kinematic map is necessary in angular-ordered
showers, for both final and initial state radiation.
Ref.~\cite{Forshaw:2020wrq} introduced a shower algorithm shown
analytically to reproduce the NLL results for the $e^+e^-$ thrust
distribution and the subjet multiplicity (with soft-collinear
subleading colour effects in Ref.~\cite{Holguin:2020joq}).
Ref.~\cite{Nagy:2020dvz} showed that the so-called $\Lambda$-ordered
(effectively time-ordered) Deductor shower reproduces the NLL thrust
distribution in $e^{+}e^{-}$ collisions~\cite{Nagy:2020dvz}.

Almost all of the above work has focused on the accuracy of
final-state showers.
In this work, we show how to extend and adapt the PanScales showers to
include initial-state radiation, concentrating on the case of
colour-singlet production, i.e.\ $q\bar q \to Z^{0}$ and $gg\to H$ at
hadron colliders.
Aside from the many hadronic observables that have already been tested
in the final state, a particularly important observable in
hadron-hadron collisions is the transverse momentum of the colour
singlet system, whose resummation is well
established~\cite{Parisi:1979se,Collins:1984kg}.
The approaches that we develop build on observations both from the
PanScales work and from earlier parton-shower work that specifically
considered the question of colour-singlet transverse momentum recoil
in a
shower~\cite{Nagy:2009vg,Platzer:2009jq,Hoche:2015sya,Cabouat:2017rzi}.

This paper is organised as follows. In Section~\ref{sec:basic-formulation} we
provide the basic building blocks for designing a dipole shower in hadronic
collisions and give  an explicit example of a transverse-momentum ordered shower
based on the standard colour dipole approach, which we dub
``Dipole-$k_t$''.
In Section~\ref{sec:fo-test} we review the
general methodology for performing a fixed-order study of a dipole shower in the
limit of interest for NLL accuracy. We show that Dipole-$k_t$, variants of which are currently available in all the major Monte Carlo event generators, fails to
reproduce the correct soft radiation pattern already at the two emissions level.
Building on this knowledge, in Section~\ref{sec:panshowers} we introduce two
families of PanScales showers for hadronic collisions, one with a local and
another with global recoil schemes. The choice of the ordering variable and the recoil scheme are crucial to satisfy the fixed-order shower requirement. 
In Section~\ref{sec:colour}, we adapt the subleading-colour
prescriptions of Ref.~\cite{Hamilton:2020rcu} to address initial-state
radiation, and show a number of associated matrix-element tests.
The inclusion of spin correlations following
Refs.~\cite{Karlberg:2021kwr,Hamilton:2021dyz} is explained in
Section~\ref{sec:spin}. We conclude in
Section~\ref{sec:conclusions}. The appendices contain details on the
splitting functions we use (Appendix~\ref{app:splitting-func}), the
kinematic mappings (Appendix~\ref{app:shower-coefficients}), the
analytic expectations for our colour tests (Appendix~\ref{app:colour})
and the derivation of the spin branching amplitudes (Appendix~\ref{app:spin}). 
The validation of our approach at all-orders across many observables
and a presentation of the associated all-order testing methodology are
to be found in a separate publication~\cite{vanBeekveld:2022ukn}.

\section{Basics of hadron-collision dipole showers}
\label{sec:basic-formulation} 

In this section we will highlight common features of dipole showers and
formulate a generic standard dipole shower, which will be used as a
convenient reference for a LL-accurate shower throughout this work and
our companion article~\cite{vanBeekveld:2022ukn}.
We will concentrate on colour-singlet production in proton-proton
collisions, specifically $q\bar q \to Z$ and $gg\to H$, with a
hadron-hadron centre-of-mass energy $\sqrt s$ and a colour-singlet
Born four-momentum $Q^{\mu}$.

\subsection{Generic formulation of a hadron-collider shower}

Standard dipole showers and the PanScales hadron-collider showers that
we develop later in Section~\ref{sec:panshowers} have a number of
characteristics in common.
These include the final and initial-state splitting probabilities, as
well as the generic structure of recoil for emission of a parton from
a dipole.
In this work, all partons are considered to be massless and we will
often refer to the colour singlet as the ``hard system".

First, we consider a final-state parent parton ${\itilde}$ that radiates a collinear
emission $k$. The post-branching momentum of the parent is denoted by
$i$. The phase-space of the emission $k$ is parameterised by its
transverse momentum $k_\perp$, 
its longitudinal momentum fraction $z$
(relative to the pre-branching parent)
and its azimuthal angle $\varphi$.
In the collinear limit ($\theta_{ik} \ll 1$), the differential
branching probability then reads 
	\begin{align}
		\label{eq:p-final-state}
		\begin{gathered}
			\includegraphics{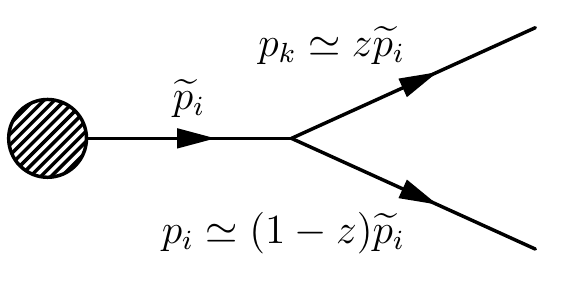}
\end{gathered}
			\hspace{0.1cm} \rightarrow  \quad 
	{\rm d} \mathcal{P}^{\rm FS}_{ \itilde \to ik} = \frac{\as(k_{\perp}^2)}{2\pi} \frac{{\rm d} k_\perp^2}{k_\perp^2} \frac{{\rm d}z}{z} \frac{{\rm d}\varphi}{2\pi} N^{\rm sym}_{ik} \left[zP_{ \itilde \to ik}(z) \right] ,\,\,\,
	\end{align}%
with $\as$ the strong coupling and $N^{\rm sym}_{ik}$ a symmetry factor that
is equal to $1/2$ for $g \rightarrow g g$ splittings, and $1$
otherwise.
We use symbols with a tilde to indicate pre-branching partons and
their momenta, and symbols without any decoration to indicate
post-branching partons.
The DGLAP splitting functions $P_{\itilde \rightarrow ik}$ are
provided in Appendix~\ref{app:splitting-func}.
A well-known feature of Eq.~\eqref{eq:p-final-state} is its singular
behaviour in the soft ($z\to 0$) collinear limit for
flavour-conserving emissions (i.e.~$P_{g\rightarrow gg}$ and
$P_{q\rightarrow qg}$), and in the hard ($z\sim 1$) collinear limit
for every type of emission.
The soft and collinear singularities compensate the smallness of $\as$
in the corresponding regions of phase space, resulting in the large
logarithms that the shower resums.

In hadronic collisions, final-state radiation is to be supplemented with emissions
from the incoming partons.
Over three decades ago, it was realised that
a backwards evolution from the hard scale $Q^2$ down to the hadronic
scale ($\mathcal{O}(1)$~GeV) provides an efficient way of simulating such initial-state
radiation~\cite{Sjostrand:1985xi}.
Defining $z$ as the longitudinal momentum fraction carried by the
emission relative to the post-branching incoming parton, the differential
collinear branching probability is given by\medskip
	\begin{align}
		\label{eq:p-initial-state}
		\begin{gathered}
		\includegraphics{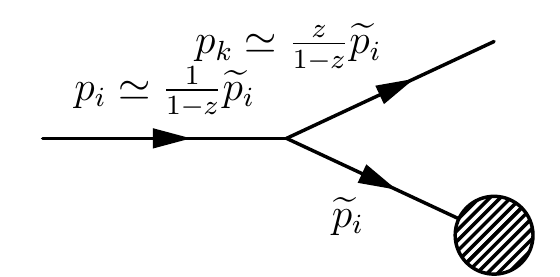}			
        \end{gathered}
		\hspace{0.1cm} \rightarrow  \quad 
	{\rm d}\mathcal{P}^{\rm IS}_{i \to \itilde k} =
	    \frac{\as(k_{\perp}^2)}{2\pi} \frac{{\rm
	        d}k_{\perp}^2}{k_{\perp}^2} \frac{{\rm d}z}{z}
	    \frac{d\varphi}{2\pi} zP_{i \rightarrow \itilde k}(z) \frac{x_i
	      f_i(x_i,k_{\perp}^2)}{\tilde{x}_i
	      f_{\itilde}(\tilde{x}_i,k_{\perp}^2)}\,.\qquad
	\end{align}%
The main
difference with respect to Eq.~\eqref{eq:p-final-state} is the
additional presence of a ratio
of parton distribution functions (PDFs) $f_i(x_i,k^2_\perp)$, denoting
the density per unit momentum fraction of partons of flavour $i$,
carrying a momentum fraction $x_i$ inside a proton, at a factorisation
scale $k_\perp$.
Its
inclusion follows directly from the DGLAP evolution equation with
evolution variable $k_{\perp}^2$, and accounts for the change in the
hadronic momentum fraction from $\tilde x_i$ to $x_i \equiv
\tilde{x}_i/(1-z)$.

The branching probabilities given by Eqs.~\eqref{eq:p-final-state}
and~\eqref{eq:p-initial-state} describe $1 \to 2$ collinear
branchings.
In this paper we will work with dipole showers in which the fundamental
building block is instead a $2 \to 3$ branching
kernel~\cite{Gustafson:1987rq}.
In the limit where the number of colours ($\nc$) is large, dipole
showers make it straightforward to reproduce the matrix element for an
arbitrary number of emissions in the soft wide-angle limit, and to
achieve the associated resummation of non-global logarithms.
In dipole showers, the particle that branches, $\itilde$, is
associated with a colour-connected 
spectator $\jtilde$, such 
that the branching is $\itilde \jtilde \to i j k$, where $i$ and $j$
are the post-branching counterparts of $\itilde$ and $\jtilde$
respectively, and $k$ is the radiated parton. Each of the dipole legs
can be either an initial (I) or a final (F) state particle. As such,
four types of dipoles exist: II, IF, FI and FF. By symmetrising
Eqs.~\eqref{eq:p-final-state} and~\eqref{eq:p-initial-state} we obtain
a generic dipole differential splitting probability\footnote{This specific form of the branching probability is not unique. Non-singular terms may be added, and the assignment of the soft singularity to different dipole ends (achieved by $g(\bar{\eta})$) may be accomplished in various ways. }
\begin{multline}
\label{eq:dipole-prob}
     {\rm d} \mathcal{P}_{\itilde \jtilde \to ijk} =
    \frac{\as(\mu_\text{\textsc{r}}^2)}{2\pi}\left(1 +\frac{\as(\mu_\text{\textsc{r}}^2) K}{2 \pi}\right) \frac{{\rm d} v^2}{v^2} {\rm d}
    \bar{\eta} \frac{{\rm d}\varphi}{2\pi}
    \times
    \\
      \times \frac{x_i f_i(x_i, \muF^2)}{\tilde{x}_i f_{\itilde}(\tilde{x}_i,\muF^2)} \frac{x_j f_j(x_j, \muF^2)}{\tilde{x}_j f_{\jtilde}(\tilde{x}_j,\muF^2)  } \left[ g(\bar{\eta}) z_i P^{\rm {IS/FS}}_{ik} (z_i) + g(-\bar{\eta}) z_j P^{\rm {IS/FS}}_{jk} (z_j) \right] \,.
\end{multline}
We will consider two standard approaches to generating the kinematics
associated with the branching:
so-called ``dipole''
showers and ``antenna'' showers.
In the former, the two terms in square brackets are associated with
distinct kinematic maps.
In the latter one uses a common kinematic map for both
terms.\footnote{
  \label{foot:antenna-partition}
  Eq.~(\ref{eq:dipole-prob}) suggests a natural way of assigning
  branching to one end or other of the dipole.
  In antenna showers, while this division does not affect the
  kinematics of the branching, it can affect the subsequent choice of
  splitting channel and associated spin correlations, e.g.\ concerning the choice of $g \to gg$ v.\
  $g \to q\bar q$, which in the context, say, of a $gq$ dipole might
  occur only at one end.
  In the implementations of antenna showers used here and in our
  companion work~\cite{vanBeekveld:2022ukn}, the division is slightly
  different in a region
  where one of the $z_i$ or $z_j$ is not soft and additionally
  $g(\bar\eta)$ differs substantially from $0$ or $1$.
  This region affects logarithmic accuracy only for terms that are
  beyond NLL, and will be re-examined in future work when considering
  higher logarithmic accuracy.
}
The phase-space has been transformed to two specific shower variables:
an evolution scale $v$ and a pseudo-rapidity-like variable
$\bar{\eta}$.\footnote{Note that we use the notation
  $\bar{\eta},k_{\perp}$ to denote shower variables, whereas $\eta$
  and $k_t$ are reserved to denote the physical pseudorapidity and
  transverse momentum measured with respect to the incoming hadron beams.} The exact relation
between ($v,\bar{\eta}$) and ($z,k_\perp$) is shower-dependent, and we
thus postpone its discussion. However, in this work the following
relation is always satisfied
\begin{equation}
\frac{\dd k_{\perp}^2}{k_{\perp}^2} \frac{\dd z}{z} = \frac{\dd v^2}{v^2} \times 
\begin{cases}
    \dd\bar{\eta} & \text{(final state),} \\
    (1-z) \, \dd\bar{\eta} & \text{(initial state)},
\end{cases}
\end{equation}
such that we can use the compact notation
\begin{equation}
P^{\text {IS}}_{ik}(z) = (1-z) P_{i \to \itilde k}(z),~\quad  P_{ik}^{\text {FS}}(z) = P_{\itilde \to i k}(z)\,,
\label{eq:PIS}
\end{equation}
for initial-state (IS) and final-state (FS) splittings, respectively.
As in Eq.~(\ref{eq:p-initial-state}), a PDF ratio for each of the
dipole legs enters as a multiplicative factor.
For final-state branchings the $x$ and $\tilde x$ values are either
identically equal, or very close to each other, so the ratio tends to
one.
There is limited freedom in choosing the factorisation and
renormalisation scales, $\muF$ and $\muR$.
For a hard-collinear initial-state branching, $\muF$ should be
commensurate with the emission transverse momentum.
As concerns the evaluation of the coupling, the requirement of
achieving NLL accuracy brings several constraints, notably for
soft-collinear emissions: the running of the coupling should be
performed at two loops or higher, and with $\muR$ chosen to 
coincide with the emission transverse momentum, the soft-collinear
gluon emission probability must include an $\as K/(2\pi)$ correction term
with $K=(67/18-\pi^2/6)C_A-5\nf/9$.
As usual we have $C_A=3$, $T_R=1/2$ and we will work with $\nf=5$ light
flavours. 
Finally, the function $g(\bar{\eta})$ partitions the soft singularity among
the two contributions of the dipole, avoiding any double
counting.
This function, which we will specify below, needs to satisfy
three main requisites: (i) $ 0\leq g(\bar{\eta}) \leq 1$, (ii) $g(\bar{\eta}) +
g(-\bar{\eta}) = 1$ and (iii) $g(\bar{\eta}) = 0\; (1)$ for very
negative (positive) $\bar{\eta}$.

Every time an emission is generated according to Eq.~\eqref{eq:dipole-prob},
momentum conservation has to be restored through a choice of a
suitable recoil scheme.
We will explore two classes of shower, one where the kinematic map
conserves momentum locally within the dipole that is branching and
another that performs an overall global momentum-conservation
procedure.
Both share the feature that the emission's
momentum $p_k$ may be Sudakov-decomposed in terms of the parent dipole
momenta  $(\tilde p_i, \tilde p_j)$ and a transverse component:
\begin{equation}
\label{eq:sudakov-decom}
p_k^\mu = a_k \tilde p_i^\mu + b_k \tilde p_j^\mu + k_\perp^\mu\,,
\end{equation}
Within a given shower, the $a_k$, $b_k$ and
$|k_\perp|$ magnitudes are fixed by the value of the shower
ordering variable (which may, for example, directly set the value of
$|k_\perp|$), by a longitudinal variable (i.e.\ $\bar{\eta}$ in Eq.~(\ref{eq:dipole-prob})) and by
requiring
Eq.~\eqref{eq:sudakov-decom} to be on the mass 
shell, $|k_\perp|^2= 2 a_k b_k\, \tilde p_i \cdot \tilde p_j \equiv a_k b_k
\tilde{s}_{ij}$, where $\tilde{s}_{ij}$ is the dipole mass. The
direction of the transverse component is
given by
\begin{equation}
  \label{eq:kperp-decomposition}
k_\perp^\mu = |k_\perp|( \hat{n}_1^\mu \sin \varphi + \hat{n}_2^\mu \cos \varphi),
\end{equation}
where $\hat{n}_{1,2}^2 = -1$, $\hat{n}_1\cdot\hat{n}_2=0$ and
$\hat{n}_{1,2}\cdot {\tilde p}_{i,j}=0$.

\paragraph{Local recoil schemes.}
The momentum map is generically  given by 
\begin{subequations}
    \label{eq:general-local-map}
  \begin{align}
    & \bar p^{\mu}_i = a_i \tilde{p}_i^{\mu} + b_i \tilde{p}_j^{\mu} \pm f k_{\perp}^{\mu} \\
    & \bar p^{\mu}_j = a_j \tilde{p}_i^{\mu} + b_j \tilde{p}_j^{\mu} \pm (1-f) k_{\perp}^{\mu} \\
    & \bar p^{\mu}_k = a_k \tilde{p}_i^{\mu} + b_k \tilde{p}_j^{\mu} + k_{\perp}^{\mu}\,,
  \end{align}
\end{subequations}
where for each of $i$ and $j$ we use a plus (minus) sign for the
transverse component if that particle is in the initial (final) state.
The coefficients $(a_{i,j}, b_{i,j})$ can be determined as a function of
$a_k$ and $b_k$ by imposing longitudinal momentum conservation plus on-shell
conditions.

The function $f$ in
Eq.~\eqref{eq:general-local-map} determines how transverse recoil is shared
between the two parent legs of the dipole. 
In the case of dipole showers, one assigns a dedicated ``emitter" and ``spectator"
parton, corresponding to setting $f=1(0)$ if $p_i(_j)$ is the
emitter.
The $g(\bar\eta)$ function in Eq.~(\ref{eq:dipole-prob}) determines
the relative weight for one end or other of the dipole to be the
emitter.
Following
Ref.~\cite{Dasgupta:2020fwr} we take
\begin{equation}
    g(\bar{\eta}) = g^\text{dip.}(\bar{\eta}) \equiv 
    \begin{cases} 
        0 &\text{ if } \eta < -1 \\ 
        \frac{15}{16} \left(\frac{\bar{\eta}^5}{5} - \frac{2\bar{\eta}^3}{3} + \bar{\eta} + \frac{8}{15} \right) & \text{ if } -1 < \bar{\eta} < 1 \\
        1 &\text{ if } \bar{\eta} > 1.
    \end{cases}
    \label{eq:geta}
\end{equation}

Conversely, antenna showers do not identify a dedicated emitter and
spectator (i.e.~$f\neq 1, 0$ in general).
Instead, they use a smooth function for $f$, so as to assign an
$\bar{\eta}$-dependent fraction of the transverse momentum to each of the
dipole legs.
We choose to make the function for $f$ coincide with the $g(\bar{\eta})$
that we use in Eq.~(\ref{eq:dipole-prob}), and set both of them as
follows:
\begin{equation}
  f(\bar{\eta}) = g(\bar{\eta}) = g^\text{ant.}(\bar{\eta}) \equiv \frac{e^{\bar{\eta}}}{e^{\bar{\eta}}+e^{-\bar{\eta}}}=\frac{e^{2\bar{\eta}}}{e^{2\bar{\eta}}+1}.
  \label{eq:fantenna}
\end{equation}
For the splitting of a final--final dipole, the above equations are
sufficient and we write the ultimate post-branching momenta as
$p_{i,j,k} = \bar p_{i,j,k}$.
When one or other of the pre-branching momenta is in the initial
state, the resulting $\bar p$ will no longer be aligned with the beam
axis.
Therefore, one needs a Lorentz transformation, $\Lambda^{\mu\nu}$, to
realign it after the splitting.
This transformation consists of a boost and a rotation, but is an
under-constrained system, whose precise form we will discuss later on.
It is applied to all particles in the event:
\begin{equation}
  \label{eq:local-final-boost}
  p^\mu_a = \Lambda^{\mu\nu} \,\bar p_{\nu,a}\quad  \forall a \in i,j,k\,,\qquad
  p^\mu_a = \Lambda^{\mu\nu} \,\tilde p_{\nu,a}\quad  \forall a \notin i,j,k\,.
\end{equation}

\paragraph{Global recoil schemes.}
We explore also global recoil schemes in which the transverse-momentum
imbalance is shared among a subset of the particles in the event.
The global kinematic map for the particles in the dipole reads
\begin{subequations}
  \label{eq:general-global-map}
  \begin{align}
    & \bar p^{\mu}_i = a_i \tilde{p}_i^{\mu}\,,  \\ 
    & \bar p^{\mu}_j = b_j \tilde{p}_j^{\mu}\,,  \\
    & \bar p^{\mu}_k = a_k \tilde{p}_i^{\mu} + b_k \tilde{p}_j^{\mu} + k_{\perp}^{\mu}\,.
  \end{align}
\end{subequations}
The details of which event particles' momenta are subsequently
modified to achieve overall momentum conservation are shower specific.

Recall that we distinguish $k_{\perp}$, a transverse momentum used in
the mapping and defined relative to the parent dipole, and $k_t$, a
transverse momentum defined with respect to the beam directions.
In general, the two do not coincide. 
Furthermore, across both local and global recoil prescriptions, the
boosts and rotations that we perform to realign initial-state
particles and achieve global momentum conservation may alter the
transverse momentum ($k_t$) of the emission relative to the beam
directions. 

To sum up this section, designing a dipole shower for hadronic collisions
involves: (i) the definition of a branching kernel, see
Eq.~\eqref{eq:dipole-prob}, (ii) a choice of ordering variable $v$, (iii) a prescription
on how to partition the dipole, i.e. the definition of $\bar{\eta}$ in
Eqs.~(\ref{eq:dipole-prob}) and \eqref{eq:geta}, and (iv) a recoil scheme, see
Eqs.~\eqref{eq:general-local-map}, \eqref{eq:general-global-map}. The choices
made for each of the four ingredients affect the logarithmic accuracy of the
shower, as was shown in Refs.~\cite{Dasgupta:2018nvj} for final-state showers.

\subsection{A standard transverse-momentum-ordered dipole shower}
\label{sec:dipolekt}
To provide a reference for our discussions of logarithmic accuracy, it
is useful to introduce two concrete realisations of ``standard'' dipole
showers for hadronic collisions, which we generically call ``Dipole-$k_t$".
All modern dipole showers are based on the pioneering ideas set out in
Refs.~\cite{Gustafson:1987rq,Catani:1996vz}.
The specific shower that we use takes the ordering variable and
kinematic maps of the Dire-v1 shower~\cite{Hoche:2015sya} and
partitions the two halves of the dipole according to
Eq.~(\ref{eq:dipole-prob}), with the midpoint between the two halves,
$\bar{\eta} \equiv \bar{\eta}_{\rm dip} =0$, corresponding to zero rapidity in the dipole centre of mass
frame.
The resulting shower shares substantial similarities with the showers
available in all major Monte Carlo event generators, such as
Pythia~\cite{Sjostrand:2004ef,Cabouat:2017rzi},\footnote{Additional
  issues related to colour coherence have recently been raised
  regarding the standard Pythia
  shower~\cite{Cabouat:2017rzi,Jager:2020hkz,Hoche:2021mkv} in the
  context of deep-inelastic scattering and vector-boson fusion, in its
  default option with global recoil for the space-like (initial-state)
  shower. 
  While we do not explore this question here, it would deserve further
  study so as to understand whether it has implications for logarithmic
  accuracy more generally, especially as colour coherence issues can
  in some cases induce problems for leading logarithmic terms.}
Sherpa~\cite{Schumann:2007mg} and
Herwig~\cite{Platzer:2009jq}.
It is also expected to give similar logarithmic structure to the
Vincia shower~\cite{Fischer:2016vfv,Brooks:2020upa}, even though the
latter is antenna rather than dipole based.

All details of the shower are given in App.
\ref{app:shower-coefficients-dipole-kt} and here we limit ourselves to
outlining its main physical features.
Its ordering variable $v$ is transverse-momentum-like.
It is convenient to relate the longitudinal shower variable $\bar{\eta}_\text{dip}$
used in Eq.~(\ref{eq:dipole-prob}) to a collinear momentum fraction
$z$ carried by the emission $p_k$, defined relative to the
pre-branching momentum, $\tilde p_i$, in the final-state case and the
post-branching momentum, $p_i$, in the initial-state case.
The relation reads
\begin{equation} \label{eq:eta-dipolekt}
\bar{\eta} \equiv \bar{\eta}_{\rm dip}  = 
\begin{cases}
    \frac{1}{2} \ln\frac{z^2 \, \tilde{s}_{ij}}{v^2} \,  & \text{(final state),} \\
    \frac{1}{2}\ln\frac{z^2 \tilde{s}_{ij}}{(1-z)^2v^2} & \text{(initial state)}.
\end{cases}
\end{equation}
This makes it manifest that Eq.~\eqref{eq:geta} partitions the dipole
in its rest frame, 
a property common to all widely-used dipole showers, and whose adverse
consequences are expanded upon below.

The kinematic maps depend on the dipole type.
In case both dipole constituents are in the final state (FF), we use a
local recoil scheme, following the kinematic map given by
Eq.~\eqref{eq:general-local-map} with $f=1$.
This choice implies that
all the other particles in the event, including the colour singlet,
are not modified.

For an initial-initial (II) dipole, we use a global recoil scheme
followed by a Lorentz transformation.
The recoil scheme is as given
by Eq.~\eqref{eq:general-global-map} with $b_j=1$, i.e.\ the spectator
momentum is preserved.
All other final-state particles (excluding $p_k$), are then boosted to
achieve momentum conservation. Due to this boost, the transverse
momentum of the colour singlet is modified.
Several choices are possible as concerns the longitudinal component of
the boost.
A common prescription, which we adopt for our Dipole-$k_t$ shower,
is to preserve the longitudinal momentum fraction of the spectator
parton, following the same logic as the use of $b_j=1$ in
Eq.~\eqref{eq:general-global-map}. 
Other options are, for example: (i) preserving the rapidity of the
colour singlet, or (ii) preserving the longitudinal momentum of the
colour singlet.
We believe the specific choice is immaterial in terms of logarithmic
accuracy, at least up to and including NLL, because the difference
between them vanishes in the limit where the branching transverse
momentum is small.

In the initial--final (IF) case two main options have been used in the
literature.
The simplest choice is a fully local map like
Eq.~\eqref{eq:general-local-map} with $f=0$, i.e.\ transverse recoil
is always assigned to the final-state dipole end.
Taking $pp\to Z$ as an example, this has the consequence that after
emission of a first gluon, no further transverse recoil is taken by
the $Z$-boson.
It was appreciated some time ago~\cite{Nagy:2009vg,Platzer:2009jq}
that this is unphysical (for example it is not expected to reproduce
the structure of the Parisi-Petronzio Drell-Yan $p_t$
resummation~\cite{Parisi:1979se}).
Still, such an option is available (or even default) in various public
showers, and it is useful to include such an option in our studies.
We refer to it as \textbf{``Dipole-$k_t$ (local)"}.

Additionally, various authors have explored the possibility of a
global map for IF dipoles~\cite{Platzer:2009jq,Hoche:2015sya}.
We therefore include such a shower in our studies here, named
\textbf{``Dipole-$k_t$ (global)"}.  It is identical to the local
variant with the exception of the IF kinematic map, which is given by
\begin{align}
    \label{eq:if-global-map}
  & \bar p^{\mu}_i = a_i \tilde{p}_i^{\mu} + b_i   \tilde{p}_j^{\mu} +  k_{\perp}^{\mu} \nonumber \\
  & \bar p^{\mu}_j = b_j \tilde{p}_j^{\mu}\nonumber \\
  & \bar p^{\mu}_k = a_k \tilde{p}_i^{\mu} + b_k \tilde{p}_j^{\mu} + k_{\perp}^{\mu}\, .
\end{align}
Note that this global map differs from the generic one given by
Eq.~\eqref{eq:general-global-map} since the initial-state leg acquires
a $k_{\perp}$ component.
Momentum conservation is then achieved by performing a boost and a
rotation on the entire event.
The
longitudinal degree of freedom of the boost is constrained by imposing
that the momentum of the other incoming parton that does not
participate in the splitting is unchanged. This Lorentz transformation
is provided in App.~\ref{app:shower-coefficients-dipole-kt}. Lastly,
the FI mapping follows a local recoil scheme as in
Eq.~\eqref{eq:general-local-map} with $f=1$.%
\logbook{Discussion on 2022-03-02 ARGGS}{%
  We believe that Pythia's
  global recoil will be logarithmically similar: after splitting a
  dipole in the centre-of-mass, for the initial-state part we
  understand it shares the transverse recoil across all final-state
  particles, and when all particles are soft, that corresponding to
  giving it all to the $Z$ boson.
  But we have not tested this.
  Still, see \cite{Hoche:2021mkv} for concerns about how the IF
  part of the dipole is split ($\pi/2$ in the lab frame), which
  affects DIS and VBF processes and our understanding of this is not
  currently sufficient to exclude a potential logarithmic impact that
  may be worse than the NLL effects we are discussing here.
}

\section{Methodology for fixed-order tests}
\label{sec:fo-test}
%

Eq.~\eqref{eq:dipole-prob} provides the correct singular limit for a
single emission.
Within the context of our NLL shower requirements, however, we require
a shower to be able to reproduce the squared matrix element for
producing multiple emissions whenever each emission is well-separated
from all others in a Lund diagram~\cite{Andersson:1988gp}.
Specifically, for a separation by some increasing distance $d$ (e.g.\
the sum of the rapidity and of the $\ln k_t$ separations), we expect
deviations from the true matrix element to vanish as a power of
$e^{-|d|}$.
One of the crucial characteristics of the full matrix elements is that
when there is a large separation in rapidity between multiple
soft-collinear emissions, all emitted from the Born partons, the
squared matrix element and phase space for $n$ emissions can be
written as a product of $n$ independent emission factors,
schematically
\begin{equation}
  \label{eq:indep-emission}
  d\mathcal{P}_{\text{Born} + n} = d\mathcal{P}_{\text{Born}}
  \times
  \frac1{n!} \prod_{j=1}^n \frac{2C \as}{\pi} \frac{dk_{tj}}{k_{tj}}
  \frac{d\phi_j}{2\pi} d\eta_j\,.
\end{equation}
Most showers effectively implement that structure of the matrix
elements, generating momenta one at a time.
As was observed in Ref.~\cite{Dasgupta:2018nvj}, such a procedure only
reproduces the matrix element of Eq.~(\ref{eq:indep-emission}) if each
emission leaves all prior emissions' kinematics unchanged.
To understand this requirement, consider the following sequence of
steps.
Firstly, emission $1$ is generated with a certain momentum
$\tilde k_{1}$ using a matrix element corresponding to that
$\tilde k_{1}$.
Then as a second step particle $2$ is emitted, taking significant
recoil from particle $1$ such that the new momentum for particle $1$,
$k_{1}$, differs substantially from $\tilde k_{1}$.
The result will be a configuration where the first emission has
momentum $k_{1}$, but generated with a matrix element corresponding
to $\tilde k_{1}$, i.e.\ that does not match the final kinematics.
When this phenomenon occurs over a logarithmically enhanced region, it
results in a failure to reproduce certain classes of NLL terms.
This type of analysis leads to the PanScales condition that in order
for a shower to correctly reproduce the matrix element, a given
emission that is well separated (e.g.\ in rapidity) from other prior emissions
should not alter the kinematics of those prior
emissions. 
It is mainly this condition that we will be testing here and in
Section~\ref{sec:panshowers}. 

\begin{figure}[tb]
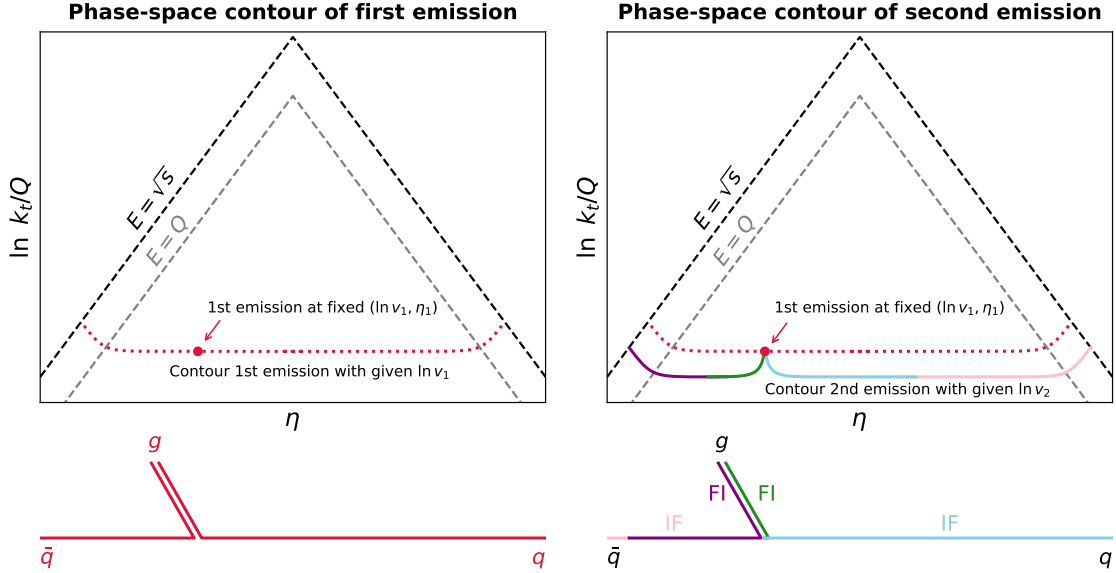

    \includegraphics[page=3,width=0.49\textwidth]{plots/contour_plots.pdf}
    \includegraphics[page=4,width=0.49\textwidth]{plots/contour_plots.pdf}
    \caption{
          Illustration of the hadron-collider primary Lund plane,
          where the rapidity and transverse-momentum coordinates are
          defined with respect to the hadron beam directions.
          The left-hand plot shows the contour for the first emission
          from a Dipole-$k_t$ shower for a fixed value $v_1$ of the ordering
          variable (local and global are identical here).
          The red point indicates the specific value of $\eta_1$ that
          is used in subsequent plots when adding a second emission.
          The right-hand plot additionally shows the contour for such a second emission
          at a fixed value of $v_2 < v_1$ .
          The contour is colour-coded to roughly represent the
          partitioning of the two dipoles into IF and FI regions, as
          illustrated also below the plot.
          Note that the momentum of the second emission is shown as
          projected onto the primary Lund plane, to avoid having to represent
          the 3-dimensional secondary Lund plane.
          See text for further details.
        }
    \label{fig:fixed-order-test-cartoons}
\end{figure}

To verify this PanScales condition, it is useful to represent emission
phase space on the Lund plane, i.e.\ in terms
of the rapidity $\eta$ and transverse momentum $k_t$ of the emission,
as shown in Fig.~\ref{fig:fixed-order-test-cartoons}.
Recall that $\eta$ and $k_t$ (as used in
Fig.~\ref{fig:fixed-order-test-cartoons}) are always defined with
respect to the beam directions, while $\bar\eta$ and $k_\perp$ as
used in Eq.~(\ref{eq:dipole-prob}) and in the various kinematic maps
are defined with respect to the dipole that is branching.
Here, for illustration, we choose the Dipole-$k_t$ shower of
section~\ref{sec:dipolekt}, which evolves following horizontal
contours in the bulk of the Lund plane.
We consider the process $q\bar{q}\to Z$, where $Q$ is the $Z$-boson
mass, $y_Z = 0$ is its rapidity and the proton--proton centre-of-mass
energy is denoted by $\sqrt{s}$.
The dashed black line illustrates the phase space boundary
corresponding to the radiated parton having energy $\sqrt{s}$, while
the grey one delineates an inner Lund plane where the maximum energy
of the radiated parton is $Q$ (the grey and black boundaries would
coincide in the $e^+e^-$ case).
Had we chosen $y_Z\neq 0$, this would have led to a relative shift of
the two planes.
The phase space in between the two Lund planes corresponds to region
where the PDF factors in Eq.~(\ref{eq:p-initial-state}) can play a
significant role in the branching probabilities.\footnote{ When
  $\sqrt{s} \gg Q$, there can be additional logarithmic enhancements
  associated with so-called ``small-$x$'' terms $\as^n \ln^m s/Q^2$
  for both $t$-channel gluon~\cite{Kuraev:1977fs,Balitsky:1978ic}
  (BFKL) and quark~\cite{Kirschner:1983di} exchange.
  The inclusion of such terms in the context of a parton shower is a
  theoretically interesting question that has been explored by the
  CASCADE~\cite{Jung:2010si,Baranov:2021uol} and
  HEJ~\cite{Andersen:2009nu,Andersen:2009he,Andersen:2011hs} groups,
   but we do not consider these terms in our work here.}

For our tests, we generate a first gluon emission from the (II)
$q\bar q$ dipole at a fixed value of the evolution variable $\ln v_1$.
Scanning over $\eta_1$ yields the dotted red contour shown in
Fig.~\ref{fig:fixed-order-test-cartoons} left.
Given an emitted momentum fraction $z$ (relative to the post-branching
incoming momentum), the transverse momentum of the emission is given by
\begin{equation}
  \label{eq:kt-II-dipolekt}
  |k_{t,1}| = \frac{v_1}{\sqrt{1-z}} + \order{v_1^2/\sqrt{s_{ij}}}\,.
\end{equation}
When the emission is soft $z \ll 1$, this means that the emitted
transverse momentum coincides with $v$, while in the hard collinear
regions the emitted transverse momentum curve bends upwards.
    
Going forwards, to keep the discussion relatively simple, we constrain
the kinematics of the first emission, by fixing its rapidity to be in
the soft-collinear region (here, $\eta_1 = -10$), as illustrated by
the red dot. 
Note that we have also carried out tests with large-angle soft and
hard-collinear choices for $\eta_1$, and we will highlight any
relevant issues as they occur.

Next (Fig.~\ref{fig:fixed-order-test-cartoons} right), we consider a
second gluon being emitted from either of the $\bar{q}g$ or $qg$
dipoles at a commensurate scale, i.e.
$\ln v_2 = \ln v_1 - \delta \ln v$, with $\delta \ln v$ a number of
order $1$.
For a fixed value of the shower evolution variable $\ln v_2$, we show
the emission contour, scanning over $\eta_2$.
We choose a specific value of $\varphi_2$  such that the second emission
is in the same plane as the first emission, and that it points in the same
direction when far away in rapidity, though the issues that we will
see below are relevant for essentially all values of $\varphi_2$.
The 2nd-emission contour in Fig.~\ref{fig:fixed-order-test-cartoons}
follows the shape of the 1st-emission contour except at rapidities
close to the first emission, where the upwards bending reflects the
fact that it takes momentum from that 1st emission.
The fact that elsewhere the 2nd emission contours runs parallel to the
first, at a distance of order $\delta \ln v$, ensures that the shower
properly fills the double-logarithmic phase space.
The contour is colour-coded to reflect the specific part (IF or FI) of
the given dipole from which the 2nd emission is dominantly emitted,
reflecting the use of the $g(\bar{\eta}_{\rm dip} )$ function in
Eq.~(\ref{eq:dipole-prob}) to partition the dipole and the specific
definition of $\bar{\eta}_{\rm dip} $ in Eq.~(\ref{eq:eta-dipolekt}) for the
Dipole-$k_t$ showers.

\begin{figure}[tb]
    \centering
    \includegraphics[page=1, width=0.49\textwidth]{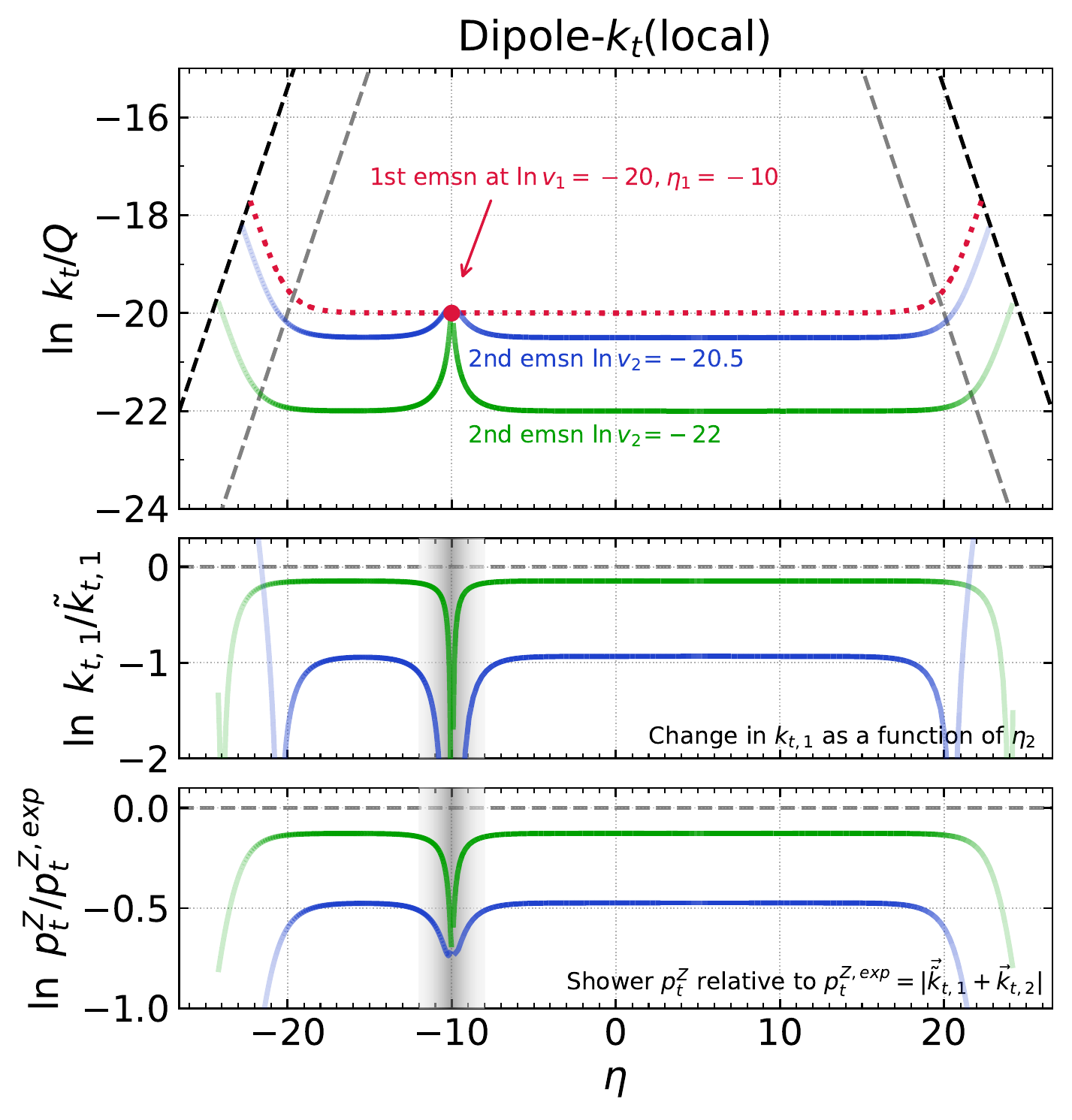}
    \includegraphics[page=7, width=0.49\textwidth]{plots/contour_plots.pdf}
    \caption{Double-emission contours for \emph{Dipole-$k_t$ (local)} (left) and
      \emph{Dipole-$k_t$ (global)} (right).
      The phase-space contours are shown as a function
    of $\ln k_{t}/Q$ and $\eta$. A red dot indicates the kinematics of the first
    emission which has $\eta_1\approx-10$ and $\ln v_1/Q=-20$ (in the
    plot labels, values of $v_i$ are always expressed in units of $Q$).
    For illustrative
    purposes, the contour of the first emission is shown with a red dotted line,
    whereas that of the second emission at $\ln v_2/Q=-20.5$ ($\ln v_2/Q=-22$)
    is drawn as a blue (green) solid line.
    The colour shading of the lines indicates the
    branching probability.
    In the middle panels we show the logarithm of the ratio between the
    transverse momentum of the first emission before ($k_{t,1}$) and after the
    second emission took place ($k'_{t,1}$), which is expected to be
    zero (dashed line) except when the two emissions are close in
    rapidity (shaded vertical band). 
    The bottom panels show the ratio of the observed to the expected
    $Z$ transverse momentum (where again, the expectation is valid
    when the two emissions are well separated in rapidity, i.e.\
    well outside the grey vertical band).  }
    \label{fig:contour-dipole-kt-soft}
\end{figure}

Next we turn to Fig.~\ref{fig:contour-dipole-kt-soft}, concentrating first
on the left-hand plot, which is for the variant of Dipole-$k_t$ with
dipole-local recoil in IF dipoles. 
The upper panel is a zoomed-in version of
Fig.~\ref{fig:fixed-order-test-cartoons} (right), but now showing
contours for two different values of $\ln v_2$.
The colour coding has changed and the intensity of the colour reflects the
emission probability.
The vanishing of the parton distribution function at large $x$ causes
the contours to fade in the hard collinear regions (the specific
details of the PDF are not critical here).

The middle panel shows the logarithmic change in the first emission's
momentum, which we write as $\ln k_{t,1}/\ktilde_{t,1}$.
Recall our discussion at the beginning of this subsection where we
emphasised that when the second emission is well separated in rapidity
from the first (i.e.\ everywhere except the grey-shaded region), the
first emission momentum should not change, otherwise we break the
independent-emission picture of Eq.~(\ref{eq:indep-emission}).
This condition is clearly violated in
Fig.~\ref{fig:contour-dipole-kt-soft} (left).
The reason is that $\vec{\tilde k}_{t,1}$ becomes
$\vec{k}_{t,1} = \vec{\tilde k}_{t,1}-\vec{k}_{t,2}$ after the branching, as
can be deduced from the kinematic map in
Eq.~\eqref{eq:general-local-map}.
This is worse than the analogous situation for final-state showers
discussed in Ref.~\cite{Dasgupta:2018nvj}, because the
transverse-momentum recoil is \emph{always} taken from the previous
emission, rather than just half of the time.

In the case of initial-state branching, an alternative way of
visualising the issue is given in the bottom panel, which shows the
logarithmic deviation of the $Z$-boson transverse momentum from the
expectation $\vec p_t^{\,Z} = -\vec {\tilde k}_{t,1} - \vec k_{t,2}$.
There is no location in rapidity where $k_{t,1}$ or the $Z$ transverse
momentum agrees with the physical expectation (which, again, is
fundamental to obtaining NLL accuracy).
Note that the NLL impact at $\as^2 L^2$ ($L = \ln m_Z/p_{t}^Z$, $Q
\equiv m_Z$) is
zero, owing to azimuthal averaging, as was found for the vector $p_t$ sum in
the final-state case, Table~1 of Ref.~\cite{Dasgupta:2018nvj}.
That same analysis identified a non-vanishing discrepancy from order
$\as^3 L^3$ onwards.
A further point to note is that in the Dipole-$k_t$ (local) shower,
the $Z$ transverse momentum is always
given by the transverse momentum of the first emission at the moment
of its creation.
This means that the mechanism for obtaining low $Z$ transverse momenta
identified long ago by Parisi and Petronzio~\cite{Parisi:1979se},
i.e.\ the vector 
cancellation between the momenta of subsequent emissions, is missing
with local IF recoil.
We will further explore the consequences of this in our companion
paper~\cite{vanBeekveld:2022ukn}.%
\footnote{For gluon--fusion Higgs production,
  which involves two II dipoles, there can be independent emission
  from each of the II dipoles, and so the vector cancellation can
  still occur. }

The right-hand plot shows analogous results for the Dipole-$k_t$
shower in its variant where IF dipoles use global recoil.
The main difference is that the unphysical shift in $\ln k_{t,1}$ and
in $\ln p_{t}^Z$ is restricted to the rapidity region
\begin{equation}
  \label{eq:dipole-kt-global-bad-rap-region}
  \frac{1}{2} \left(\eta_1+\ln \frac{k_{t,1}}{Q}\right) <
  \eta_2 <  \frac{1}{2} \left(\eta_1-\ln \frac{k_{t,1}}{Q}\right),
\end{equation}
reflecting the fact that, inside this region, dipole-local recoil is
used, while outside the region event-wide recoil is used (which
implies that the transverse recoil is mainly assigned to the $Z$
boson). 
Note that the rapidity extent,
Eq.~(\ref{eq:dipole-kt-global-bad-rap-region}), $\sim \ln Q/k_{t,1}$,
of the region in which transverse recoil is incorrectly assigned is
the same as for the pure final-state dipole showers discussed in
Ref.~\cite{Dasgupta:2018nvj}.
This suggests that the deviations from NLL accuracy will be the same
as for those final-state showers.
Concerning the vector cancellation of recoil for the $Z$ transverse
momentum, this can now occur in at least some of the phase space.

Overall, we have shown that the generic transverse-momentum ordered
dipole showers of Section~\ref{sec:dipolekt} do not pass the
fixed-order requirements needed to achieve NLL accuracy.
In the next section, we propose a new family of
showers that solve the observed issues.

\section{The PanScales showers}
\label{sec:panshowers}

As we have seen, the issues discussed in Section~\ref{sec:fo-test} are
qualitatively the same as those identified in final-state dipole
showers in Ref.~\cite{Dasgupta:2018nvj}.
As such it is natural to explore solutions similar to the PanScales
$e^+e^-$ showers of Ref.~\cite{Dasgupta:2020fwr}.
There are, however certain important differences in initial-state
showers.
One concerns the choice of the conserved quantity during the shower:
in the $e^+e^-$ context it was essential to preserve the partonic
centre of mass momentum; in contrast, an initial-state emission must
induce a transverse recoil in the hard-system momentum in order to be
consistent with momentum conservation (and, for multiple initial-state
emissions, with Parisi-Petronzio resummation for the hard-system
transverse momentum~\cite{Parisi:1979se}).
Another difference relates to the fact that with incoming beams an emission can
be significantly more energetic than any of the pre-existing partons
in the event.
We will address these issues below, as they arise.

\subsection{Aspects common to all showers}

The PanScales showers need a reference momentum $Q^\mu$, which defines
a centre-of-mass frame.
We set it equal to the four-momentum ($p_x,p_y,p_z,E$) of the hard system prior to
showering,
\begin{equation}
    Q^\mu =  m_X (0,0, \sinh y_X, \cosh y_X)\,,
\end{equation}     
where
$X$ is a colour-singlet hard system (for example a $Z$ or a Higgs
boson), $y_X$ is the hard-system rapidity, and $m_X^2\equiv Q^2 =
(\tilde{p}_a+\tilde{p}_b)^2$, where ($\tilde{p}_a, \tilde{p}_b$) are the initial
four-momenta of the incoming partons.
We will make the choice to keep $Q^\mu$
fixed during the shower
evolution, even as the momentum of the hard system evolves, for
example by acquiring a transverse momentum recoil.\footnote{The choice
  of $Q^\mu$ for processes where the hard system contains coloured
  particles, like $Z+{\text{jet}}$ production, will be addressed in
  future work, as will the extension to the deep-inelastic and vector-boson-fusion
  processes.}
As in the final-state PanScales showers, we will consider a family of
ordering variables parametrised by a variable $0\le \betaps < 1$.
Our ordering variable will be labelled $v$.
We will design our kinematic map such that, in a frame where $Q^\mu$
is at rest, for a soft-collinear emission at an angle $\theta$ and
with a transverse momentum $k_\perp$ relative to the emitter, we will
have
\begin{equation}
  \label{eq:v-soft-collinear}
  v \simeq k_\perp (\theta/2)^{\betaps}\,.
\end{equation}
Using an auxiliary longitudinal variable $\bar\eta_Q$, and a
transverse momentum scale for emissions $\kappa_\perp$, this will be
achieved by adopting the following definitions
\begin{equation}
\label{eq:v-def}
\kappa_\perp \equiv \rho v e^{\betaps|\bar\eta_Q|} \,,
\end{equation}
with 
\begin{equation}
  \rho =\left(\frac{\tilde s_i\tilde s_j}{\tilde
      s_{ij}Q^2}\right)^{\betaps/2}\, ,
  \quad
  \tilde{s}_{i} = 2 \tilde p_{i} {\cdot} Q\,,\quad
  \tilde{s}_{j} = 2 \tilde p_{j} {\cdot} Q\,,\quad
  \tilde{s}_{ij} = 2 \tilde p_{i} {\cdot} \tilde p_{j}\,.
\end{equation}
Setting $\betaps = 0$ corresponds to a
transverse-momentum ordered shower.
The choice $\betaps=1$ would result in a time-ordered shower, but as
we shall explain below for the showers that we consider, that choice
is not consistent with NLL accuracy.

Aside from the parametric form of the evolution variable $v$, the main novelty of
the PanScales showers is the definition of an alternative pseudorapidity-like
variable $\bar\eta_Q$ that enters the dipole partitioning function
$g(\bar\eta_Q)$ in Eq.~\eqref{eq:geta}.
It represents the rapidity of the
emission $p_k$ with respect to the parent dipole, in the frame where
the reference vector $Q^\mu$ is at rest, i.e. 
\begin{equation}
\label{eq:etabar-def}
\bar\eta_Q = \frac{1}{2}\ln\frac{p_k\cdot\tilde p_j}{p_k\cdot\tilde p_i} - \frac{1}{2}\ln\frac{\tilde s_j}{\tilde s_i}.
\end{equation}
That is, $\bar\eta_Q\!=\!0$ corresponds to a direction equidistant to $\tilde p_i$
and $\tilde p_j$ in the rest frame of $Q^{\mu}$.
This is an important difference when
compared to the standard dipole showers which, as explained in the previous
section, split the dipole in the dipole rest frame (see Eq.~\eqref{eq:eta-dipolekt}).

When formulating the kinematic maps, we find it useful to define
intermediate variables
\begin{align}
    \label{eq:definition-alphak-betak}
    \alpha_k &\equiv \sqrt{\frac{\tilde s_j}{\tilde s_{ij}\tilde s_i}}\kappa_\perp e^{\bar\eta_Q}\,, &
    \beta_k  &\equiv \sqrt{\frac{\tilde s_i}{\tilde s_{ij}\tilde s_j}}\kappa_\perp e^{-\bar\eta_Q}\,.
\end{align}
The specific relation between $\alpha_k$, $\beta_k$, and the $a_k$,
$b_k$ as used in Sudakov decomposition for the emitted momentum
Eq.~\eqref{eq:sudakov-decom} will depend on the shower.
One common property for all our showers is that for emissions that are
soft and collinear, $a_k = \alpha_k$ and $b_k = \beta_k$ (from which
one can verify that Eq.~(\ref{eq:v-soft-collinear}) is reproduced).
A further common property is that for final-state branchings, the $z$
values used in the splitting functions, Eq.~(\ref{eq:dipole-prob}),
are given by
\begin{equation}
  \label{eq:z-final-state}
  z_i = \alpha_k,\qquad z_j = \beta_k,\qquad\qquad\qquad\quad\text{(final state),}
\end{equation}
while for initial-state branchings they are
\begin{equation}
  \label{eq:z-initial-state}
  z_i = \frac{\alpha_k}{1+\alpha_k},
  \qquad
  z_j = \frac{\beta_k}{1 + \beta_k},
  \qquad\quad\text{(initial state).}
\end{equation}
Note that for initial-state branchings, $\alpha_k$ and $\beta_k$ can
grow larger than $1$. Uniform generation of the $\bar\eta_Q$
variable ensures logarithmic sampling of both small $z$ and small
$1-z$.

In what follows, we present two recoil schemes and show their fixed-order behaviour. 

\subsection{PanGlobal}
\label{sec:panglobal}
%

The hadron-collider PanGlobal shower is an antenna shower.
For the splitting probability, Eq.~(\ref{eq:dipole-prob}), it uses
$g(\bar\eta_Q)=g^\text{ant.}(\bar\eta_Q)$ (cf.\ Eq.~(\ref{eq:fantenna})).
Its kinematic map can be viewed as follows:
\begin{enumerate}
\item
  \label{item:pg:dipole}
  Apply the ``global'' dipole map of
  Eq.~(\ref{eq:general-global-map}) with $a_k = \alpha_k$ and $b_k =
  \beta_k$ as defined in Eq.~(\ref{eq:definition-alphak-betak}).
\item
  \label{item:pg:transBoost}
  The preceding step breaks momentum conservation for the event as
  a whole.
  First the component that is transverse to the beams is restored by applying
  a boost to the hard system such that the hard system
  rapidity remains unchanged and that the sum of the transverse
  momenta (with respect to the beams) of the boosted hard system and
  all final-state particles adds up to zero.
\item
  \label{item:pg:ISR}
  To restore conservation of the components of momentum that are
  longitudinal with respect to the beams, evaluate the sum of the
  light-cone ($p_+ = E + p_z$ and $p_{-} = E - p_z$) momenta of all
  final-state particles including the boosted hard system.
  Set the momentum of the incoming parton on side $a$ to be $p_+/2$ and
  that on side $b$ to be $p_-/2$.
\end{enumerate}
Note that step \ref{item:pg:ISR} affects the momenta of the incoming
partons regardless of whether the dipole is II, IF, FI or FF.
In cases where one end of the dipole is in the initial state, it
causes the incoming momentum to differ from that of the global map in
Eq.~(\ref{eq:general-global-map}).
Detailed equations for all steps are given in
Appendix~\ref{app:panglobal-map}.

As compared to the $e^+e^-$ PanGlobal shower~\cite{Dasgupta:2020fwr}
there are both points in common and differences.
The use of the global map, Eq.~(\ref{eq:general-global-map}), is for
example very similar.
However the way in which we restore momentum conservation after that
map is different: in particular the $e^+e^-$ shower balances the
momentum across all final-state particles, while the hadron-collider
shower defined above leaves all final-state particles untouched
(unless they belong to the hard system) and relies on adjusting the
incoming particle momenta to conserve energy.\footnote{We also
  explored options in which all final-state particles get boosted.
  However, such options are delicate.
  Specifically, a transverse boost that modifies the hard-system
  transverse momentum by an amount of order $k_t$ also modifies any
  energetic but collinear initial-state emissions by a comparable
  amount.
  This introduces long-distance correlations between soft-collinear
  and hard-collinear initial-state emissions, in violation of the
  the PanScales conditions outlined in Section~\ref{sec:fo-test}.
  Such problems have also been commented upon in the context of the
  Deductor work, which introduces a specific Lorentz
  transformation to work around the issue~\cite{Nagy:2009vg}.
}

\subsubsection{Tests for $\betaps = 0 $ and $1/2$}

In Fig.~\ref{fig:contour-panglobal-working}, we perform the same
fixed-order analysis as in Fig.~\ref{fig:contour-dipole-kt-soft} for
the PanGlobal shower using $\betaps = 0$ (left) and $\betaps = 1/2$
(right).
Let us first consider the situation where the second emission is far in
rapidity from the first.
Step~\ref{item:pg:dipole} of the PanGlobal recoil scheme leaves an
overall momentum imbalance in the direction transverse to the (IF)
dipole.
It can be shown that a unit vector transverse to any IF dipole
($\hat n_{1,2}$ in Eq.~(\ref{eq:kperp-decomposition})) has a unit component
transverse to the beam.
\logbook{}{It's obvious for the out-of-plane transverse vector, less
  so for the in-plane one. When the first emission is at rapidity $y$,
  and assuming its transverse component is alone the $x$ direction,
  the explicit four-vector for the in-plane $\hat n$ is $\hat n =
  (1,0,e^y,e^y)$.}%
As a result, step~\ref{item:pg:transBoost} assigns the shower
transverse momentum to the hard system, i.e.\ the $Z$ boson, as is
physically correct, thus reproducing the pattern needed for NLL
accuracy.
When the 2nd emission is close in rapidity to the first, it is
arguably less physically correct to take the transverse recoil from
the $Z$ boson.
However, the assignment of transverse recoil only has a significant
impact when $k_{\perp,2} \sim k_{\perp,1}$, and the region of
commensurate rapidity and commensurate transverse momentum only
affects terms at NNLL accuracy, the correct treatment of which would
in any case require the inclusion of the full double-soft matrix
element.\footnote{Note that $k_{t,1}$ is also affected by recoil that
  is longitudinal with respect to the dipole when emission $2$ is
  collinear to $1$.
  This is not the case for $p_t^Z$.
}
A similar discussion can be extended to subsequent emissions.
%
Our conclusion, therefore, is that the PanGlobal showers with $\beta=0$
and $\beta=1/2$ satisfy the fixed-order NLL accuracy requirement.

\subsubsection{Discussion of $\betaps=1$ case (time ordering)}

\begin{figure}[tb]
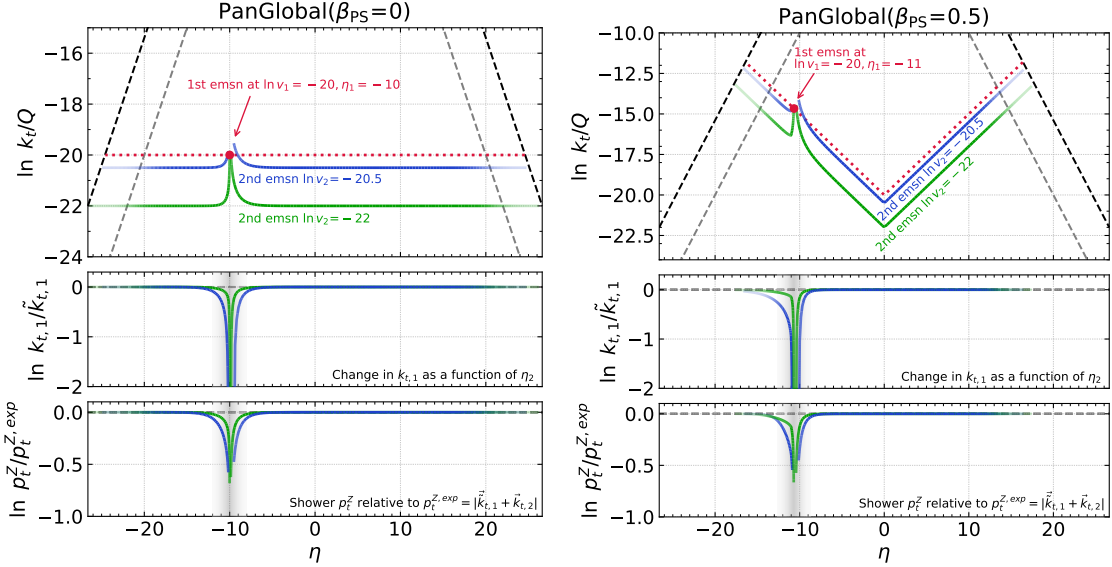

    \centering
    \includegraphics[page=10, width=0.49\textwidth]{plots/contour_plots.pdf}
    \includegraphics[page=15, width=0.49\textwidth]{plots/contour_plots.pdf}
    \caption{Same as Fig.~\ref{fig:contour-dipole-kt-soft}, but for
      PanGlobal with $\betaps = 0$ (left) and $\betaps = 0.5$ (right).}
    \label{fig:contour-panglobal-working}
\end{figure}

We close our discussion of the PanGlobal shower with an explanation of
why that shower requires $\betaps < 1$ and an illustration of the
issues that arise with $\betaps=1$.
The choice of $\betaps=1$ is of interest because, physically, it
corresponds roughly to a time ordering.
This can be relevant, for example, in a heavy-ion context where one
may wish to relate individual steps of the shower with the
time-dependent evolution of a quark--gluon plasma.
That $\betaps=1$ corresponds roughly to time-ordering can be seen as
follows.
Firstly, consider a soft large-angle emission with transverse momentum
$k_\perp$ with respect to the parent dipole.
The uncertainty principle tells us that the formation time is roughly
$1/k_\perp$.
Next, observe that a soft-collinear emission, with energy $E$ and
transverse momentum $k_\perp$ is equivalent to a soft
emission that has been boosted along the parent dipole direction.
The boost factor is roughly $E/k_\perp$ and so the formation time
acquires a Lorentz dilation by that same factor, giving a net
formation time of order $t \sim E/k_\perp^2 \sim 1/(k_\perp \theta)$,
where $\theta$ is the emission angle.
Inspecting Eq.~(\ref{eq:v-soft-collinear}), one sees that with
$\betaps=1$, $t \sim 1/v$.

\begin{figure}[tb]
    \centering
    \includegraphics[page=17, width=0.49\textwidth]{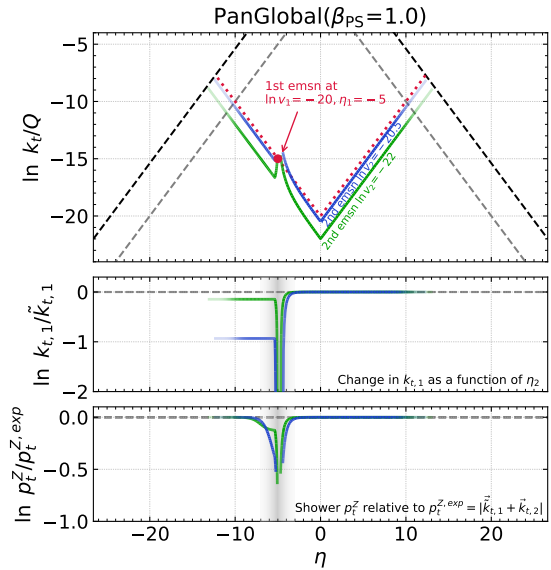}
    \caption{Same as Fig.~\ref{fig:contour-dipole-kt-soft}, but for
      PanGlobal with $\betaps = 1$ and $\eta_1 = -5$. }
    \label{fig:contour-panglobal-beta1}
\end{figure}

Fig.~\ref{fig:contour-panglobal-beta1} repeats the analysis of
Fig.~\ref{fig:contour-panglobal-working} for PanGlobal with
the ``time-ordered'' $\betaps = 1$ choice.
Contrary to the $\betaps = 0$ and $\betaps = 1/2$ cases, we see that
the second emission modifies the transverse momentum of the first when
that second emission is in the same hemisphere and at more forward
rapidities than the first.
One way of understanding why $\betaps=1$ might cause problems is to
observe that, for $v_2 \simeq v_1$, the second emission is created
close to the kinematic boundary of its parent dipole (recall that
kinematic boundaries have slope $\pm1$ in the soft-collinear region of
the Lund plane).
Close to that kinematic boundary, it is almost inevitable that the
parent dipole momenta will be substantially modified.

To obtain a more analytic understanding of our observations, let us
consider a sequence of two soft-collinear emissions with commensurate
$\ln v$ values, working with a general value of $\betaps$.
We start with an $a{-}b$ initial state, where particle $a$'s momentum,
$p_a$, has a negative $z$ component.
Since both emissions will be soft and collinear, the momenta of $a$ and
$b$ will not be affected by those emissions and we need not worry about
distinctions between their pre- and post-branching momenta (i.e.\ we
will never write tildes over $p_a$ and $p_b$).
We place a first soft-collinear emission at negative physical
rapidity, which, given $a$'s negative $z$ momentum, is obtained by
choosing a large positive $\bar\eta_{Q,1}$,
\begin{equation}
  \tilde p_{1}^{\mu} = a_{1} p_a^{\mu} + b_{1} p_b^{\mu} + k_{\perp,1}^{\mu}\,,
\end{equation}
where, using Eqs.~(\ref{eq:v-def})--(\ref{eq:definition-alphak-betak}),
$a_1 = \alpha_1$, $b_1 = \beta_1$, and $s_{ab}=s_a = s_b = Q^2$, we have
\begin{equation}
  \label{eq:abk1-betaps1-analysis}
  a_{1} = \frac{v_1}{Q}{\rm e}^{(\betaps+1) \bar\eta_{Q,1}}\,, \qquad 
  b_{1} = \frac{v_1}{Q}{\rm e}^{(\betaps-1) \bar\eta_{Q,1}}\,,
  \qquad
  (\bar\eta_{Q,1} > 0)\,.
\end{equation}
Note that we have written the momentum of gluon $1$ as
$\tilde p_{1}^{\mu}$, i.e.\ with a tilde, in anticipation of a second
soft-collinear emission to come.

For that second emission, we first consider the case where it comes
from the $a-1$ IF dipole, with $v_2$ only moderately smaller than
$v_1$, but a large positive value of $\bar{\eta}_{Q,2}$.
The latter choice causes emission $2$ to be substantially more
forward than emission $1$ (i.e.\ at a more negative physical rapidity).
The global map, Eq.~(\ref{eq:general-global-map}), with $i=a$, $j=1$
and $k=2$ for this IF dipole, results in a change in the momentum of
particle $1$,
\begin{equation}
  p_{1}^{\mu} = (1-b_{2}) \tilde p_1^{\mu}\,,\qquad
  b_{2} =
  \frac{{s}_a}{{s}_{1a}Q}\left(\frac{{s}_a{s}_1}{{s}_{1a}Q^2}\right)^{\!\!\frac{\betaps-1}{2}}
  \!\!v_2\,
  {\rm e}^{(\betaps-1)\bar\eta_{Q,2}}\,,
  \qquad
  (\bar\eta_{Q,2} > 0)\,.
\end{equation}
Using 
\begin{equation}
  s_a = Q^2,
  \qquad
  s_1 = (a_{1} + b_{1})Q^2,
  \qquad
  s_{1a} = b_{1} Q^2\,.
\end{equation}
Exploiting our assumption of a large value of $\bar\eta_{Q,1}$ so
that $b_{1} \ll a_{1}$, we have
\begin{equation}
  b_{2} =
  \frac{1}{b_{1}}
  \left(\frac{a_{1}}{b_{1}}\right)^{\!\!\frac{\betaps-1}{2}}
  \frac{v_2}{Q}
  {\rm e}^{(\betaps-1)\bar\eta_{Q,2}}
  = \frac{v_2}{v_1} {\rm e}^{(\betaps-1)\bar\eta_{Q,2}}\,.
\end{equation}
We observe that for $\betaps=1$, $b_{2}$ is just given by $v_2/v_1$,
independently of $\bar\eta_{Q,2}$, i.e.\ 
\begin{equation}
  \label{eq:pk1-rescale}
  p_{1}^{\mu} = \left(1-\frac{v_2}{v_1}\right) \tilde p_1^{\mu}\,,
  \qquad
  (\betaps = 1)\,,
\end{equation}
no matter how far in rapidity emission $2$ is from emission $1$.
Note that this scaling of the momentum gluon $1$ changes its
transverse momentum as measured with relative to the beam.
This is a long-distance side effect of emission $2$ on the momentum of
emission $1$, which prevents the shower from being NLL
accurate.\footnote{Ref.~\cite{Nagy:2020dvz} commented that a standard
  final-state dipole-local map with so-called $\Lambda$ ordering, i.e.\
  effectively time ordering, results in NLL violations for the thrust
  observable.
  We believe that the issue observed there is effectively identical
  to that discussed here.
} 
The change in transverse momenta of the first emission after the
second splitting (Eq.~\eqref{eq:pk1-rescale}) is precisely what is
observed in the ratio plot of Fig.~\ref{fig:contour-panglobal-beta1}.
One can carry out a similar analysis for an emission from the $b{-}2$
dipole, and the findings are consistent with
Fig.~\ref{fig:contour-panglobal-beta1}.
The main further point that one observes is that if emission $1$ is at
large angles, then it is affected by a second soft-collinear emission
regardless of which hemisphere that second emission is in.

To conclude, we have shown that using $\betaps = 1$ for our ordering
variable, we introduce a longitudinal rescaling of the first-emitted
parton after emitting a second one.
This rescaling alters the transverse momentum of the first-emitted
parton in the event frame, even when the two emissions are well
separated in rapidity.
Therefore, the PanGlobal shower with $\betaps = 1$ does not
satisfy our fixed-order logarithmic accuracy requirement.
It is reasonable to suppose that this behaviour of $\betaps = 1$
showers can be cured by assigning the recoil only to the side of the
dipole that is emitting.\footnote{The maps of the showers proposed in
  Refs.~\cite{Nagy:2009vg} and \cite{Forshaw:2020wrq} indeed satisfy this
  requirement, though only the former has explored time (``$\Lambda$'')
  ordering.}
This would require another form of the boost.
We will leave further modifications and tests of the $\betaps = 1$
showers for future work.

It is useful to be aware that $\betaps=1$ ordering differs from
virtuality ordering in the sense of Ref.~\cite{Hartgring:2013jma}.
Working in a frame where the fixed reference vector $Q^\mu$ is at rest, for a soft-collinear
emission $k$ with energy $E_k$ and angle $\theta$ from a parent with
energy $E_i$, we have $v_{\betaps=1} \simeq E_k \theta^2/2$,
independently of $E_i$, cf.\ Eq.~(\ref{eq:v-soft-collinear}). 
Virtuality, as defined in Ref.~\cite{Hartgring:2013jma} for Vincia,
specifically the $m_D^2$ variable, corresponds to $2E_i E_k \theta^2$
in the soft-collinear limit, which does depend on $E_i$.
\logbook{}{Cf. Eq. 2.23 in 1303.4974, together with $2.6$; note that
  $I$ is the emitter, $K$ the ``spectator'' (insofar as that has a
  meaning for Vincia) and $j$ is the emission}%
Since one should consider a broad range of (soft and hard) parent
energies $E_i$, ordering in $v_{\betaps=1}$ is not the same as
virtuality ordering.
On the other hand, the ordering $v_{\betaps=1}$ is similar to what is
referred to as virtuality-based ordering in
Deductor~\cite{Nagy:2014nqa,Nagy:2017dxh}, where, in our notation, the
ordering variable is
$\Lambda^2 \sim \frac{2p_i.p_k}{2\ptilde_i.Q} Q^2 \simeq E_k \theta^2
Q/2$ (for $Q^\mu$ at rest), i.e.\ independent of $E_i$.
\logbook{}{See Eq.~(5) of \cite{Nagy:2017dxh} (or 2.13 of 1401.6366,
  with special care to read the paragraph below it that defines the
  various quantities differently in numerator and denominator).}%
Note that our conclusions about logarithmic accuracy still do not
apply to Deductor.
Their recoil treatment differs from ours, and it is our understanding
that this is crucial to their observation of NLL accuracy for
thrust~\cite{Nagy:2020dvz}.

\subsection{PanLocal dipole and antenna}
\label{sec:panlocal}

Here we consider two closely related variants of local shower, one of
the dipole type, the other of antenna type, both intended to be used
with an ordering choice $0<\betaps<1$.
The two non-trivial choices that we need to make in order to obtain
showers that have NLL accuracy concern the kinematic map and the
generation variables in the hard-collinear initial-state region.
Let us start with the former:
\begin{enumerate}
\item For all dipole types (FF, II, etc.), we apply a dipole-local map
  Eq.~\eqref{eq:general-local-map}.
  The relation that we use to obtain $a_k$ and $b_k$ from $\alpha_k$
  and $\beta_k$ depends on the type of dipole and will be discussed
  below.
  When the dipole involves one or more initial-state particles, and at
  least one of them is assigned transverse recoil from the map, the
  local map results in the new incoming particle acquiring a
  transverse momentum.
  
\item When an initial-state particle acquires transverse momentum, we
  perform a Lorentz transformation (boost and a rotation) to all event
  particles, outgoing and incoming, so as to realign the initial-state
  parton with the original beam axes.
  The Lorentz transformation is constrained by the following requirements:
  after its application, both incoming particles should be aligned
  along the original beam axes, and the rapidity of the hard system,
  as defined with respect to the proton beams, should coincide with
  the pre-splitting hard-system rapidity.
\end{enumerate}
The functional form of the transformation, together with the mapping
coefficients, are provided in Appendix~\ref{app:panlocal-map}.
The choice to boost all particles in the case of initial-state
splittings differs from that of the Dipole-$k_t$ algorithm (recall
that for II dipoles, it boosts all final-state particles \emph{except}
for the new emission). 

Our choice of Lorentz transformation leads to an important subtlety for
hard-collinear initial-state branchings, notably as concerns the
relation between the generated transverse momentum, $k_\perp$, and the
final emission transverse momentum with respect to the beam, $k_t$.%
\footnote{The argument that follows is based on a physical picture of
  the angles involved in the branching.
  Some readers may instead prefer to consider a Lorentz-invariant
  definition of the transverse momentum with respect to reference
  directions $a$ and $b$,
  $k_{ti}^2 = \frac{2(p_a \cdot p_k) (p_b \cdot p_k)}{p_a \cdot p_b}$,
  and to use the explicit kinematic map in
  Eq.~(\ref{eq:general-local-map}) to derive the result that is shown
  below in Eq.~(\ref{eq:kt-v-kperp-panlocal}).
}
Let us consider such a branching, concentrating just on the
part of the dipole close the initial-state emitter $\itilde$, which
emits a particle $k$ and produces a new incoming particle $i$.
Representing the angles as follows
  \begin{eqnarray}
      \label{eq:panlocal-angles}
		\begin{gathered}
  	\includegraphics{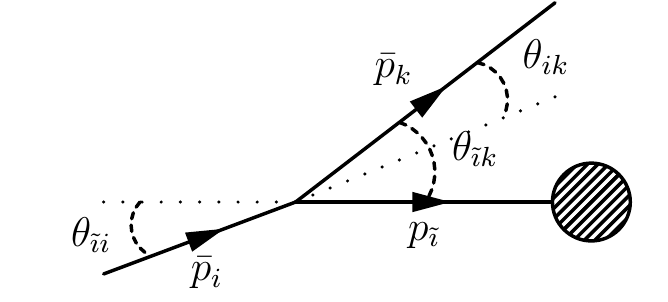}
\end{gathered}
	\end{eqnarray}%
we note that the generated transverse momentum, defined with
respect to the $\itilde$ direction, is given by
$k_\perp \simeq E_k \theta_{\itilde k}$, where $E_k$ is the energy of
$k$.
To determine the transverse momentum, $k_t$, of the emission with respect to the beam, note
that in the limit where the ordering variable is small, the boost
component of the Lorentz transformation is also small, and it essentially
preserves the angles between all particles.
As a result, we can write $k_t \simeq E_k \theta_{i k}$.
Keeping in mind that $E_i \theta_{\itilde i} = E_k \theta_{\itilde k}$
(by conservation of transverse momentum),
$\theta_{ik} = \theta_{\itilde k} - \theta_{\itilde i}$,
$E_k = a_k E_\itilde$ and $E_i = (1 + a_k)E_\itilde$ (with $a_k$ as in
the kinematic map, Eq.~(\ref{eq:general-local-map})), one has the
following relation between the generated transverse momentum $k_\perp$
and the final transverse momentum with respect to the beam, $k_t$:
\begin{equation}
  \label{eq:kt-v-kperp-panlocal}
  k_t = \frac{k_\perp}{1 + a_k}\,.
\end{equation}
The critical feature to note is that $k_t$ becomes much smaller than
$k_{\perp}$ when $a_k \gg 1$.
Physically this is a consequence of the fact that, in that limit,
$E_k \simeq E_i$ and so the angles $\theta_{\itilde i}$ and
$\theta_{\itilde k}$ almost coincide, resulting in a small value of
the difference between them, $\theta_{ik}$, which is the angle
relevant for calculation the transverse momentum with respect to the
beam.

\begin{figure}
  \centering
  \includegraphics[width=0.48\textwidth,page=1]{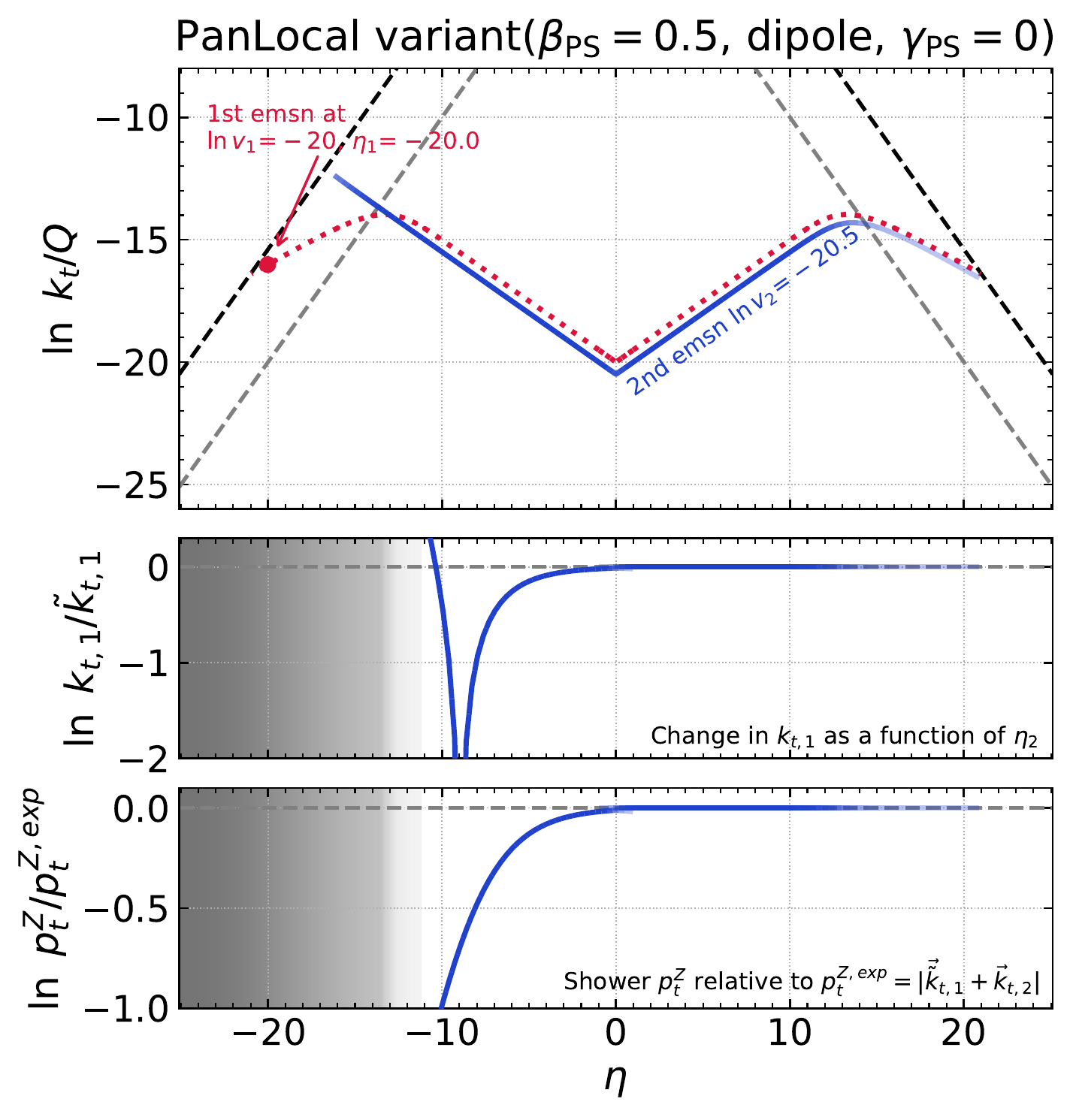} 
  \includegraphics[width=0.48\textwidth,page=2]{plots/contour_plots_panlocal_variants.pdf} 
  \caption{As in Fig.~\ref{fig:contour-dipole-kt-soft}, showing
    non-default PanLocal variants with (left)
    the choice $b_k=\beta_k$ (corresponding to $\gamma=0$ in
    Eq.~(\ref{eq:panlocal-bk-relation})) and (right) the choice
    $\gamma=1$.
  }
  \label{fig:panlocal-variants}
\end{figure}

A degree of freedom that we have in the kinematic map is the relation
between the generation $v$ and $\bar\eta$ (or equivalently $\alpha_k$
and $\beta_k$ in Eqs.~(\ref{eq:definition-alphak-betak})) and the
variables $a_k$ and $b_k$ of the kinematic map.
At first sight, the simplest choice would appear to be to use
$a_k = \alpha_k$ and $b_k = \beta_k$, as we did for the PanGlobal
shower.
Considering the ordering parameter $\betaps=0.5$, we obtain the
contours shown in Fig.~\ref{fig:panlocal-variants} (left).
Observe that in the soft-collinear region, i.e.\ the inner Lund
triangle, the (red-dotted) contour for the first emission slopes upwards as
one moves out from central rapidities, as expected for $\betaps=0.5$.
However, beyond the edge of the inner-Lund triangle, where
$a_k = \alpha_k$ becomes much larger than $1$, the behaviour changes
and the first emission contours bends down, a consequence of the
$1+a_k$ denominator in Eq.~(\ref{eq:kt-v-kperp-panlocal}).
Next, we fix the first emission to be hard and at negative
rapidities (red dot) and examine a second emission at $v_2$ slightly
smaller than $v_1$, shown by the blue contour.
When the second emission, $k_2$, satisfies
$\eta_1 \lesssim \eta_2 \lesssim 0$, it is emitted from the FI dipole
that stretches between emission $1$ and the right-going beam, and so
takes its transverse recoil from emission $1$.
When $k_{t2} \gtrsim k_{t1}$, this induces a significant recoil on
particle $1$, as is visible from the middle panel of that figure.
How problematic is this?
Strictly it only matters when the first emission has
$\alpha_{k_1} \gg 1$, because only then is there a substantial region
part of the inner Lund plane (i.e.\ soft and collinear region) with
$k_{t2} \gtrsim k_{t1}$.
In practice the region $\alpha_{k_1} \gg 1$ is suppressed by
the PDF ratio in Eq.~(\ref{eq:p-initial-state}).
From a logarithmic point of view it is relevant only to small-$x$
resummations, which are beyond what we aim to control.
Still, one might argue it would not be within the spirit of our
overall approach to have a long-distance correlation between hard and
soft-collinear emissions.

To address this concern, we consider a generalisation of the relation
between the coefficients in the kinematic map and the ordering and
auxiliary variables.
Specifically, if the emitter is an initial-state particle, we maintain
$a_k = \alpha_k$, but replace the $b_k = \beta_k$ relation with
\begin{equation}
  \label{eq:panlocal-bk-relation}
  b_k = \beta_k (1 + \alpha_k)^{2\gamma}\,,
\end{equation}
where $\gamma$ is a parameter to be chosen.
This has a substantial effect when $\alpha_k \gg 1$, and in that
region one finds%
\logbook{e51e82c19b7}{See kt2HardCollinear in
  2022-03-panlocal-pp-constraints.nb, with $z_1 \to \alpha_k$}
\begin{equation}
  \label{eq:panlocal-kt-gamma}
  \ln k_t = \left(\gamma - \frac{1}{1+\betaps}\right) \ln \alpha_k
  + \left(\frac{1}{1+\betaps}\right) \ln v + \order{1}.
\end{equation}
If we are to prevent the contour from decreasing in $k_t$ outside the
inner Lund plane, then we need
\begin{equation}
  \label{eq:gamma-condition-no-kt-dip}
  \gamma \ge \frac{1}{1+\betaps}\,.
\end{equation}
This requirement ensures that a soft-collinear emission does not
affect a prior hard-collinear, even when that hard collinear one has
large $\alpha_k$.

\begin{figure}
  \centering
  \includegraphics[width=0.48\textwidth,page=3]{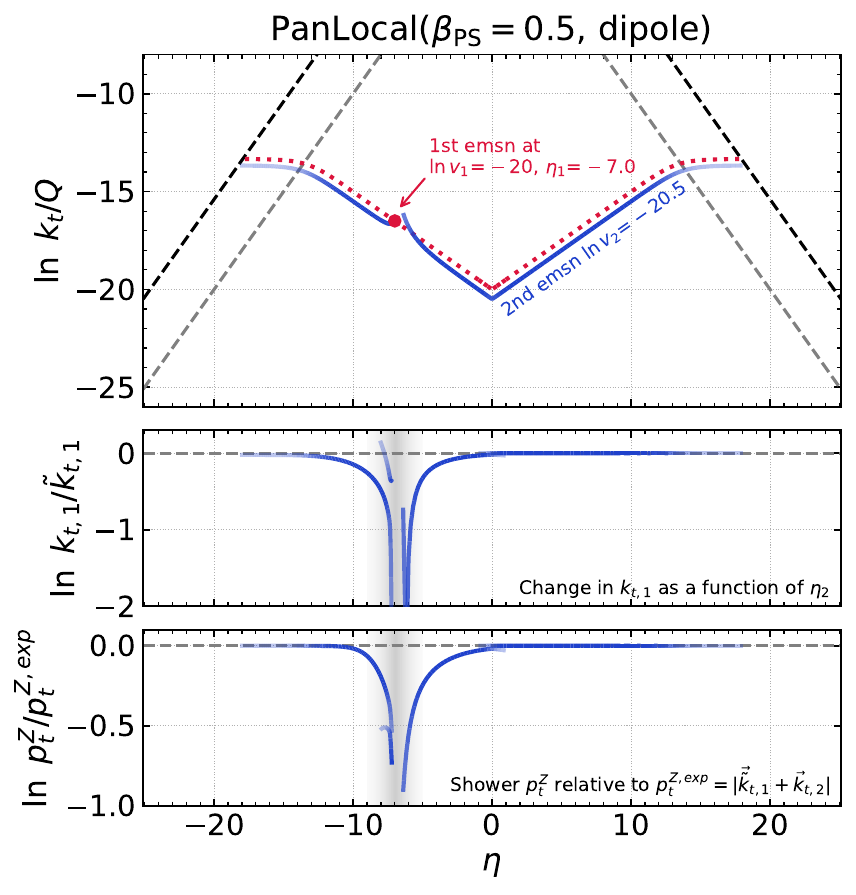} 
  \includegraphics[width=0.48\textwidth,page=4]{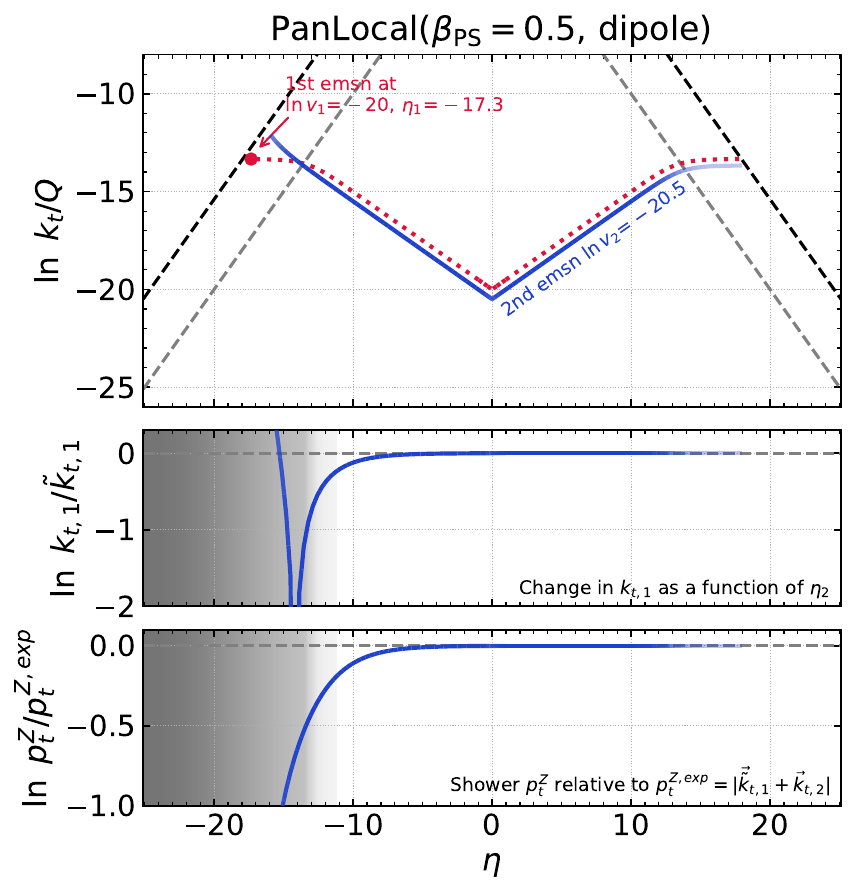} 
  \caption{PanLocal contours for two first-emission configurations
    similar to those of Fig.~\ref{fig:panlocal-variants}, but with our
    default choice of the parameter $\gamma=1/(1+\betaps)$.
  }
  \label{fig:panlocal-final}
\end{figure}

A second consideration concerns the effect of a hard-collinear
emission on a prior soft-collinear emission.
Let us examine what happens if we take $\gamma = 1$.
Such a choice has the effect of cancelling the denominator in
Eq.~(\ref{eq:kt-v-kperp-panlocal}) so that $k_t$ coincides with
$\kappa_\perp$ in Eq.~(\ref{eq:v-def}).
As can be seen in Fig.~\ref{fig:panlocal-variants} (right), it gives a
straight contour for the first emission.
Next, we fix the first emission to be soft and collinear (as
represented with the red dot) and examine how its momentum is modified
by a second emission, as a function of the rapidity of that second
emission, $\eta_2$.
We see from the middle panel of the figure that, sufficiently far into
the left hard-collinear region, the transverse momentum of the
first emission receives a large modification.\footnote{With
  sufficiently large $\sqrt{s}/Q$ such an effect would be visible also
  in the right-hand collinear region.}
The origin of this problem requires an examination of the detailed
kinematic map, as given in Appendix~\ref{app:panlocal-map}.
Concentrating on the left-hand part of the plot, with
$\eta_2 < \eta_1 < 0$, consider the $b_j$ coefficient in
Eq.~(\ref{eq:PanLocal-dipole-IF-bj}) for an IF dipole with the
initial-state parton as the emitter ($f=1$),
\begin{equation}
  \label{eq:PanLocal-dipole-IF-bj-simplified}
  b_j = 1 - \frac{b_k}{1+a_k}\,.
\end{equation}
The essential point to understand is that the large-$\alpha_k$
enhancement of $b_k$ that we introduce with $\gamma=1$ in
Eq.~(\ref{eq:panlocal-bk-relation}) can cause $b_k$ to become
commensurate with $a_k$, which results in a substantial modification of
$b_j$.
Considering the limit $a_{k} = \alpha_{k} \gg 1$, this occurs when
$b_k/a_k$, given by
\logbook{}{see ratioSubz2 in
  logbook/2020-12-09-pp-schemes/gavin-mathematica/2022-03-panlocal-pp-constraint.nb,
  replacing $z_2$ with $\alpha_k$.
  Note also: that section of the mathematica file shows that for
  $\betaps=1/2$ the region of cross talk is limited by
  $z_1 > 1/\alpha_k^2$, i.e.\ the extent into soft collinear region is
  commensurate with the extent into the hard collinear region.  }
\begin{equation}
  \label{eq:PL-bk-ak-ratio}
  \frac{b_k}{a_k} \sim
  \alpha_k^{2\gamma -\frac{2}{1+\betaps}}
  z_1^{\frac{2}{1+\betaps} - 1}
  \left(\frac{v_2}{v_1}\right)^{\frac{2}{1+\betaps}}
  \,, \qquad
  \text{valid for } \alpha_k \gg 1\,,\; z_1 \ll 1\,,
\end{equation}
becomes of order $1$.
Since we started with a first soft-collinear emission, $z_1 \ll 1$,
and since $v_2 < v_1$, by construction, the ratio in
Eq.~(\ref{eq:PL-bk-ak-ratio}) is guaranteed to be smaller than one
if
\begin{equation}
  \label{eq:gamma-condition-no-bj-issue}
  \gamma \le \frac{1}{1+\betaps}\,.
\end{equation}
In particular, this shows that if we are to avoid non-trivial
cross-talk between hard-collinear and soft-collinear emissions,
$\gamma=1$ is not suitable for any $\betaps >0$.
Taken together, Eqs.~(\ref{eq:gamma-condition-no-kt-dip}) and
(\ref{eq:gamma-condition-no-bj-issue}) tell us that there is only one
value of $\gamma$ in Eq.~(\ref{eq:panlocal-bk-relation}) that enables
us to avoid such cross-talk, specifically
\begin{equation}
  \label{eq:gamma-final-choice}
  \gamma = \frac{1}{1+\betaps}\,.
\end{equation}
Fig.~\ref{fig:panlocal-final} shows the contour plots as obtained with
this choice.
There can still be non-trivial side effects of the second emission on
the first one, but only in the region where both emissions are hard and
collinear (left), or close in rapidity (right).
In particular, the problems seen in Fig.~\ref{fig:panlocal-variants}
are absent with the choice for $\gamma$ in
Eq.~(\ref{eq:gamma-final-choice}).
We make this the default choice for our PanLocal dipole shower.

The antenna variant of PanLocal is broadly similar in construction.
The guiding principle is that for any dipole end that can be involved
in a hard-collinear initial-state branching, the mapping coefficient
for the light-cone component associated with the effective
``spectator'' end should have an enhancement as in
Eqs.~(\ref{eq:panlocal-bk-relation}) and
(\ref{eq:gamma-final-choice}).
The details are given in Appendix~\ref{sec:app-panlocal-antenna}.
The contour plots, which we omit for brevity, are very similar to
Fig.~\ref{fig:panlocal-final}.
\logbook{}{see plots/individual-contour-plots/PLa-beta0.5-1hard.pdf
  and plots/individual-contour-plots/PLa-beta0.5-1soft.pdf}

\begin{figure}[tb]
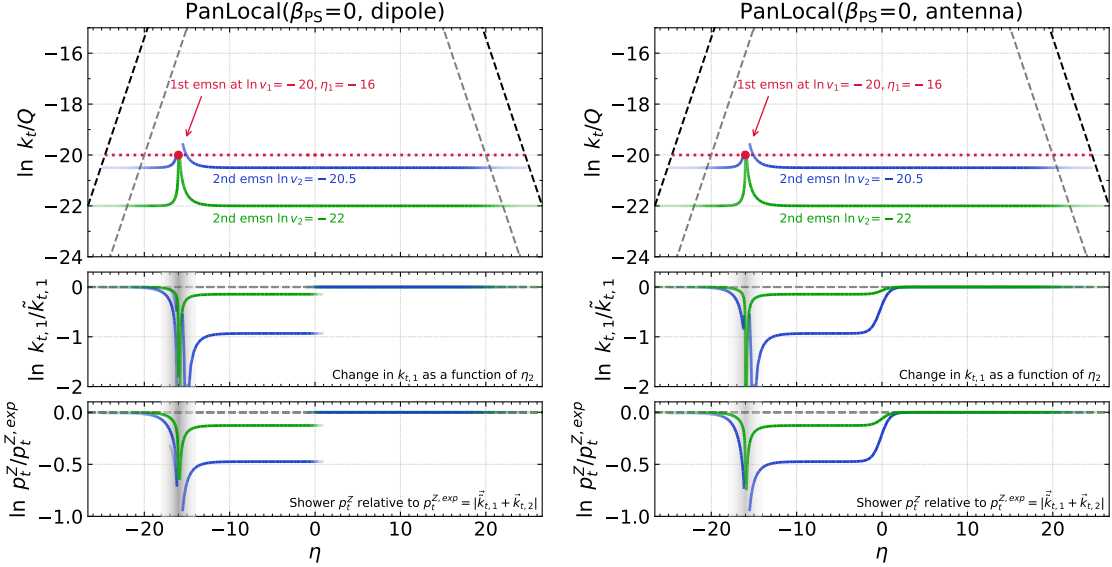

    \centering
        \includegraphics[page=30, width=0.49\textwidth]{plots/contour_plots.pdf}
        \includegraphics[page=21, width=0.49\textwidth]{plots/contour_plots.pdf}
\caption{Same as Fig.~\ref{fig:contour-dipole-kt-soft}, but for PanLocal, dipole~(left) and antenna~(right) with $\betaps = 0$.}
    \label{fig:contour-panlocal-beta0}
\end{figure}

To close this section, we note that $\betaps = 1$ for the PanLocal
showers suffers from the same problem as $\betaps=1$ PanGlobal
showers, i.e.\ a substantial effect of spectator longitudinal recoil
when two emissions are in the same hemisphere, $v_2$ is commensurate
with $v_1$ and $|\eta_2| > |\eta_1|$.
Additionally, PanLocal showers have issues of incorrect transverse
recoil for $\betaps = 0$, as in the final state case, cf.\
Fig.~\ref{fig:contour-panlocal-beta0}. 
In particular for $\eta_1 < 0$, the transverse momentum recoil from an
emission from the $g_1q$ dipole is taken by the gluon when $\eta_1<\eta_2<0$
even if $\eta_{2} - \eta_1 \gg 1$.
Requiring $\beta_{\rm PS}>0$ solves this problem:
emissions with commensurate $k_t$ values are ordered so that
smaller-angle emissions are generated later.
Then, the use of $g(\bar\eta_Q)$ to partition the dipole ensures that
when the smaller-angle emission is in the primary Lund plane, the
recoil is taken from the initial-state particle and so, after the
boost, effectively from the hard system.

\section{Subleading colour at LL accuracy and beyond}
\label{sec:colour}

Dipole showers naturally achieve the large-$\nc$ limit of QCD.
It is common practice for dipole showers to adopt a modification of
the large-$\nc$ limit in which one uses a $C_F$ colour factor for a
quark or anti-quark end of a dipole and a $C_A/2$ colour factor for
an end that corresponds to a gluon (we refer to this as
the colour-factor-from-emitter scheme, CFFE).
It has long been known~\cite{Gustafson:1992uh} that such a choice is
inconsistent with colour coherence.
Ref.~\cite{Dasgupta:2018nvj} pointed out that, for quite a range of
observables, this results in the wrong subleading-$\nc$ contributions
at LL accuracy.
Insofar as one views $1/\nc^2$ as being comparable with $\as$,
subleading-$\nc$ LL corrections are comparable to a leading-colour NLL
terms and so should, arguably, be addressed on a par with such
leading-colour NLL terms.

In this section, we extend the two efficient colour
schemes introduced~\cite{Hamilton:2020rcu} for the final-state
PanScales showers to the initial-state case.
Section~\ref{sec:segment} summarises and extends the so-called segment
colour scheme, Section~\ref{sec:NODS} does the same for the
nested-ordered-double-soft (NODS) colour scheme, and
Section~\ref{sec:colour-tests} shows the results of various
fixed-order validation tests.
These schemes achieve full-colour accuracy not just for LL terms but
also for NLL terms in the case of global event-shape type
observables~\cite{Banfi:2004yd}, and for next-to-double-logarithms
(NDLs) for jet multiplicities.
Related schemes, adequate for full-colour LL accuracy have been
explored also by other
authors~\cite{Friberg:1996xc,Holguin:2020joq}.\footnote{There has also
  been
  extensive work towards the inclusion of subleading-colour effects through amplitude-level
  evolution~\cite{Nagy:2015hwa,Platzer:2018pmd,Nagy:2019pjp,DeAngelis:2020rvq,Forshaw:2021mtj}
  and other
  schemes~\cite{Platzer:2012np,Hoche:2020pxj,Frixione:2021yim}.
  Note that amplitude evolution also has the potential to induce
  coherence-violating effects~\cite{Forshaw:2006fk,Catani:2011st}
  (sometimes called super-leading logarithms), terms specific to
  hadron colliders, whose all-order impact has recently started to be
  evaluated~\cite{Nagy:2019rwb,Becher:2021zkk}.
  The impact of coherence-violating terms on logarithmic accuracy
  depends on the observable~\cite{Banfi:2010xy,Forshaw:2021fxs},
  though the quantitative understanding of this question remains to be
  elucidated.
  In this context, one may wish to keep in mind observations
  concerning spurious super-leading logarithms generated by
  long-distance shower-recoil effects~\cite{Dasgupta:2020fwr},
  specifically terms $\as^n L^{2n-3}$ for the thrust.
  Their sum was found to vanish as $\as \to 0$ for fixed $\as L$, a
  behaviour that would be expected from NNLL terms or beyond.
  These questions clearly deserve further study.
  For now, when we refer to the full-colour NLL accuracy of our
  shower, this is to be understood as applying to terms of coherent
  origin.
}

\subsection{Adaption of segment scheme to initial-state splittings}
\label{sec:segment}

The segment colour scheme divides each dipole into distinct segments in the
underlying $\bar \eta$ generation variable, with some segments having
a $C_F=(\nc^2-1)/(2\nc)$ 
colour factor and the others having a $C_A/2
= \nc/2$ colour factor.
The guiding principle for the assignment is colour coherence.
More concretely, consider the example of a $q\bar q \to Z$
event in which the incoming $q$ backwards evolves by emitting a
final-state gluon $g_1$, i.e.\ $q_I \rightarrow q_I g_1$.
For this first emission, the colour factor is unequivocally
$C_F$ since it is emitted from a quark-like leg.
Next, a second gluon $g_2$ is emitted.
Let $\theta_{i}$ be the angle between $g_i$ and the beam direction,
and $\theta_{12}$ the angle between the two gluons.
If $\theta_{12} \ll \theta_{1}$ then the segment method assigns a
$C_A/2$ colour factor, while if $\theta_{2} \ll \theta_{1}$ or
$\theta_{2} \gg \theta_{1}$ a $C_F$ colour factor is assigned.
In the region where the angles are commensurate, the correct emission
intensity cannot be reproduced by simply using one of $C_F$ or $C_A/2$
but the segment method still uses a discrete choice between them,
engineered so as to reproduce NDL accuracy for observables such as the
multiplicity (or equivalently the correct integrated rate of emission
of $g_2$ when it has an energy or $k_t$ much lower than that of
$g_1$).

The segment colour scheme implements this reasoning throughout the shower
evolution.
A dipole can have any number of alternating $C_F$ and $C_A/2$ colour
segments.
We represent the segments with the notation
\begin{equation}
  \label{eq:schematic-dipole-segments}
  {}_{\bar 3}[-\infty, C_A, \eta_1 , C_F, \eta_2, C_A, \ldots, \infty]_3\,,
\end{equation}
This indicates that for all gluon emission rapidities (as defined
below) between $-\infty$ and $\eta_1$, one uses a $C_A/2$ colour
factor,\footnote{In the list of segments, it is labelled $C_A$ rather than
  $C_A/2$ for ease of notation.} for emissions between $\eta_1$ and
$\eta_2$, one uses a $C_F$ colour factor, etc.
Negative rapidity is always associated with the end of the dipole that
has outgoing anti-triplet colour.
In our notation we always refer to the effective outgoing colour-flow,
so, for example, an initial-state quark is to be thought of as having
outgoing anti-triplet colour.
We establish where an emission $k$ is positioned in the sequence of
Eq.~(\ref{eq:schematic-dipole-segments}) based on a specific
approximate determination of its rapidity, which for PanScales showers
and a parent dipole $ij$ reads
\begin{equation}
    \label{eq:eta-approx}
    \eta_\text{approx} \equiv \left\{
    \begin{array}{lll}
        \bar \eta_Q - \frac12 \ln \left(\frac{1- \cos\theta_{ij}}{2}\right)\,,
        && \bar\eta_Q > 0\,,
        \\
        \bar \eta_Q + \frac12 \ln \left(\frac{1- \cos\theta_{ij}}{2}\right)\,,
        && \bar\eta_Q < 0\,.
    \end{array}
  \right.
\end{equation}
In the soft and collinear limit, this variable coincides with the
physical rapidity of the emission with respect to the closer of $i$
and $j$ (in this section, all angles are evaluated in the frame where
$Q^\mu$ is at rest).
The sign of $\bar\eta_Q$ (and $\eta_\text{approx}$) is such that
emission from the triplet (anti-triplet) end of a dipole has positive
(negative) $\bar\eta_Q$.

After an emission, one needs to update the set of segments on the
dipoles involved in the splitting.
Returning to our earlier example of $q_I\bar q_I \to Z$, the
initial-state $q_I$ is at the $\bar 3$-end of the dipole, and the
starting dipole segment structure is
\begin{align}
    {}_{q_I \!}\left[-\infty, C_F, \infty \right]_{\bar{q}_I}\,.
\end{align}
When we radiate a gluon from this $q_I\bar{q}_I$ dipole, we obtain two
new dipoles $q_I g_{F}$ and $g_{F} \bar{q}_I$, each of which now has a
$C_F$ and a $C_A$ region.
\begin{equation}
    \label{eq:colour-ii}
    {}_{q_I \!}[-\infty,C_F,
    \eta_{g_F}^L, C_A, \infty]_{g_{F}}
    + {}_{g_{F} \!}[-\infty, C_A, \eta_{g_F}^R,
    C_F, \infty]_{\bar{q}_I}\, .
\end{equation}
For a gluon emission $k$, the new transition points are given by
\begin{eqnarray}
        \label{eq:colour-transition points}
    \eta_{k}^L = \max(0, \eta_{k})\,, \quad      \eta_k^R = \min(0, \eta_{k})\,.
\end{eqnarray}
To determine $\eta_k$, we evaluate the angle of the emission
$\theta_k$, with respect to the
triplet (anti-triplet) end of the dipole if $\eta_{\rm approx} > 0$
($<0$) and then take $\eta_{k} = \pm |\ln\tan\theta_k/2|$, where a
positive (negative) sign is used when $\eta_{\rm approx} > 0$
($<0$). 

So far, the discussion has largely followed that of the final-state
case in Ref.~\cite{Hamilton:2020rcu}. 
We now summarise the non-trivial extensions needed for the new
radiation channels that open up in hadron collisions:
\begin{itemize}
\item $g_I \to q_I \bar{q}_F$: the backwards evolution of an initial-state quark $q_I$
  to an initial-state gluon $g_I$ and a final-state anti-quark $\bar{q}_F$ results in the
  following  configuration of segments
  \begin{align}
      {}_{q_I\!}\left[-\infty, C_F, \dots \right]_{\dots} \to  {}_{g_I\!}\left[-\infty, C_A, \eta_{\bar{q}_F}^R, C_F, \dots \right]_{\dots } + {}_{\bar{q}_F\!}\left[-\infty, C_F, \eta_{\bar{q}_F}^L, C_A, \infty \right]_{g_I}
  \end{align}
  The case in which $g_I \to \bar{q}_I q_F$ is similar.
  Relative to the default
  large-$\nc$ splitting probability, no rejection factor needs to be
  applied because the splitting function involves a
  $T_R$ colour factor (this is the same as for the treatment of a
  final state $g\to q\bar q$ splitting).

\item$q_I \to g_I q_F$: the backwards evolution of an initial-state gluon $g_I$
  to an initial-state quark $q_I$ and a final-state quark $q_F$ is slightly more subtle.
  Two dipoles exist before the splitting: one with the $C_A$ colour factor on the
  $3$-end, and another on the $\bar{3}$-end, i.e.
  \begin{align}
      {}_{g_I\!}[-\infty,C_A,\dots]_{\dots} + {}_{\dots\!}[\dots,C_A,\infty]_{g_I}\,.
  \end{align}
  After the evolution to an initial-state quark,
  which becomes the new $\bar{3}$-end of a dipole, both dipoles need to be updated to
  include a new $C_F$ segment, resulting in 
  \begin{align}
      {}_{g_I\!}[-\infty,C_A,\dots]_{\dots} &\to {}_{q_I\!}[-\infty,C_F, -|\eta_{q_I q_F}|, C_A,\dots]_{\dots}  \nonumber \\
       {}_{\dots\!}[\dots,C_A,\infty]_{g_I}&\to  {}_{\dots\! }[\dots,C_A, |\eta_{q_I q_F}|, C_F, \infty]_{q_F}\,,
  \end{align}
  where $\eta_{q_I q_F} = - \ln \tan \theta_{q_I q_F}/2$.
  Instead, for the $\bar q_I \to g_I \bar{q}_F$  case one has
  \begin{align}
      {}_{g_I\!}[-\infty,C_A,\dots]_{\dots} &\to {}_{\bar{q}_F\!}[-\infty,C_F, -|\eta_{\bar q_I \bar q_F}|, C_A,\dots]_{\dots}  \nonumber \\
      {}_{\dots\!}[\dots,C_A,\infty]_{g_I}&\to  {}_{\dots\! }[\dots,C_A,|\eta_{\bar q_I \bar q_F}|, C_F, \infty]_{\bar{q}_I}\,. 
  \end{align}
  Note that even though the emission is in a
  $C_A$ segment, relative to the default dipole emission strength of
  $C_A/2$ one must apply a rejection factor of $(1-2C_F/C_A)$,
  because the underlying splitting function is $P_{q\to qg}$.
\end{itemize}
All other channels and the precise implementation of the algorithm are
as discussed in Ref.~\cite{Hamilton:2020rcu}.

\subsection{NODS}
\label{sec:NODS}

%
Despite capturing the dominant subleading-colour correction for the
integrated rate of soft gluon emissions at NLL, as
shown in Ref.~\cite{Hamilton:2020rcu}, the segment method fails in the
limit where two soft and energy-ordered emissions occur at
commensurate angles.
Thus, a second method was explored in Ref.~\cite{Hamilton:2020rcu},
referred to as NODS.
This method extends the segment method so as to provide the correct
full-colour branching probability to produce any number of
energy-ordered commensurate-angle pairs, as long as each pair is well
separated in rapidity from all others (``NODS accuracy'').
The method works by accepting a given emission with a
colour acceptance probability that is the ratio of the full-colour
soft matrix element, $|M^2|$, to the leading-colour one, $|M_{\rm LC}^2|$.

To illustrate the method, we again consider the case in which a gluon
$g_2$ is emitted from a $q\bar qg_1$ system. The colour acceptance
probability for emission of a softer gluon $g_2$ is given by
\begin{equation}
\label{eq:colour-paccept}
p^{\rm accept} \equiv \frac{|M|^2}{|M_{\rm LC}|^2} =
1 + \left( \frac{2C_F-C_A}{C_A}\right) \frac{(\bar q q)}{(\bar q
  g_1)+(g_1 q)},
\quad(ab)\equiv \frac{p_a\cdot p_b}{(k \cdot p_a) (k\cdot p_b)}\,.
\end{equation}
This expression has full colour accuracy and one should recall that
$(2C_F-C_A)/C_A = -1/\nc^2$.
The discussion of Ref.~\cite{Hamilton:2020rcu} showed how to
generalise this formula so that it delivers the NODS accuracy for
events with arbitrary numbers of gluons and quarks.
If we remain with a single $\bar q q$ pair and an arbitrary number of
gluons that are well-separated in rapidity, the correction factor
relative to the large-$\nc$ emission probability is given exactly by
\begin{equation}
  \label{eq:colour-paccept-n}
  p^{\rm accept}(\bar q, 1, 2, \ldots, n, q)
  = 1 + \left( \frac{2C_F-C_A}{C_A}\right) \frac{(\bar q q)}{(\bar q1)+(12)+\cdots +(n q)}\,,
\end{equation}
which holds for each dipole.
One of the observations of Ref.~\cite{Hamilton:2020rcu} was that for a
given \mbox{individual} large-$\nc$ $ij$ dipole, one can drop most of the
terms in the denominator of Eq.~(\ref{eq:colour-paccept-n}) and
instead use $p^{\rm accept}([\bar a], i, j, [a])$.
In this context, $[\bar a]$ and $[a]$ are suitably chosen ``auxiliary''
momenta, which can in practice be any of the momenta along the
$\bar q, 1, \ldots, q$ respectively to the left and right of $ij$.
The auxiliaries are shown in square brackets, because if one or other of $i$ and
$j$ is at the end of the dipole chain (with a resulting infinite
extent of the $C_F$ segment), there is no auxiliary momentum and the
corresponding argument is omitted in $p^{\rm accept}$.

One additional aspect of the full algorithm is the choice of what to
do when there is more than one $\bar q q$ pair.
Ref.~\cite{Hamilton:2020rcu} uses a product of $p^{\rm accept}$
factors, one for each $C_F$ segment in the dipole (each finite
extremity of a segment has an associated auxiliary
momentum).\footnote{In practice, in NODS limit, i.e.\ where all 
  pre-branching momenta are well separated in the rapidity direction
  in the Lund diagram, only one of the $p^{\rm accept}$ factors ever
  differs noticeably from $1$.}
Another aspect is that of how to update the auxiliary momenta after a
dipole branching.
Suppose we have an $ij$ dipole with a single $C_F$ segment, whose
auxiliaries are $\bar a$ and $a$.
When a gluon $k$ is emitted from that $C_F$ segment, each of the two
new dipoles, $ik$ and $kj$, will inherit part of that $C_F$ segment.
Then the auxiliaries are updated so that the $C_F$ segment of the $ik$
dipole is assigned auxiliaries $\bar a$ and $j$, while the $C_F$
segment of the $kj$ dipole is assigned auxiliaries $i$ and $a$.
We represent this as
\begin{equation}
  \label{eq:nods-example}
  \begin{gathered}
    \includegraphics[scale=0.7]{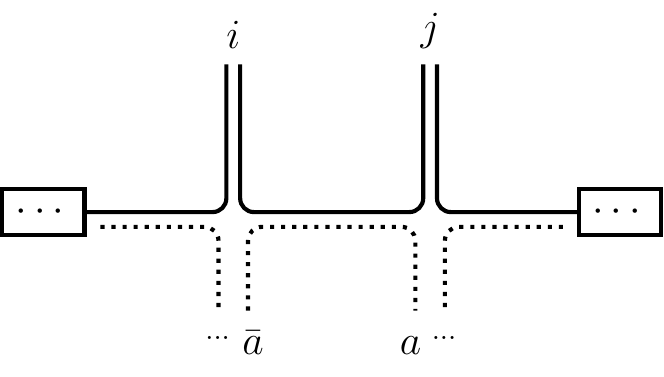}
  \end{gathered}
\quad
  \longrightarrow
  \quad
  \begin{gathered}
    \includegraphics[scale=0.7]{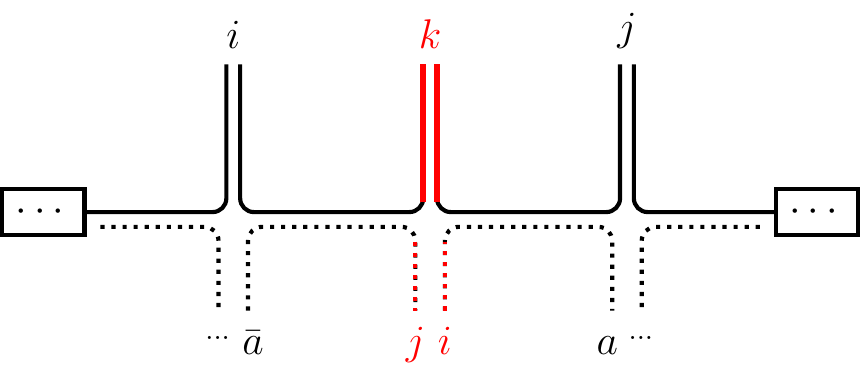}
  \end{gathered}
\end{equation}
where the solid lines represent large-$\nc$ dipoles and the dotted
lines represent the relative $-1/\nc^2$ correction to the
leading-$\nc$ emission probability.
Those dotted lines are labelled with the corresponding auxiliary
momenta.
If one or other of the $\bar a$ and $a$ auxiliaries is absent from the
parent dipole, it remains absent from the child dipoles.

For backwards evolution of an initial-state parton, there are some 
additional configurations to consider beyond those studied in
Ref.~\cite{Hamilton:2020rcu}.
Backwards evolution of an initial-state quark into an initial-state
quark and a final-state gluon behaves precisely as in
Eq.~(\ref{eq:nods-example}). 
Backwards evolution of an initial-state gluon into an initial-state
quark and a final-state quark can be accounted for with the following update
\begin{equation}
  \label{eq:nods-g-backwards}
  \begin{gathered}
    \includegraphics[scale=0.7]{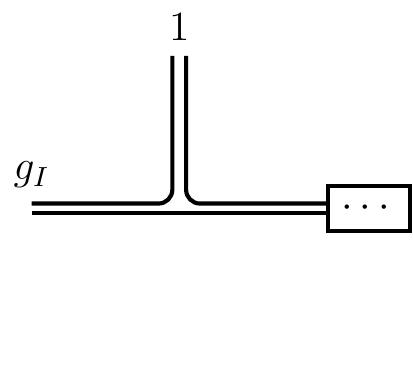}
  \end{gathered}
  \quad\longrightarrow\quad
  \begin{gathered}
    \includegraphics[scale=0.7]{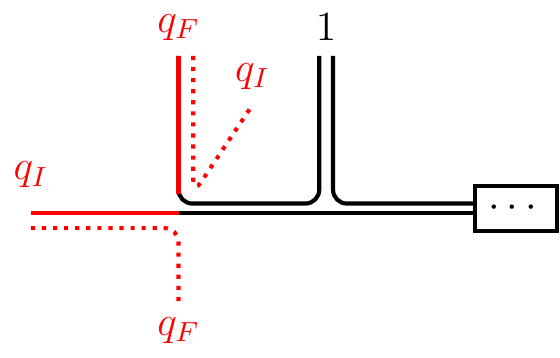}
  \end{gathered}
\end{equation}
where the dipoles associated with the initial and final-state quarks
each acquire a new $C_F$ segment, whose (single) auxiliary is given by
the final and initial-state quark respectively.
Finally we have the situation where an initial-state quark backwards
evolves to give an initial-state gluon and a final-state anti-quark\,,
\begin{equation}
  \label{eq:nods-q-backwards}
  \begin{gathered}
    \includegraphics[scale=0.7]{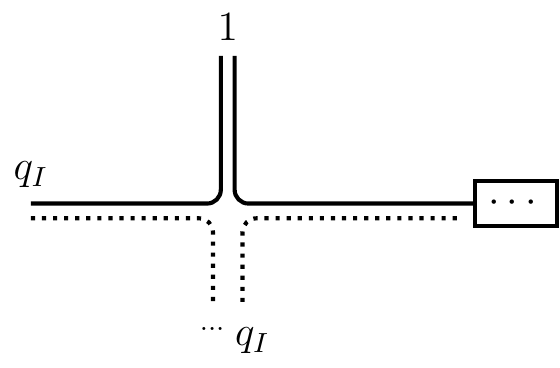}
  \end{gathered}
  \quad\longrightarrow\quad
  \begin{gathered}
    \includegraphics[scale=0.7]{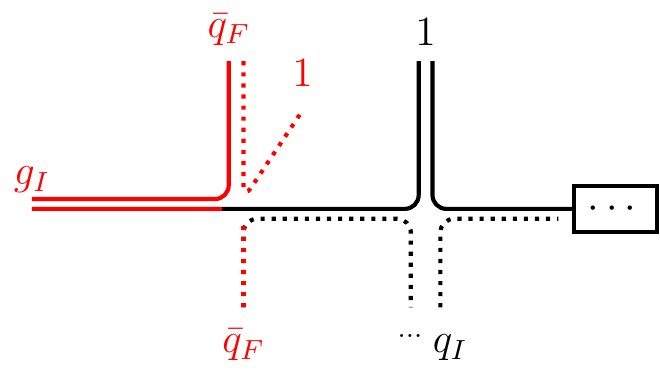}
  \end{gathered}
\end{equation}
Here the existing $C_F$ segment of the $q_I1$ dipole is inherited by the
$g_I1$ dipole, with an additional auxiliary momentum $\bar q_F$, while
the new $g_I \bar q_F$ dipole has a $C_F$ segment with auxiliary $1$.

\subsection{Matrix-element tests}
\label{sec:colour-tests}

We perform two kinds of fixed-order tests of the colour schemes
outlined above.
Both involve configurations with a fixed number of energy-ordered
branchings, at least one of which is an initial-state branching.
In a first set of tests we fix the kinematics of a first emission and
examine the branching probability for a second emission differentially
in its rapidity and azimuth.
In a second set of tests we consider several kinematic configurations
for one or two soft emissions and verify that the integral over the
rapidity and azimuth of an additional soft emission, at a fixed much
smaller transverse momentum, reproduces the analytic expectation, as
required for NDL accuracy for the observables such as the
multiplicity.
The analytic expectation coincides with the result that one would
obtain by assuming an exact angular-ordered pattern for
emissions.
We show results just for the PanGlobal $\betaps=0$
(transverse-momentum ordered) shower, keeping in mind that the
underlying implementation is common to all showers.
\logbook{170fa75baf3d6}{see other pages of plots/plot-colour-integral.pdf for
  checks both at $y_Z!=0$ and for PanLocal}

\subsubsection{Differential tests}
\label{sec:differential-colour-tests}

We fix the kinematics of a first emitted parton (which can be either a
quark or a gluon) to be soft and collinear using 
\begin{equation}
  \ln \frac{v_1}{Q} = -10\,,\quad \bar{\eta}_{Q,1} = 5\,, \quad \varphi_1 = 0\,,
\end{equation}
which corresponds to $z_{1} = 6.69\cdot 10^{-3}$, with a measured
transverse momentum $\ln k_{t,1}/Q = -10$ and rapidity $\eta_{1} = 5$.
The second emission, a soft gluon, is emitted at $\ln v_2/Q = -60$,
and we examine the shower branching probability differentially as a
function of its direction, i.e.\ we sample over $\bar{\eta}_{Q,2}$ and $\varphi_2$.

Specifically, we sample over the generation variables $\bar\eta_Q$ and
$\varphi$ for emission $2$. 
Then, we compute the angular separation between the two emissions
 \begin{equation}
\Delta R_{12} = \sqrt{\Delta \eta_{12}^2 + \Delta \psi_{12}^2}\,,
\end{equation}
where $\Delta \eta_{12}$ and $\Delta \psi_{12}$ are their difference in
rapidity and azimuth.
If $\Delta R_{12}>1$, the second emission is considered to be primary,
and we examine its distribution differentially in the rapidity $\eta$
and azimuth $\psi$ defined with respect to the beam
direction.
Otherwise it is secondary,
i.e.\ emitted predominantly from the first emission, and we examine its
distribution differential in a rapidity $\eta$ and an azimuthal angle
$\psi$ defined with respect to emission 1.
The relevant analytic expectations are
provided in Appendix~\ref{app:colour}.

\begin{figure}[t]
    \centering
    \begin{subfigure}{0.50\textwidth}
        \includegraphics[page=1, width=\textwidth]{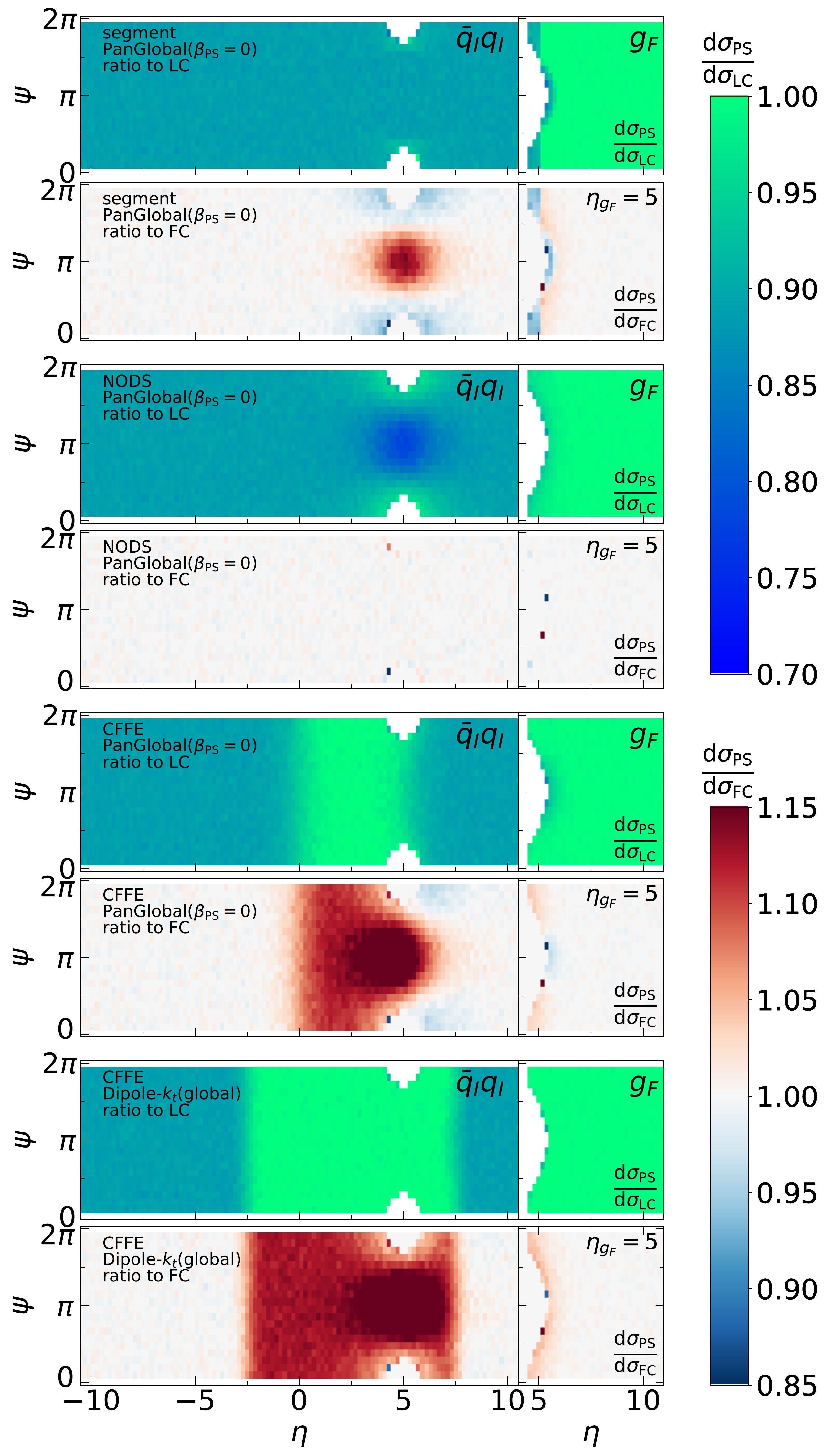}
    \end{subfigure}\hfill
    \begin{subfigure}{0.48\textwidth}
      \includegraphics[page=2, width=\textwidth]{plots/plot-2dcolour.pdf}
    \end{subfigure}
    \caption{Density for the emission of a soft gluon from a
      $q_I\bar{q}_I\rightarrow Z g_F$ (left) or $g_I\bar{q}_I\rightarrow Z
      \bar{q}_F$ (right) system in the rapidity-azimuth plane.
      From top to bottom, the rows show the segment, NODS and CFFE
      methods with the PanGlobal shower, and the CFFE method with the
      Dipole-$k_t$ shower.
      For each plot, the upper panels display the ratio
      between the parton shower differential cross section and the
      leading colour result, $\dd\sigma_{\rm
        PS}/\dd\sigma_{\rm LC}$.
      The LC parton-shower result is obtained setting
      $C_A = 2C_F = 3$, i.e.\ an emission strength $C_A/2 = 3/2$ for
      each dipole.
      The lower panels show the deviation from the FC
      differential matrix element, given in
      Appendix~\ref{app:colour},  $\dd\sigma_{\rm
        PS}/\dd\sigma_{\rm FC}$.
      For each plot, the left-hand panels correspond to emissions from
      the incoming partons (in the sense of a Cambridge/Aachen
      algorithm with jet radius
      $R=1$~\cite{Dokshitzer:1997in,Wobisch:1998wt}), whereas the
      right-hand panels correspond to emissions from the first
      emitted parton (either $g_F$ or $\bar{q}_F$).}
    \label{fig:colour-2dcolour-qq}
\end{figure}

\begin{figure}[t]
    \centering
    \begin{subfigure}{0.48\textwidth}
        \includegraphics[page=3, width=\textwidth]{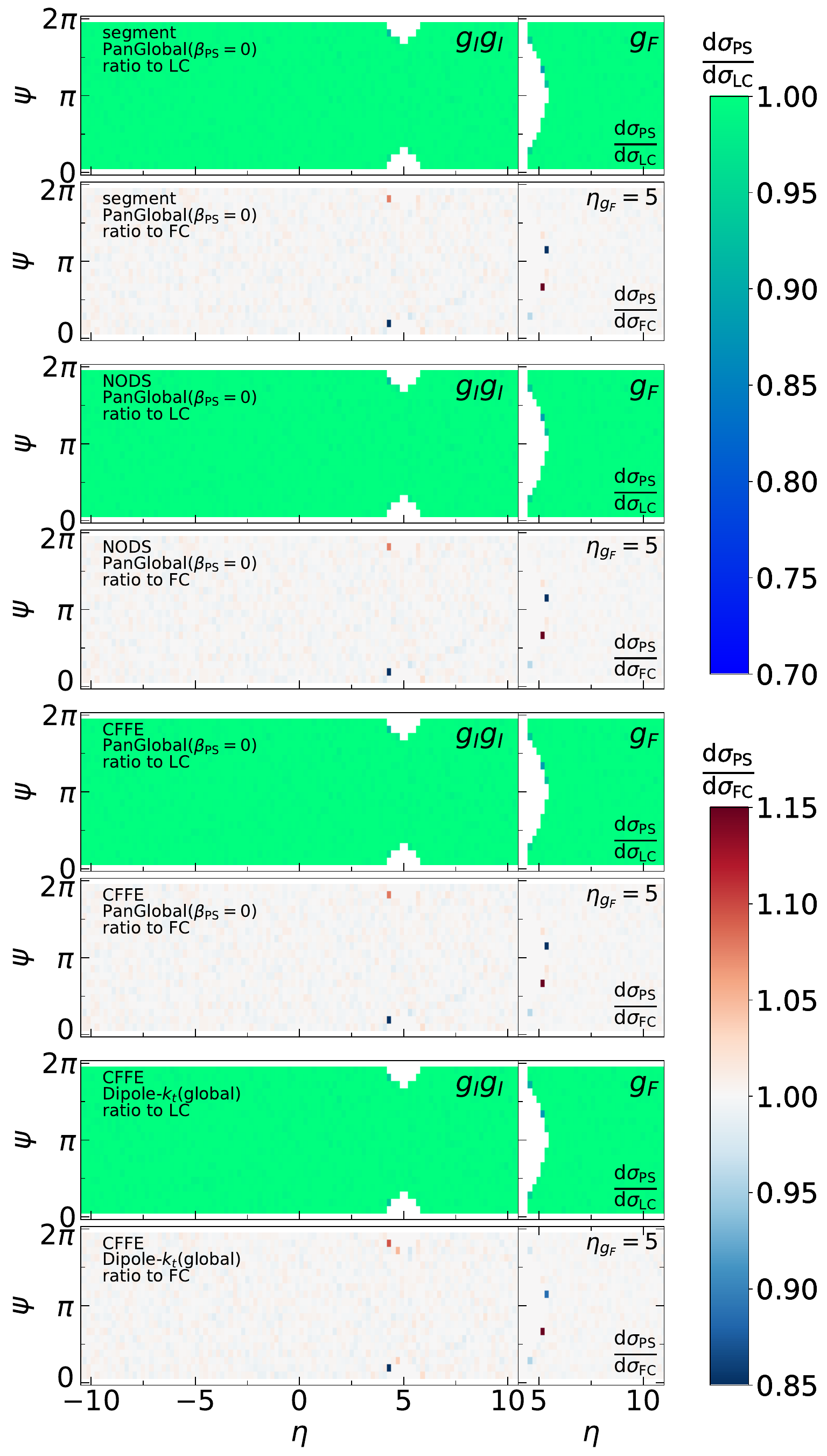}
    \end{subfigure}
    \begin{subfigure}{0.48\textwidth}
        \includegraphics[page=4, width=\textwidth]{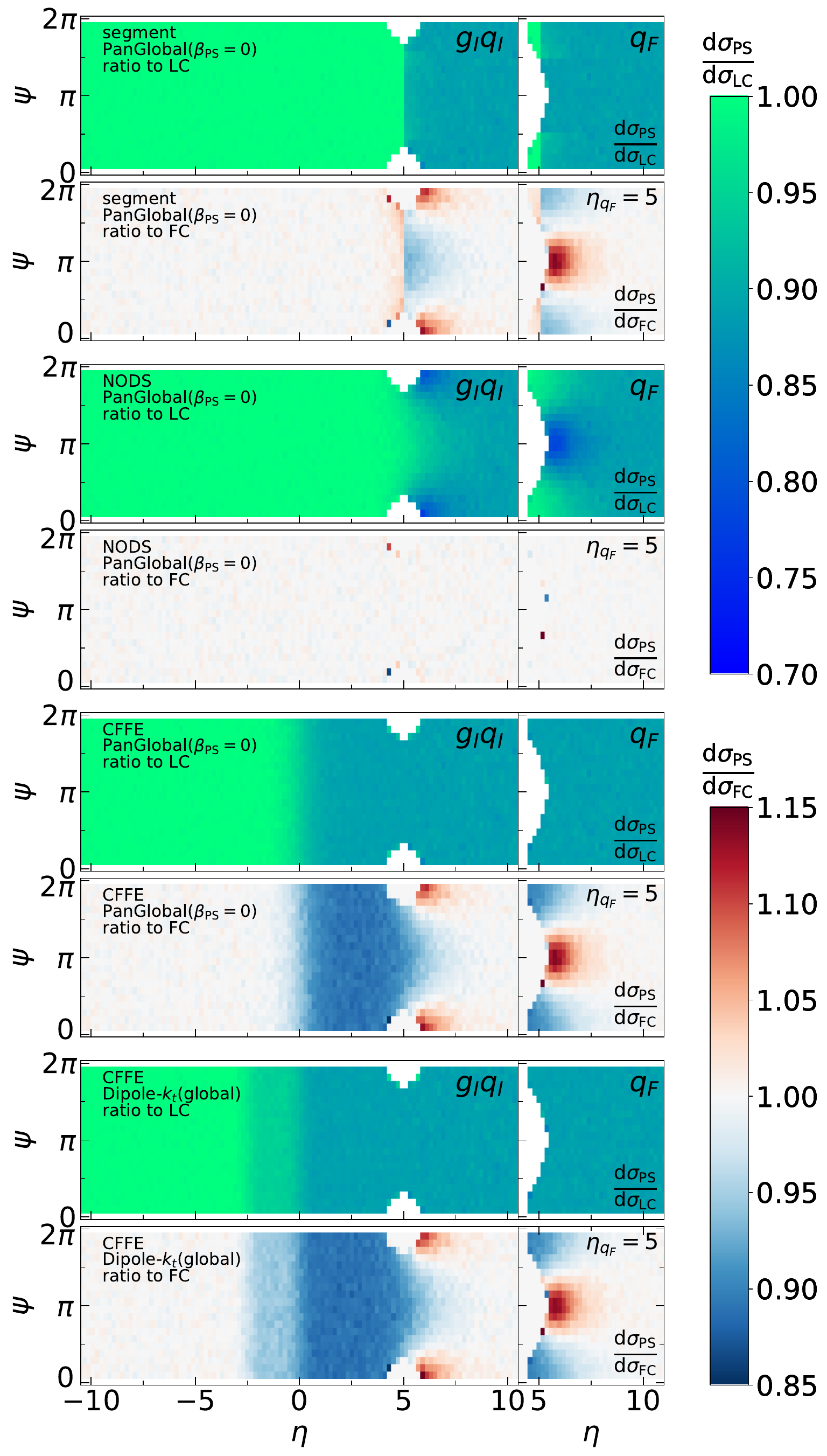}
    \end{subfigure}
    \caption{Same as Fig.~\ref{fig:colour-2dcolour-qq} but for
      emission of a soft gluon from $g_Ig_I\rightarrow H g_F$ (left)
      and $g_Iq_I\rightarrow H q_F$ (right) systems.}
    \label{fig:colour-2dcolour-gg}
\end{figure}

The results for the segment and NODS colour schemes are shown in the
first two rows of Figs.~\ref{fig:colour-2dcolour-qq}
and~\ref{fig:colour-2dcolour-gg}, for $q\bar q \to Z$ and $g g \to H$
production respectively (the bottom two rows are discussed further
below).
We separate the cases in which the second emission is
either primary (left-hand panels of each plot) or secondary (right-hand
panels).
Let us first focus on the ratio between the parton shower and the
analytic LC result (obtained using $C_A = 2C_F = 3$, upper panels of
each plot).
In most regions, the value of this ratio for all cases is either $1$,
when the effective colour factor is $C_A/2$, or $8/9$ if the emission
occurs with an additional $2C_F/C_A$ relative weight.
The other quantity that is displayed is the deviation between the
shower and the analytic result at full colour (lower panels of each
plot).
The segment method fails to describe the colour pattern if the two
emissions have commensurate angles.
The size of the deviation depends on the azimuthal angle $\psi$.
However, as we will see below, the deviation vanishes after
integrating over angular phase space.
The NODS scheme reproduces the exact full-colour matrix element in all
double-emission configurations.

For completeness, in the bottom two rows,
Figs.~\ref{fig:colour-2dcolour-qq} and \ref{fig:colour-2dcolour-gg}
also show results for the colour-factor-from-emitter (CFFE)
approach.
This scheme is implemented in standard dipole showers, and the colour
factor that multiplies the splitting function is determined according
to whether the emitter (as determined by the shower) is a (anti-)quark
or a gluon.
The results with the CFFE approach depend on how one decides which
of the two dipole ends is the emitter, and so we show CFFE with both
the PanGlobal shower (penultimate row) and the Dipole-$k_t$ shower (last
row), with the latter corresponding to the behaviour expected in
standard dipole showers. 
Labelling the rapidity of the first final-state emission as $\eta_1$
(equal to $5$), CFFE with the PanGlobal showers gives the wrong answer
on the primary plane (left-hand panels) when the second emission has
$0 \lesssim \eta \lesssim \eta_1$ (except in the $g_Ig_I\to Hg_F$ case).
For a $q\bar q$ initial state, i.e.\ Fig.~\ref{fig:colour-2dcolour-qq}
(left), CFFE with the Dipole-$k_t$ shower gives wrong answers in the
logarithmically extended region
\begin{equation}
  -2.5 =
    \frac12\left(\eta_1 - \ln \frac{Q}{k_{t,1}}\right)
  < \eta
  <
  \frac12\left(\eta_1 + \ln \frac{Q}{k_{t,1}}\right)
  =7.5\,,
\end{equation}
as expected from the analysis in
Refs.~\cite{Dasgupta:2018nvj,Hamilton:2020rcu}.
Note that this is the same region where recoil is misassigned for this
configuration, cf.~Eq.~(\ref{eq:dipole-kt-global-bad-rap-region}).
Where the initial state is $\bar q g$ or $gq$ (the right-hand
columns, respectively, of Figs.~\ref{fig:colour-2dcolour-qq} and
\ref{fig:colour-2dcolour-gg}), there are rapidity regions with two
dipoles simultaneously in play, each of which have different regions
of incorrect colour, resulting in a more complex overall structure.
\logbook{}{
  For the $\bar q g$ case (Fig.~\ref{fig:colour-2dcolour-qq})
\begin{verbatim}
            eta=0
   CF         :          CA  
   ------------------------
                   .------- CA
                   |   :
                   |  eta=7.5
                   CF
\end{verbatim}
  while for the $gq$ case (Fig.~\ref{fig:colour-2dcolour-gg}) we have
\begin{verbatim}
            eta=0
   CA         :          CF
   ------------------------
   ----------------.
   CA     :        | 
       eta=-2.5    | 
                   CF
\end{verbatim}
}

\subsubsection{Integrated tests}
\label{sec:integrated-colour-tests}

\begin{figure}[t]
    \centering
    \begin{subfigure}{\textwidth}
        \includegraphics[page=1, width=\textwidth]{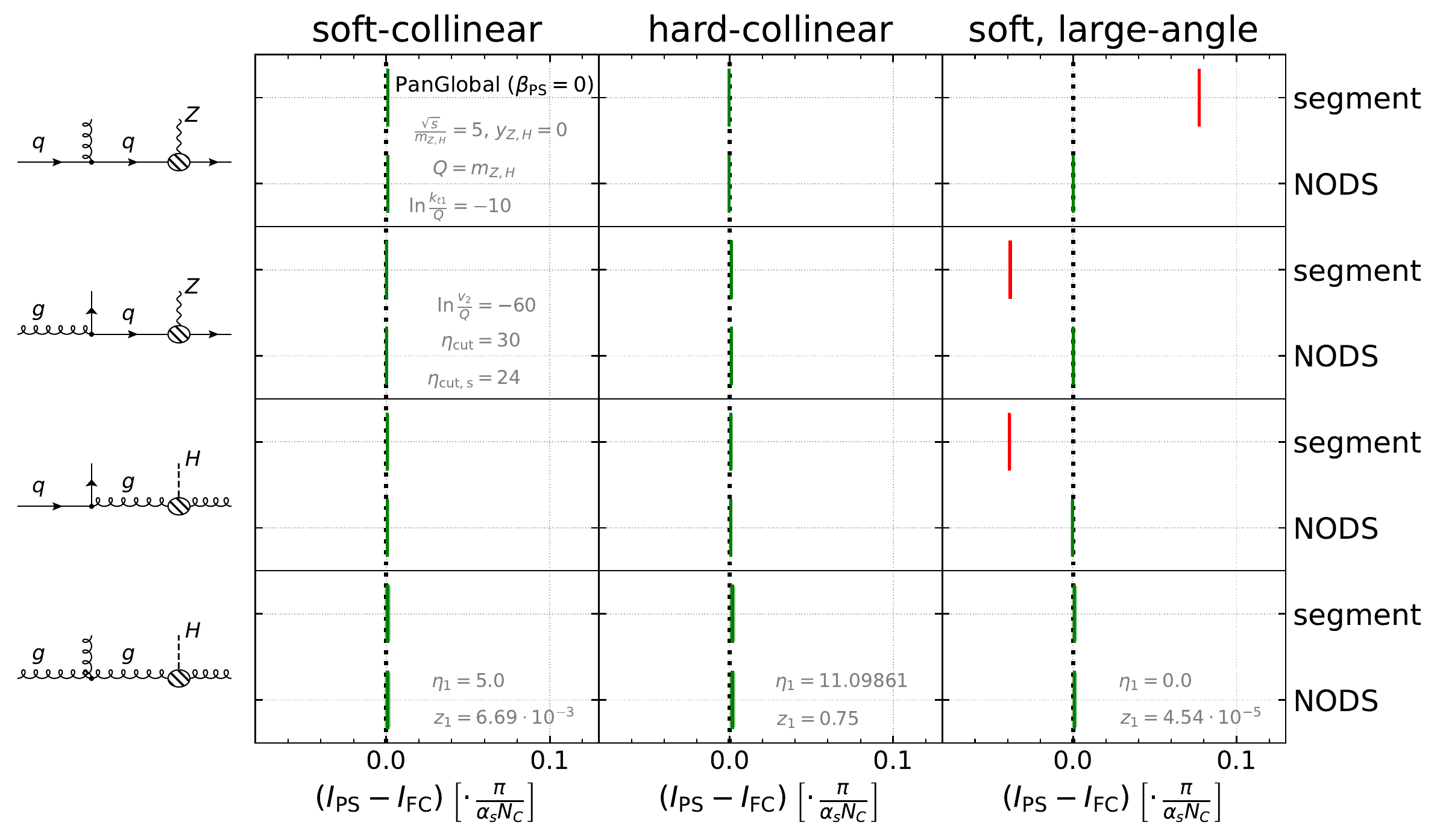}
    \end{subfigure}
    \caption{Normalised deviation of the integrated rate of a single
      soft emission (emission $2$), as produced by the parton shower $I_{\rm PS}$ from
      the configurations in the left-hand column,
      relative to the exact analytic results $I_{\rm FC}$ quoted in
      Appendix~\ref{app:colour}.
      We show four configurations, two for the $q\bar{q}\rightarrow Z$
      event, and one for the $gg\rightarrow H$ event, and only
      consider initial-state splittings (the results for final-state
      splittings may be found in Ref.~\cite{Hamilton:2020rcu}).
      For each configuration, we show three kinematics regimes for the
      first emission, as
      labelled at the top of the plot (and detailed at the bottom). 
      For each configuration and kinematic regime, we show results for
      both the segment and NODS colour schemes.
      The colours of the lines
      indicate whether the result is consistent with zero.
    }
    \label{fig:colour-integral}
\end{figure}

Next, we integrate the differential cross section over the emission
angles for the soft gluon to obtain the overall rate of emissions from
the shower, $I_{\rm PS}$.
We compare the result to the analytic full-colour expectation,
$I_{\rm FC}$.
We perform these comparisons for configurations where the first
emission (whose direction is \emph{not} integrated over) is either
soft and collinear, hard and
collinear, or soft and large-angle with respect to the initial-state emitting
system.%
\footnote{Explicitly, we set $\ln v_1/Q = -10$ and take
 $\bar{\eta}_{Q,1} = 5$ (soft collinear), $\bar{\eta}_{Q,1} = 11.09861$ (hard collinear), 
 and $\bar{\eta}_{Q,1} = 0$ (soft, large angle). 
 The corresponding $z_1, \eta_1$ values are quoted in Fig.~\ref{fig:colour-integral}.}
The second emission is radiated at a fixed
$\ln v_2/Q=-60$ and we impose a collinear regulator that takes a different value
depending on whether the emission is primary $\eta_{\rm cut}=30$ or secondary
$\eta_{\rm cut,s}=24$.\footnote{The reason for having two different regulators
    is to avoid any type of artificial cancellations that could lead to an
    apparent agreement between the shower and the analytics even if the
    implementation is not correct.} The integrated rate is computed for all
possible branching histories: starting with $q\bar{q}\rightarrow Z$, (i) the
backwards evolution of a quark into a gluon while emitting a quark ($q_1 g$),
(ii) the backwards evolution into a quark while emitting a gluon ($g_1 g$);  and
starting with $gg\rightarrow H$, (iii) the backwards evolution of a gluon into a
gluon while emitting a gluon ($g_1 g$), and (iv) the backwards evolution into a
quark while emitting a quark ($q_1 g$). The analytic expectations for $I_{\rm
FC}$ are derived in Appendix~\ref{app:colour}.

We show the comparison between $I_{\rm PS}$ and $I_{\rm FC}$ in
Fig.~\ref{fig:colour-integral} where the specific values for the kinematics of
the first emission, i.e.\ $z_1$ and $\eta_1$, are indicated on the plot. The NODS colour
scheme reproduces the FC analytic integral in all cases.
The
segment method fails to describe the analytic result for the soft large-angle
configuration except in the $g\to gg$ case, where there are no $C_F$
segments.
This failure was already observed for
final-state splittings in Ref.~\cite{Hamilton:2020rcu} and it is due to the
discrepancy between the physical $\eta$ and $\eta_{\rm approx}$, given by
Eq.~\eqref{eq:eta-approx}, in the large-angle regime. The disagreement vanishes
in the case of $g\to gg$ since in this case the leading and full-colour results
are identical. 

\begin{figure}[t]
    \centering
    \begin{subfigure}{\textwidth}
        \includegraphics[page=1, width=\textwidth]{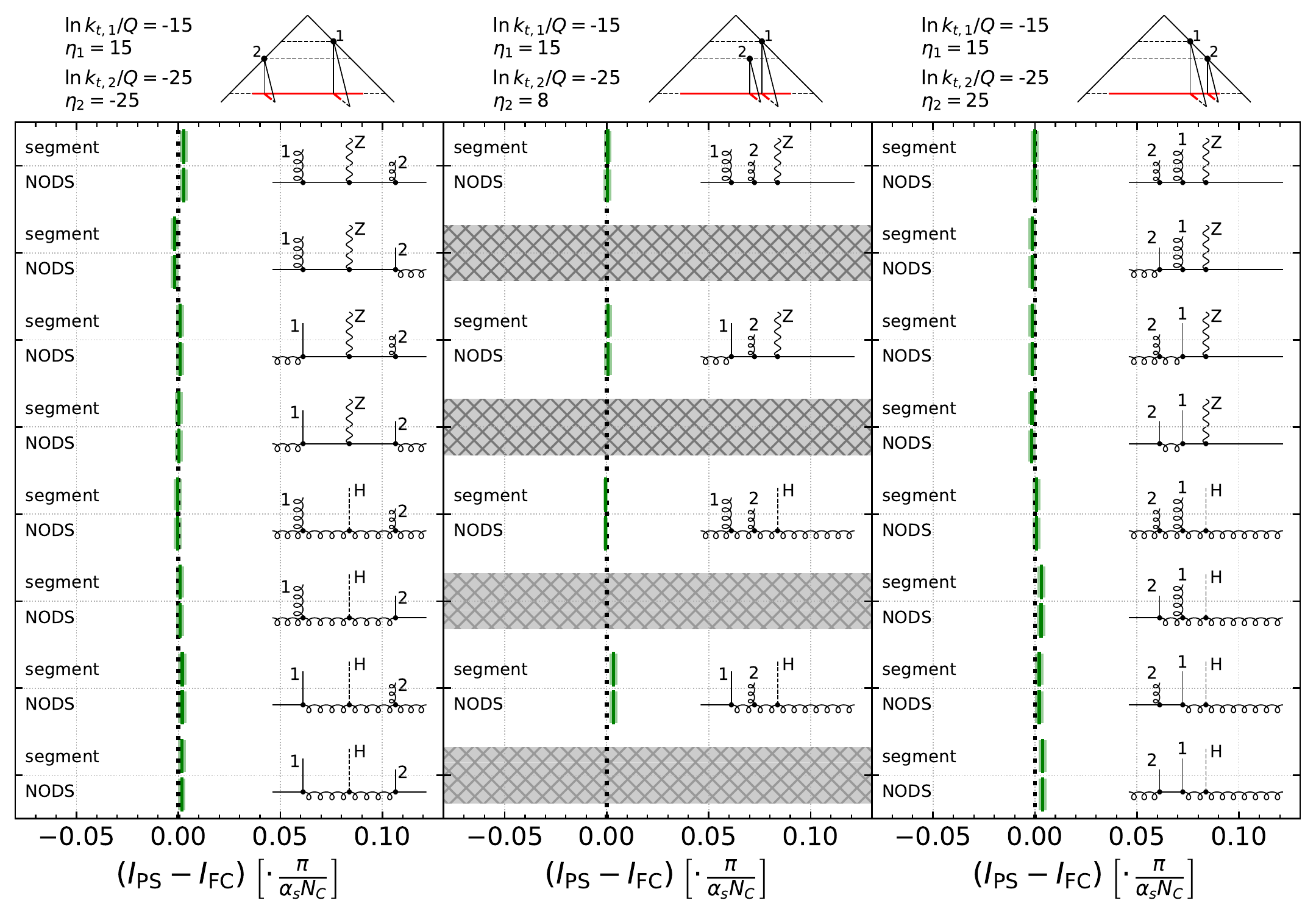}
    \end{subfigure}
    \caption{
      Similar to Fig.~\ref{fig:colour-integral}, but for soft emission
      from configurations with two prior initial-state
      emissions (well separated in rapidity, at least one of which is
      hard). 
    } 
    \label{fig:colour-integral-3em}
\end{figure}

Our last colour test considers three emissions. We fix the first emission, which
can be either a quark or a gluon, at 
\begin{align}
    \ln \frac{v_{1}}{Q} = -15\,,\quad \bar\eta_{Q,1} = 15\,,\quad \varphi_1 = 0\,.
\end{align}
Then, the shower generates a second lower-$\ln v$ emission (quark or gluon) with $\ln
v_{2}/Q = -25$ and $\varphi_2 = \pi/3$. To reduce the number of possibilities, we impose that this
second emission is a primary one. We consider three
kinematic configurations: (i) opposite hemisphere, i.e.\ $\bar{\eta}_{Q,2} < 0$,
(ii) same hemisphere with $0 < \bar{\eta}_{Q,2} < \bar{\eta}_{Q,1}$, and (iii) same
hemisphere with $0 < \bar{\eta}_{Q,1} < \bar{\eta}_{Q,2}$. At this stage, taking all the
possible flavor combinations into account, each of the previous kinematic
configurations correspond to a total of $8,4$ and $8$ cases, respectively. 

The third emission is then fixed to be a soft-collinear gluon with $\ln v_{3}/Q
= -45$, and arbitrary $\bar{\eta}_{Q,3}$, $\varphi_3$.

This gluon may effectively
be a primary 
emission or radiated off either the first or second emission. As for the
double-emission case, the third gluon contributes to the total integrated rate if it satisfies $\eta_3 < \eta_{\rm cut}(\eta_{\rm cut,s})$ depending on whether
it is a primary or secondary emission.
The analytic expectations are presented in Appendix~\ref{app:colour}. The difference between the shower and the
analytic result is shown in Fig.~\ref{fig:colour-integral-3em}. We find a
perfect agreement between the shower and
the analytic calculation with either of the two colour schemes. Notice that the segment method satisfies this test
in all cases since we are imposing that the first and second emissions
are collinear.
For the segment scheme, we would have encountered the same problem as
in the double-emission case if either had been radiated at a large
angle.

We have thus thoroughly tested the extension of both colour schemes to
initial-state radiation, up to $\order{\as^3}$, and shown that they
behave as expected.

\section{Spin correlations}
\label{sec:spin}
The last ingredient that we incorporate into our showers is an
algorithm to include spin correlations.
These are essential to
correctly reproduce the azimuthal structure of strongly angular-ordered collinear splittings, and so achieve full NLL accuracy
according to the matrix-element part of the PanScales
conditions outlined on p.~\pageref{item:panscales-1}.
In 
Refs.~\cite{Karlberg:2021kwr,Hamilton:2021dyz}, spin correlations were included
in the PanScales $e^+e^-$ showers using an approach based on the Collins-Knowles
algorithm~\cite{Collins:1987cp,Knowles:1987cu,Knowles:1988vs,Knowles:1988hu}.
While Collins~\cite{Collins:1987cp} originally proposed an algorithm to include
spin correlations for final-state showers,
Knowles~\cite{Knowles:1987cu,Knowles:1988vs,Knowles:1988hu} extended the
procedure to initial-state radiation and backwards evolution.
In this work, we extend the PanScales spin-correlation implementation
to hadron-hadron collisions, and validate it, as in the colour case,
by comparison to analytic matrix elements up to $\mathcal{O}(\as^3)$.
Relative to the Collins-Knowles algorithm, the PanScales
spin-correlation algorithm also accounts for the (dominant)
large-$\nc$ part of the spin correlations in soft
emissions~\cite{Hamilton:2021dyz}.

\subsection{Spin-correlation algorithm and extension to initial-state branching}
\label{sec:spin-correl-alg}

The starting point is the procedure outlined in
Refs.~\cite{Karlberg:2021kwr}, which successfully includes spin
correlations in any dipole or antenna shower.
The fundamental building blocks 
are the collinear branching amplitudes $\mathcal{M}^{\lambda_{\itilde}
\lambda_i\lambda_k}_{i \to \itilde k}$, which can be written in terms of
helicity-dependent splitting functions $ \mathcal{F}_{i \to \itilde 
k}^{\lambda_{\itilde} \lambda_i \lambda_k}(z)$ and a spinor product
$S_{\tau}(p_i,p_k)$, i.e.
\begin{equation}
    \mathcal{M}_{i \to \itilde k}^{\lambda_{\itilde} \lambda_i \lambda_k} = \frac{1}{\sqrt{2}} \frac{g_s}{p_i{\cdot}p_k} \mathcal{F}_{i \to \itilde k}^{\lambda_{\itilde} \lambda_i \lambda_k}(z) \, S_{\tau}(p_i,p_k)\;.
    \label{eq:branch-amp-spinor-prod}
\end{equation}
The spin indices of the particles $i\to \itilde k$ are denoted
$\lambda_{\itilde,i,k} = \pm 1$.
Note that in our convention, for initial-state splittings, the order
of spin indices in the superscript of
$\mathcal{M}_{i \to \itilde k}^{\lambda_{\itilde} \lambda_i
  \lambda_k}$ differs from that of particle indices in the subscript
(similarly for $\mathcal{F}$).
This choice facilitates re-use of code between our initial and
final-state implementations of spin correlations.
The label $\tau = \pm 1$ for the spinor
product indicates the sign of the  complex phase (following the convention of
Ref.~\cite{Kleiss:1985yh}), with
\begin{equation}
    \tau = \tilde{\lambda}_{i} + \tilde{\lambda}_{k} - \tilde{\lambda}_{\itilde} \mbox{ where } \tilde{\lambda} = 
    \begin{cases}
        \lambda/2 \mbox{ for a quark}, \\ 
        \lambda \mbox{ for a gluon}.
    \end{cases}
\end{equation} 
A derivation of the branching amplitudes for final-state collinear
splittings, $\mathcal{M}_{\itilde \to i k}^{\lambda_{\itilde}
  \lambda_i \lambda_k}$,  was presented in
Ref.~\cite{Karlberg:2021kwr}. 
In Appendix~\ref{app:spin} of this work we extend this derivation to
include initial-state collinear branchings.
The resulting helicity-dependent splitting amplitudes $ \mathcal{F}_{i \to
\itilde k}^{\lambda_{\itilde} \lambda_i \lambda_k}(z)$ for backwards
initial-state splittings are summarised in Table~\ref{tab:hel-dep-splittings}.\footnote{ 
      These amplitudes coincide with the final-state ones
      in Table~1 of Ref.~\cite{Karlberg:2021kwr} after replacing $z
      \leftrightarrow (1-z)$, including an overall factor of $\sqrt{z}$,
      exchanging $\lambda_{\itilde} \leftrightarrow \lambda_i$ and,
      in the case of incoming
      gluons, accounting for a further factor of $-1$.
}
The Collins-Knowles algorithm then makes use of these branching
amplitudes to construct a binary tree following the shower history.
This data structure facilitates the efficient computation of
spin-density matrices for new shower branchings, which are used to
sample the azimuthal distribution of that branching.\footnote{In the
  case of an antenna shower, we use $g(\bar{\eta}_Q)$ to decide which
  dipole leg acts as the emitter.}

The regime of validity of the Collins-Knowles algorithm is limited to
collinear $1 \to 2$ branchings. An extension to correctly model the
azimuthal distribution of soft large-angle branchings at leading
colour was presented in Ref.~\cite{Hamilton:2021dyz} for final-state
branchings. The colour-stripped amplitude for the emission of a soft
gluon from a colour dipole follows the eikonal approximation, and, for example for an initial-initial dipole, reads
\begin{eqnarray}
\mathcal{M}^{\lambda_k}_{ij \rightarrow \itilde \jtilde k}(\dots, p_i,
  p_j,p_k\dots )
  =
  g_s \left(\frac{p_i \cdot \epsilon_{\lambda_k}^*(p_k)}{p_i \cdot
  p_k} - \frac{p_j \cdot \epsilon_{\lambda_k}^*(p_k)}{p_j \cdot
  p_k}\right)
  \mathcal{M}(\dots, \tilde{p}_i, \tilde{p}_j,\dots )
  \,,\,\,\,\,\,\,\,\,\,\,\,\,
\label{eq:spin-eikonal-factorisation}
\end{eqnarray}
with $g_s \equiv \sqrt{4\pi\as}$ and 
an implicit constraint
$\delta_{\lambda_{\itilde}\lambda_i}\delta_{\lambda_{\jtilde}\lambda_j}$.
Note that helicity flips of the parent dipole partons are suppressed in
the soft limit, and we do not account for them here.

\begin{table}
    \centering
    \begin{tabular}{ c | c | c | c | c | c  | c }
        \toprule
        $\lambda_{\itilde}$  & $\lambda_i$  & $\lambda_k$  & $q_I \to \tilde q_Ig_F$ & $g_I \to \tilde q_I \bar{q}_F$ & $g_I \to \tilde g_Ig_F$ & $q_I \to \tilde g_Iq_F$    \\
        \midrule
        $\lambda$ & $\lambda$  & $\lambda$   & $\frac{1}{\sqrt{z(1-z)}}$ & 0 & $-\frac{1}{\sqrt{z}(1-z)}$ & $-\frac{1}{1-z}$  \\[8pt]
        $\lambda$ & $\lambda$  & $-\lambda$  & $\sqrt{\frac{1-z}{z}}$ & $\sqrt{1-z}$ & $-\frac{1-z}{\sqrt{z}}$ & $0$ \\ 
        $\lambda$ & $-\lambda$ & $\lambda$   & $0$ & $0$ & $0$ & $0$ \\
        $\lambda$ & $-\lambda$ & $-\lambda$  &  $0$ &$-\frac{z}{\sqrt{1-z}}$ &                 $-\frac{z^{3/2}}{1-z}$  & $-\frac{z}{1-z}$ \\
        \bottomrule
    \end{tabular}
    \caption{The helicity-dependent Altarelli-Parisi splitting amplitudes
        $\mathcal{F}_{i \to \itilde k}^{\lambda_{\itilde} \lambda_i
        \lambda_k}(z)$ that appear in
        Eq.~(\ref{eq:spin-eikonal-factorisation-spinor}),  using the convention
        of Eq.~\eqref{eq:p-initial-state} where $\itilde$ evolves backwards to
        $i$ while emitting $k$ such that $p_k = z/(1-z) \tilde{p}_i$, $p_i =
        1/(1-z)\tilde{p}_i$.
        Aside from overall colour factors, the amplitudes satisfy the relation 
        $(1-z) \sum_{\lambda_i,\lambda_k} |\mathcal{F}_{i \to \itilde k}^{\lambda_{\itilde} \lambda_i
        \lambda_k}(z)|^2 = P_{i \to \itilde k}(z)$, with the $P_{i \to
        \itilde k}(z)$ as given in Eqs.~(\ref{eq:Pij}).
    }
    \label{tab:hel-dep-splittings}
\end{table}

Although
formulated in terms of $1\rightarrow 2$ kinematics, the
Collins-Knowles algorithm can still be modified to account for soft
wide-angle emission, because the eikonal approximation does not depend on the spin of the dipole legs $i$ and $k$. 
The branching amplitudes of
Eq.~\eqref{eq:branch-amp-spinor-prod} have to be modified such that
the $i \rightarrow \itilde k$ branching acquires a dependence on the
kinematics of the $\itilde$'s dipole partner
$\jtilde$.
 Eq.~\eqref{eq:spin-eikonal-factorisation} can be
rewritten as (see Appendix~\ref{app:spin})
\begin{eqnarray}
\mathcal{M}^{\lambda_k}_{ij\rightarrow \itilde \jtilde k}(\dots, p_i, p_j,p_k\dots ) = \sqrt{2} g_s \frac{S_{-\lambda_k}(p_i, p_j)}{S_{-\lambda_k}(p_i, p_k) S_{-\lambda_k}(p_j, p_k)} \mathcal{M}(\dots, \tilde{p}_i, \tilde{p}_j,\dots )\,.\,\,\,\,\,\,\,\,\,\,\,\,
\label{eq:spin-eikonal-factorisation-spinor}
\end{eqnarray}
This implies that soft corrections can be included in the collinear branching amplitudes in the case $\lambda_{\itilde} =
\lambda_i$, leading to 
\begin{eqnarray}
    \mathcal{M}^{\lambda \lambda \lambda_k}_{i\rightarrow \itilde k} = \sqrt{2} g_s\sqrt{z} \mathcal{F}^{\lambda \lambda \lambda_k}_{i\rightarrow \itilde k}(z) \, \frac{S_{-\lambda_k}(p_i, p_j)}{S_{-\lambda_k}(p_i, p_k) S_{-\lambda_k}(p_j, p_k)}\,,
    \label{eq:spin-eikonal-factorisation-spinor-collins-knowles}
\end{eqnarray}
and analogously for initial-final and final-final dipoles. 
Note that the identification of a colour partner is only unambiguous
in the large-$\nc$ limit.
Therefore, our results for soft-spin correlations are only correct at
LC.\footnote{Ref.~\cite{Hamilton:2021dyz} investigated the size of
  subleading colour effects in the soft spin case for $e^+e^-$
  collisions and they were never larger than a few percent. }
Collinear spin correlations are not affected by
this.
The spinor products are evaluated numerically using the techniques
explained in Appendix~A of Ref.~\cite{Karlberg:2021kwr}, where special
care is required in choosing a reference spinor direction in the soft
limit, as is detailed in Appendix~C of
Ref.~\cite{Hamilton:2021dyz}.
The algorithm itself remains unchanged when
extended to the initial state, and details are given in Section~2.2 of
Ref.~\cite{Karlberg:2021kwr}.

\subsection{Matrix-element tests}

\begin{table}
    \centering
        \begin{tabularx}{1.\textwidth}{
            >{\hsize=0.25\hsize}C
            >{\hsize=0.25\hsize}C
            >{\hsize=0.25\hsize}C
            >{\hsize=0.25\hsize}C}
        \toprule
        \multicolumn{4}{c}{One initial, then one final} \\
        \midrule
        \multicolumn{2}{c}{Primary splitting} & \multicolumn{2}{c}{Secondary splitting}\\
        \midrule
        &&\\[-1em]
        $q_I \to \tilde{q}_I g_F$        & $A(z)= \frac{2(1-z)}{1+(1-z)^2}$ & $g_F \to q_F\bar q_F$ & $B(z)= \frac{-2z(1-z)}{1-2z(1-z)}$ \\
        $g_I \to \tilde{g}_I g_F$        & $A(z)= \frac{(1-z)^2}{(1-z(1-z))^2}$ & $g_F \to g_Fg_F$ & $B(z)= \frac{z^2(1-z)^2}{(1-z(1-z))^2}$  \\
        \midrule
        \toprule
        \multicolumn{4}{c}{Two initial, same side} \\
        \midrule
        \multicolumn{2}{c}{Primary splitting} & \multicolumn{2}{c}{Secondary splitting} \\
        \midrule
        &&\\[-1em]
        $g_I \to \tilde q_I \bar q_F$ & $A(z)= \frac{-2z(1-z)}{1-2z(1-z)}$ & $q_I \to \tilde g_I q_F$  & $B(z)= \frac{2z}{1+z^2}$ \\
        $g_I \to \tilde g_I g_F$        & $A(z)= \frac{z^2(1-z)^2}{(1-z(1-z))^2}$ & $g_I \to \tilde g_I g_F$  & $B(z)= \frac{z^2}{(1-z(1-z))^2}$  \\
        \midrule
        \toprule
        \multicolumn{4}{c}{Two initial, opposite side} \\
        \midrule
        &&\\[-1em]
        & $q_I  \to \tilde g_I q_F$ & \multicolumn{2}{l}{$A(z)= B(z) = \frac{2z}{1+z^2}$} \\
        & $g_I  \to \tilde g_I g_F$ & \multicolumn{2}{l}{$A(z)= B(z) = \frac{z^2}{(1-z(1-z))^2}$  } \\
        \bottomrule
        
    \end{tabularx}
    
    \caption{The coefficients $A(z)$ and $B(z)$ in
      Eq.~(\ref{eq:spin-azimuthal-angle}) for all possible sequences
      that involve at least one initial-state branching and non-zero
      spin correlations.
    The separate cases correspond with a final-state secondary branching (top), 
    a same-side initial-state secondary branching (middle), and an
    opposite-side initial-state secondary branching (bottom).
    These results complement the corresponding final-state ones given
    in Table~2 of Ref.~\cite{Karlberg:2021kwr}
      }
    \label{tab:A-B-coefficients1}
\end{table}

We validate our implementation of spin correlations at fixed order by comparing the shower weight to
analytic results as a function of the azimuthal angle $\Delta\psi_{ij}$ between
the planes spanned by emissions $i$ and $j$.
At $\mathcal{O}(\alpha_s^2)$, the differential cross section can be written as
\begin{align}
    \label{eq:spin-azimuthal-angle}
    \frac{d\sigma}{d\Delta \psi_{ij}} \propto a_0 \left( 1 + \frac{a_2}{a_0}
    \cos(2\Delta \psi_{ij}) \right) = a_0 \left( 1 + A(z_i) B(z_{j})
    \cos(2\Delta \psi_{ij}) \right)\,,
\end{align}
where the two non-zero Fourier coefficients $a_0$ and $a_2$ depend on
the type of branching, and on the momentum fractions associated with
the first ($z_i$) and second ($z_j$) splitting. The ratio $a_2 / a_0$ is equal to $0$ in
the absence of spin correlations. The analytic
expressions for $A_i(z_i)$ and $B_j(z_j)$ are given in
Table~\ref{tab:A-B-coefficients1}.
As was the case for the colour tests, we only show results for
the PanGlobal shower with $\betaps=0$, given that the implementation is largely
identical across all showers.

\begin{figure}[t]
    \centering
    \includegraphics[page=1, width=\textwidth]{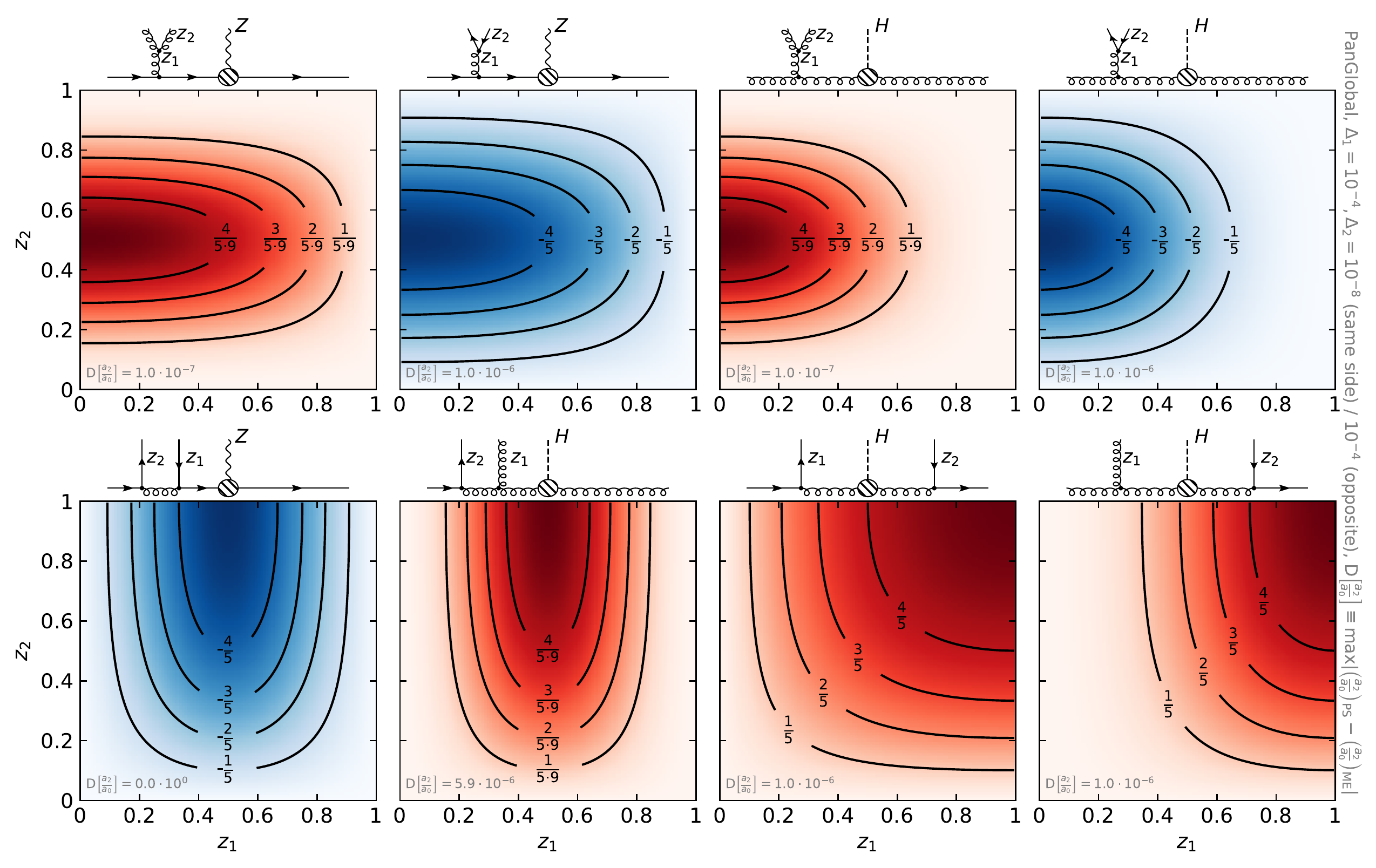}
    \caption{Size of the spin correlations $a_2/a_0$ at $\mathcal{O}(\as^2)$ for
          collinear splittings. The Feynman diagrams indicate the sequence of
          splittings under consideration. We study the azimuthal difference
          between the plane defined by the primary and secondary splittings with
          momentum fraction $z_1$ and $z_2$, respectively. The colour indicates
          the size of $a_2/a_0$ as predicted by the shower. Black lines indicate
          constant values for this ratio, and are obtained using the analytic
          predictions of Table~\ref{tab:A-B-coefficients1}.  The maximum
          deviation of the analytic prediction and the shower is given by ${\rm
          D}\big[\frac{a_2}{a_0}\big] \equiv {\rm
          max}\big|\big(\frac{a_2}{a_0}\big)_{\rm PS} -
          \big(\frac{a_2}{a_0}\big)_{\rm ME} \big|$, with
          $\big(\frac{a_2}{a_0}\big)_{\rm PS}$ the shower prediction and
          $\big(\frac{a_2}{a_0}\big)_{\rm ME}$ the analytic prediction for
          the matrix element.}
    \label{fig:plot-spin-fixed-order-collinear}
\end{figure}

To begin with, we consider purely collinear branchings at
$\mathcal{O}(\as^2)$, where a first initial-state branching is
followed by either (i) a final-state branching, (ii) another
initial-state branching of the same parton, or (iii) another
initial-state branching on the opposite site. The shower-to-analytics
comparison of the $a_2/a_0$ ratio as a function of $z_1$ and $z_2$
across a number of branching sequences is shown in
Fig.~\ref{fig:plot-spin-fixed-order-collinear}. The coloured
background shows the shower result, while the black contour lines
indicate the analytic expectation given by
Table~\ref{tab:A-B-coefficients1}.
The largest absolute magnitude of the deviation between the shower and
the analytic result, ${\rm D}[a_2/a_0]$, is indicated in the lower left
corner of each panel.\footnote{Small deviations are expected because
  the splittings angles that we use, while small, are not
  asymptotically so.}
No cases with intermediate quarks
are shown, as they have vanishing spin correlations. The top row
shows cases where an initial-state gluon emission is followed by a
final-state splitting of that same gluon. The ratio $a_2/a_0$ is
negative when the final-state gluon splits to a quark--anti-quark
pair, and positive when it splits to a gluon-gluon pair. Spin
correlations are maximal when $z_1 \rightarrow 0$ and $z_2 = 0.5$,
i.e.~the gluon is soft and the energy is shared equally between the
final-state partons. This can be deduced from
Table~\ref{tab:A-B-coefficients1}.
The spin
correlations for the $g\rightarrow gg$ final-state splittings fall off
more steeply as $z_2 \rightarrow 0$ or $1$, and its absolute maximum
size is smaller ($1/9$) than the $g\rightarrow q\bar{q}$ splittings
($1$). We also observe that spin-correlations as $z_1 \rightarrow 1$
decrease more rapidly for gluon backwards splittings than for quarks.

The two leftmost panels in the bottom row of
Fig.~\ref{fig:plot-spin-fixed-order-collinear} show the spin
correlations resulting from two subsequent initial-state branchings on
the same side. Splitting configurations with non-vanishing spin
correlations occur in both $Z$ and $H$ production processes. For
brevity, we only show the cases where an intermediate gluon is
produced, which then backwards evolves into a quark.  This time, spin
correlations are maximal when $z_1 = 0.5$, $z_2 \rightarrow 1$, but
vanish when $z_1 \rightarrow 0, 1$ and $z_2 \rightarrow 0$.\footnote{There are
two additional cases not shown, namely where the initial-state gluon
backwards evolves into a new initial-state gluon and emits a
final-state one, which carry the same sign and show the same
behaviour, except that the spin correlations drop off more sharply as
$z_1 \rightarrow 0,1$ and $z_2 \rightarrow 0$.} 

Finally, the two rightmost panels in the bottom row of
Fig.~\ref{fig:plot-spin-fixed-order-collinear} show the spin correlations in
opposite-side initial-state branchings. These peak at $z_1, z_2 \rightarrow 1$,
i.e.~the limit where the emitted partons are soft, and vanish for either $z_1
\rightarrow 0$ or $z_2 \rightarrow 0$.
The case where both gluons
backwards evolve into a quark is necessarily symmetric in the $z_1, z_2$ plane, whereas a
backwards evolution into a gluon again leads to an enhanced reduction in
$a_2/a_0$ as $z_1\rightarrow 0$. The case where both gluons backwards evolve
into a gluon is not shown in the figure.
It behaves similarly to the case where
both gluons backwards evolve into a quark, but with a stronger drop off as $z_1,
z_2 \rightarrow 0$.   

\begin{figure}[t]
    \centering
    \includegraphics[page=1, width=\textwidth]{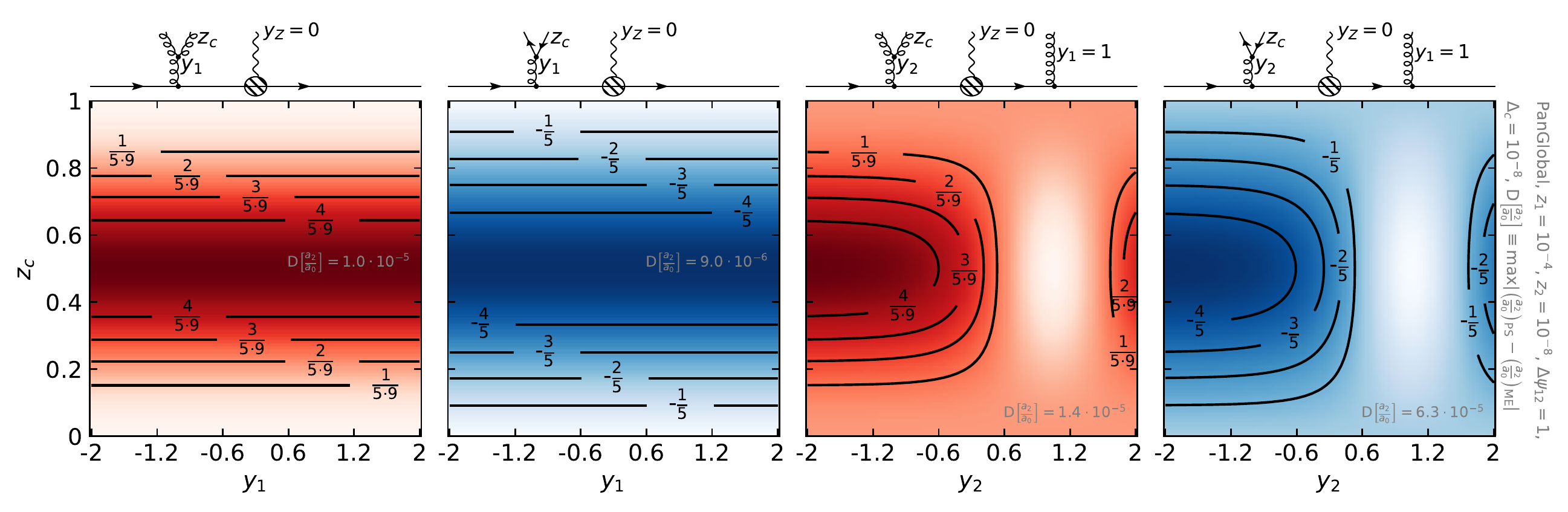}
    \caption{Size of the spin correlations for
    sequences that involve both soft and collinear splittings,
        showing $a_2/a_0$ at $\mathcal{O}(\as^2)$ 
    (two left-hand plots) and $\mathcal{O}(\as^3)$ (two right-hand
        plots).
        The Feynman diagrams indicate the
    sequence of splittings under consideration for all cases. We consider the
    azimuthal difference between the plane defined by the primary soft splitting
    with momentum fraction $z_1$ ($z_2$), and the plane defined by the second
    (third) splitting with momentum fraction $z_c$. The colour indicates the
    size of $a_2/a_0$ as predicted by the shower. Black lines indicate constant
    values for this ratio, and are obtained by using crossing relations in the
    matrix elements calculated in Ref.~\cite{Karlberg:2021kwr} for final-state
    configurations.}
    \label{fig:plot-spin-fixed-order-soft}
\end{figure}

In Fig.~\ref{fig:plot-spin-fixed-order-soft} we repeat the same
exercise as in Fig.~\ref{fig:plot-spin-fixed-order-collinear}, but
with kinematic configurations in which one or more branchings are
soft. In this case, the analytic calculation is obtained by crossing
the relevant matrix elements in the soft limit, given in Appendix~A of
Ref.~\cite{Karlberg:2021kwr}. On the two leftmost plots, we examine
configurations at $\mathcal{O}(\as^2)$, where a soft-wide angle gluon
$g_1$ with energy fraction $z_1$ is emitted from the initial-state
dipole $q\bar{q}$. This gluon then splits collinearly as
$g_1 \rightarrow gg$ (first plot) or $g_1\rightarrow q\bar{q}$ (second
plot) with momentum fraction $z_c$. We scan over the rapidity $y_1$ of
gluon $g_1$ relative to the $q\bar{q}$ system between $-2< y_1 < 2$,
and over the energy fraction $z_c$ of the emitted parton ($g$ or
$q$). Spin correlations are in this case independent of the rapidity
of the soft gluon.\footnote{A purely collinear Collins-Knowles
  algorithm would in general not correctly reproduce this pattern
  \cite{Hamilton:2021dyz}.}
The spin correlations are again maximal
in absolute size when the energy fraction of the gluon is shared equally between
the two final-state partons. 

A more interesting pattern appears at $\mathcal{O}(\as^3)$ as
displayed in the two rightmost panels of
Fig.~\ref{fig:plot-spin-fixed-order-soft}.
In this case we study the azimuthal correlations between the first and
third emission.
The
first gluon emission is now fixed at $y_1 = 1$ with an energy fraction $z_1 =
10^{-4}$, while we scan the rapidity of a second gluon emission between
$-2 < y_2 < 2$ with $z_2 = 10^{-8}$, which then splits collinearly. 
The two soft gluons are emitted at different azimuthal angles, $\Delta
\psi_{12} = 1$.
Because we fix $\Delta \psi_{12}$, the analytical form for the
azimuthal correlations needs to be extended relative to
Eq.~(\ref{eq:spin-azimuthal-angle}), and now reads 
\begin{align}
    \label{eq:spin-azimuthal-angle-13}
  \frac{d\sigma}{d\Delta \psi_{13}} \propto
  a_0 \left( 1 + \frac{a_2}{a_0} \cos(2\Delta \psi_{13}) + \frac{b_2}{a_0} \sin(2\Delta \psi_{13}) \right).
\end{align}
We plot just the ratio $a_2/a_0$ and see that it is enhanced when the
second gluon is emitted with a larger rapidity difference with respect
to the first gluon, and when its energy fraction is shared equally
between the children ($z_c = 0.5$).

Finally, we also check the spin correlations at $\mathcal{O}(\as^3)$ for
collinear splittings. We consider two configurations: (i) one backwards
splitting followed by two final-state emissions, and (ii) two backwards
splittings on opposite hemispheres followed by one final-state splitting. For
case (i) we consider both $q\bar{q}\to Z$ and $gg\to H$, whereas for case (ii)
we only consider $gg\to H$, as $q\bar{q}\to Z$ features no spin correlations due
to the intermediate quark line. The azimuthal angle of the first emission is
fixed to $\psi_1 = 0$. For the second emission we fix both the longitudinal
momentum fraction $z_2 = 0.5$ and the azimuthal angle
$\psi_2 = \pi/6$ (if emission $2$ is secondary from $1$, this is the
angle between the $1{-}2$ plane and the beam--$1$ plane; if it
is primary, it is the angle between the beam--$2$ plane and the
beam--$1$ plane).
Further, we fix the angles of these three collinear
emissions with respect to the emitter at $\Delta_1 = 10^{-4}$,
$\Delta_2 = 10^{-8}$, and $\Delta_3 = 10^{-12}$.
Finally, we integrate over the azimuthal angle $\psi_3$ to obtain the
Fourier coefficients in Eq.~(\ref{eq:spin-azimuthal-angle-13}) and scan over $z_1$ and
$z_3$.

\begin{figure}[t]
    \centering
    \includegraphics[page=5, width=\textwidth]{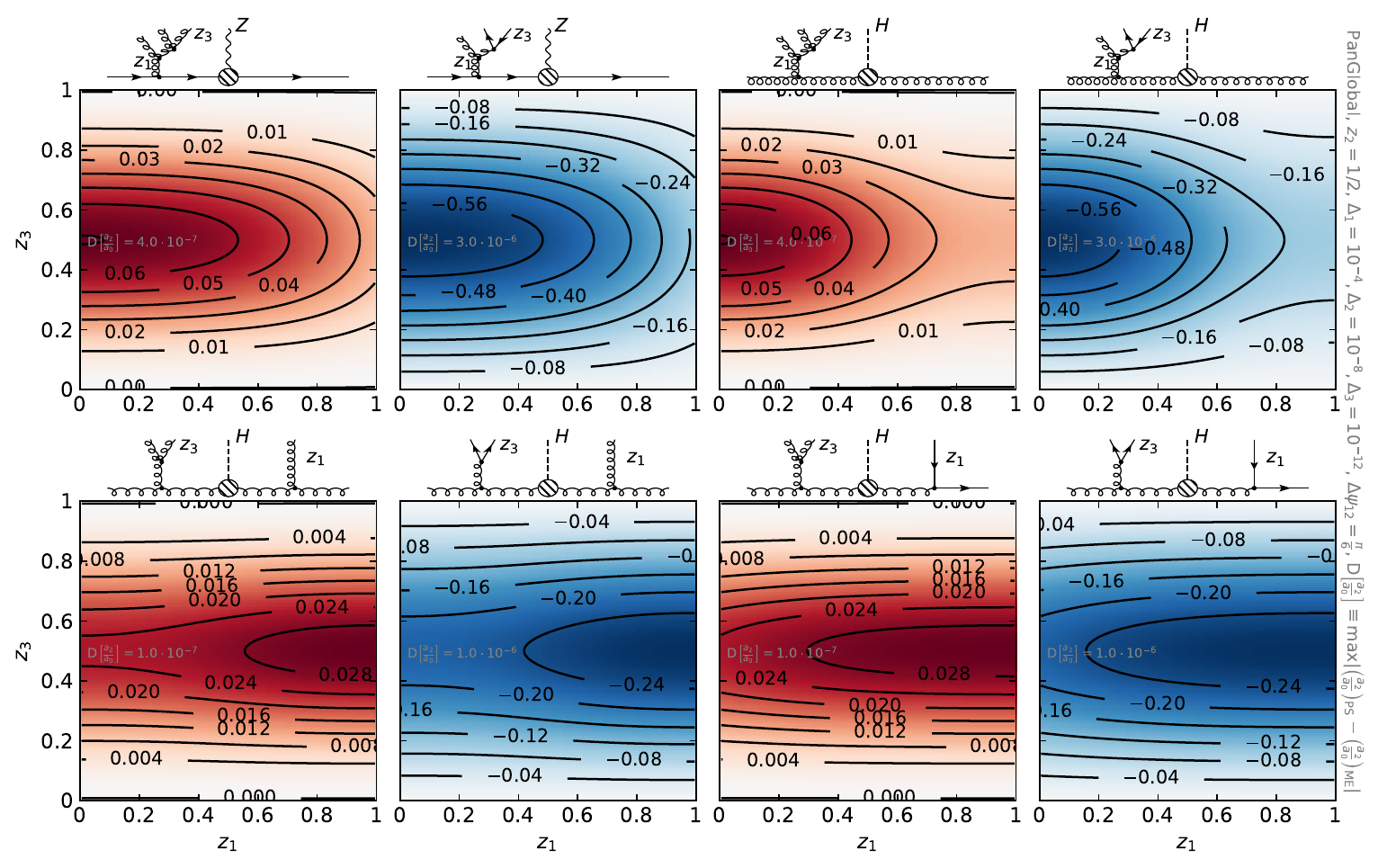}
    \caption{
      Analogue of  Fig.~\ref{fig:plot-spin-fixed-order-collinear}, but for 
    the spin correlations $a_2/a_0$ at $\mathcal{O}(\as^3)$ for
    three collinear splittings.
    The first splitting carries a longitudinal momentum fraction $z_1$, the
    third one $z_3$. We fix the longitudinal momentum fraction of the second
    emission to $z_2 = 0.5$, and its azimuthal angle such that
    $\Delta \psi_{12} = \pi / 6$.
  }
    \label{fig:plot-spin-fixed-order-triple-col}
\end{figure}

\begin{figure}[t]
    \centering
    \includegraphics[page=13, width=\textwidth]{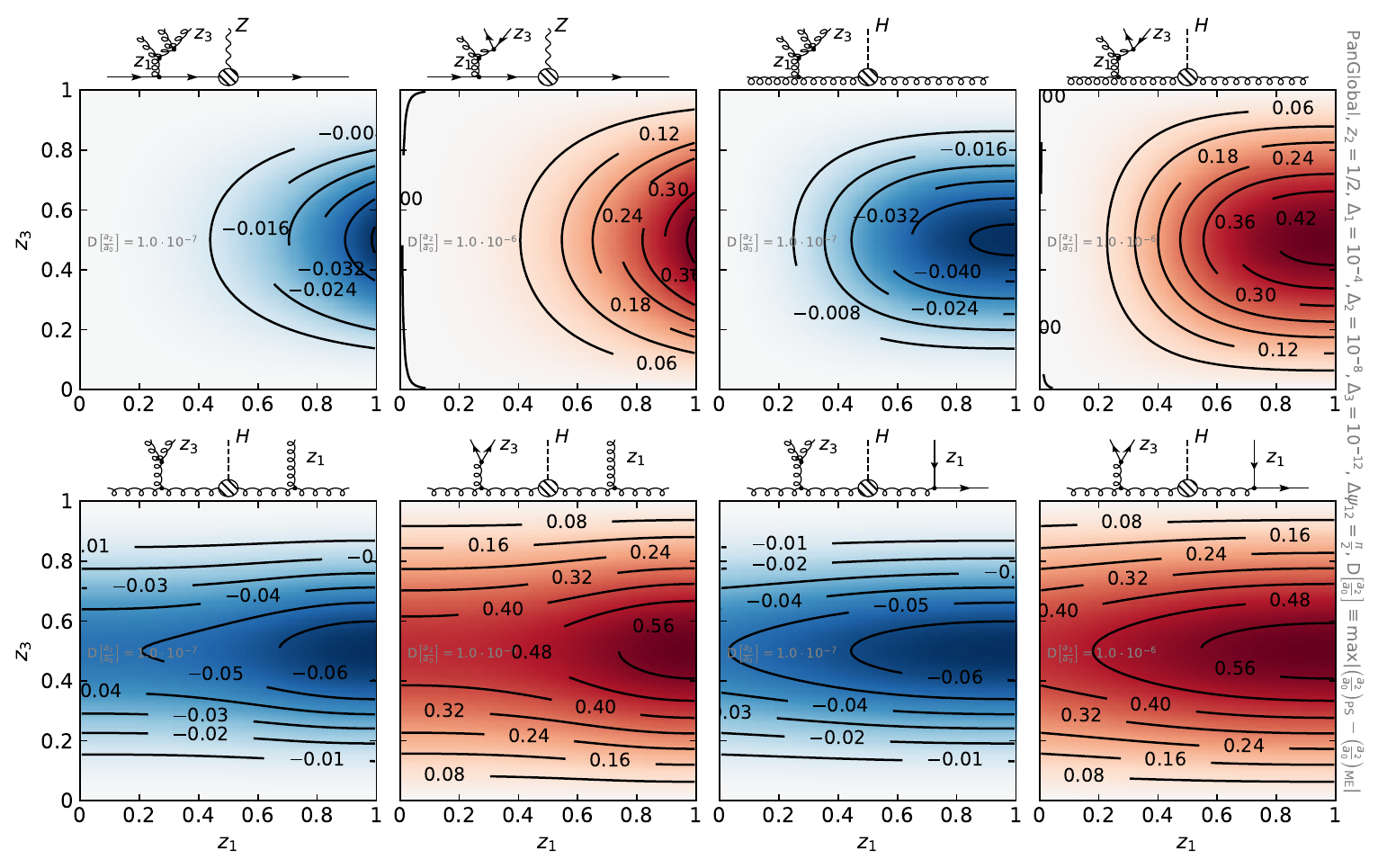}
    \caption{Same as Fig.~\ref{fig:plot-spin-fixed-order-triple-col}
      but with $\Delta \psi_{12} = \pi/2$. }
    \label{fig:plot-spin-fixed-order-triple-col-phi2}
\end{figure}

In Fig.~\ref{fig:plot-spin-fixed-order-triple-col}, we also show the resulting ratio
$a_2 / a_0$ as a function of $z_1$ and $z_3$.\footnote{The $z_2$-dependence of
$\Delta \psi_{13}$ is given as an overall normalisation, which reads
\begin{align}
    \frac{(1-z_2)^2}{(1-z_2+z_2^2)^2}
\end{align}
for a final-state splitting of a gluon into a gluon-gluon pair, relevant for the
splittings considered for configuration (i) (upper row of
Fig.~\ref{fig:plot-spin-fixed-order-triple-col}), and
\begin{align}
    \frac{(1-z_2)^2 z_2^2}{(1-z_2+z_2^2)^2}
\end{align}
for the backwards splitting of an initial-state gluon into a
final-state gluon and a new initial-state gluon, relevant for
configuration (ii) (lower row of
Fig.~\ref{fig:plot-spin-fixed-order-triple-col}). For $z_2 = 0.5$, the
overall normalisation of configurations (i) and (ii) is $4/9$ and
$1/9$, respectively.}  Overall, we find an excellent agreement between
the shower and the analytic expectations.
The shape and magnitude of the spin correlations are strongly dependent
on $\psi_2$, and for reference, in
Fig.~\ref{fig:plot-spin-fixed-order-triple-col-phi2} we show the spin
correlations if we change $\psi_2$ to $\pi /2$.
\logbook{3a8e499394d}{see even pages of
  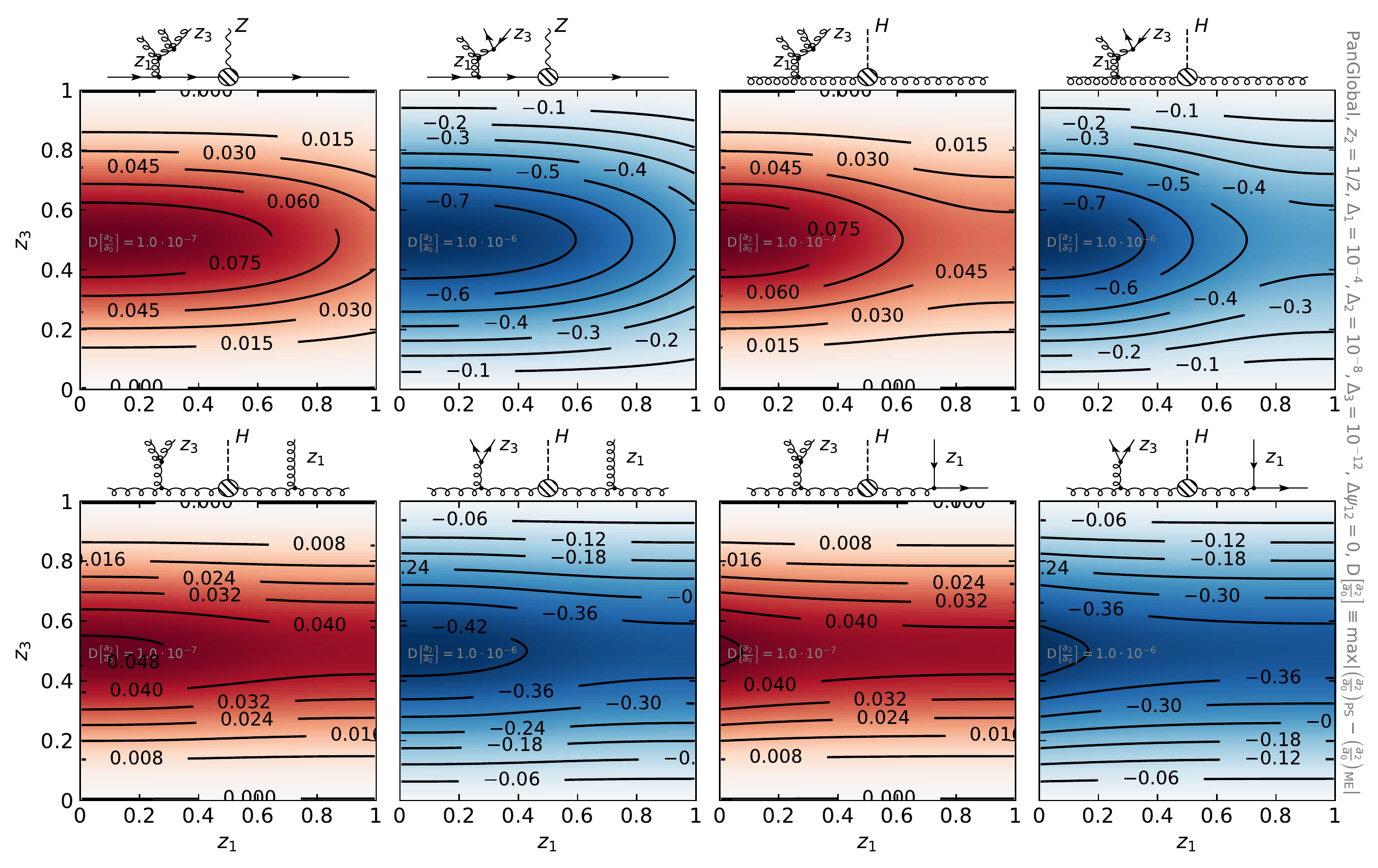 for the $b_2$ coefficients}

The tests that we have shown in this section provide solid validation
of the correctness of the spin-correlation implementation in the
PanScales showers, also for initial-state branching.

\section{Conclusions}
\label{sec:conclusions}

In this work we have introduced new dipole showers applicable to
colour-singlet production processes in hadronic collisions, extending
earlier PanScales work on final-state showers~\cite{Dasgupta:2020fwr}.

The hadron-collider PanGlobal shower, Section~\ref{sec:panglobal},
shares the characteristic of the final-state case that the dipole map
assigns only longitudinal recoil to the dipole parent particles.
The detailed mechanism to ensure energy-momentum conservation differs,
with recoil transverse to the beam being assigned to an explicit
``hard system'' (in the cases we have studied, the colour singlet) and
energy and longitudinal momentum conservation ensured by rescaling the
incoming momenta.
As in the final-state case this shower can be used with transverse
momentum ordering, $\betaps=0$ in Eq.~(\ref{eq:v-def}), and more
generally for any $0\le \betaps < 1$.

The hadron-collider PanLocal (dipole and antenna) family,
Section~\ref{sec:panglobal}, as with the final-state case, assigns all
recoil dipole locally within the dipole.
For initial-state branchings this effectively results in a change in
the direction of the incoming parton(s).
The incoming parton(s) are then realigned with the original axis
through a Lorentz transformation of the whole event, under the
constraint that the rapidity of the hard system should remain
unchanged relative to that before the branching.
As in the final-state case, the PanLocal hadron-collider showers are
expected to be NLL accurate for $0< \betaps < 1$, and the use of the
hard-system momentum to provide a fixed reference frame is critical in
order to correctly identify when to assign transverse recoil to one or
other end of the dipole.

To provide a point of comparison, we have also formulated a
``standard'' transverse-momentum ordered dipole shower, Dipole-$k_t$,
Section~\ref{sec:dipolekt}, which shares similarities with existing
widely used dipole showers.

We have carried out a number of ``matrix-element'' tests at fixed emission multiplicity,
related to the conditions needed to achieve NLL accuracy.
The core set of tests for the new showers has been to demonstrate that
they satisfy the PanScales condition that emission in one region of
phase space should not modify earlier emissions in logarithmically distant regions of
phase space, an essential condition for accuracy beyond leading
logarithms.
Subtleties that are new in the initial-state case concern potential
interplay between hard-collinear and soft-collinear emissions, in
particular because the momentum of a post-branching initial-state
particle can greatly exceed the pre-branching value.
Examining this region led to specific choices for the kinematic maps
and post-branching Lorentz transformations in the new showers, which
while not critical for NLL accuracy (at least in a regime where
$\ln s/m_Z^2$ remains finite), do help avoid unmotivated interplay
between soft-collinear and hard-collinear branchings.
Our tests also served to illustrate the way in which the PanScales
logarithmic accuracy conditions fail to be respected by standard
dipole shower schemes, as they do also for the $\betaps=1$ choice (and
$\betaps=0$ for PanLocal) for current PanScales showers.

In addition to designing new hadron-collider showers, we have also
extended the final-state schemes for subleading
colour~\cite{Hamilton:2020rcu} as well as collinear and soft spin
correlations~\cite{Karlberg:2021kwr,Hamilton:2021dyz}, introducing the
new elements needed for initial-state radiation.
These developments have been tested through comparisons to fixed-order
matrix elements with up to three emissions.

Our companion work~\cite{vanBeekveld:2022ukn} will provide
comparisons of the results of our showers to analytic all-order
resummations, demonstrating that NLL accuracy has indeed been achieved
for a wide range of observables.
Future work will examine applications to processes beyond the
colour-singlet hadro-production cases addressed here.

\section*{Acknowledgements}

We are grateful to our PanScales collaborators (%
Mrinal Dasgupta,
Fr\'ed\'eric Dreyer,
Basem El-Menoufi,
Keith Hamilton,
Alexander Karlberg,
Rok Medves,
Pier Monni,
Ludovic Scyboz%
),
for their work on the code, the underlying
philosophy of the approach and comments on this manuscript.
We also wish to thank Scarlett Woolnough for a careful reading and
helpful comments on the manuscript.

This work was supported
by a Royal Society Research Professorship
(RP$\backslash$R1$\backslash$180112) (MvB, GPS),
by the European Research Council (ERC) under the European Union’s
Horizon 2020 research and innovation programme (grant agreement No.\
788223, PanScales) (SFR, GPS, GS, ASO, RV), 
and by the Science and Technology Facilities Council (STFC) under
grant ST/T000864/1 (MvB, GPS).

\appendix
\section{Spin-averaged splitting functions}
\label{app:splitting-func}
The DGLAP splitting functions $P_{i\to jk}$  are given by 
\begin{subequations}
  \label{eq:Pij}
  \begin{align}
    P_{q\to q g}(z) &= C_F \frac{1+(1-z)^2}{z}\,,\label{eq:Pqg}\\
    P_{g\to gg}(z) &= 2C_A \left[ \frac{z}{1-z}+\frac{1-z}{z}+z(1-z)\right],\label{eq:Pggg}\\
    P_{g\to q\bar{q}}(z) &= T_R\left[z^2+(1-z)^2\right]. \label{eq:Pgqq}
  \end{align}
\end{subequations}
For final-state splitting in Eq.~(\ref{eq:dipole-prob}), 
we employ
\begin{subequations}
  \begin{align}
    P_{\tilde q\to q g}^{\text{FS}}(z) &= P_{q\to q g}(z)= C_F \frac{1+(1-z)^2}{z}\,,\\
    P_{\tilde g\to gg}^{\text{FS}}(z) &= \frac{1}{2}P_{g\to gg}^{\text{asym}}(z) = \frac{C_A}{2} \frac{1+(1-z)^3}{z},\label{eq:PgggFSR}\\
    P_{\tilde g\to q\bar{q}}^{\text{FS}}(z) &=\frac{1}{2}P_{g\to q\bar{q}}^{\text{asym}}(z)= T_R (1-z)^2. \label{eq:PgqqFSR}
  \end{align}
\end{subequations}
The last two expressions include a factor of $1/2$ to compensate for
the double counting due to the fact that a gluon belongs to two
dipoles, with a further $1/2$ symmetry factor for identical particles
in the final state in the case of $P_{\tilde g\to gg}$.
One of the factors of $1/2$ is implemented by taking an asymmetric
subset of the $z \leftrightarrow (1-z)$ symmetric terms.
The use of the
asymmetric kernels in Eqs.~(\ref{eq:PgggFSR}) and (\ref{eq:PgqqFSR}) is
possible due to the symmetry in the exchange of final-state particles.
This symmetry does not hold anymore for an initial-state branching. In that case we use
\begin{subequations}
  \begin{align}
    P_{q\to \tilde q g}^{\text{IS}}(z) &= P_{q\to q g}(z)= C_F
                                  \frac{1+(1-z)^2}{z}\,,\\
    \label{eq:PISqtoqg}
    P_{q\to \tilde g q}^{\text{IS}}(z) &=
                                  \begin{cases}
                                    P_{q\to q g}(1-z) = C_F \frac{1+z^2}{1-z}\quad& \mbox{if the gluon is the }\bar{3}\mbox{ end of the dipole,}\\
                                    0 \quad& \mbox{if the gluon is the }3\mbox{ end of the dipole,}\\
                                  \end{cases}    \\
    \label{eq:PISqbartoqbarg}
    P_{\bar q\to \tilde g \bar q}^{\text{IS}}(z) &=
                                  \begin{cases}
                                    0\quad& \mbox{if the gluon is the, }\bar{3}\mbox{ end of the dipole,}\\
                                    P_{\bar{q}\to \bar{q} g}(1-z) = C_F \frac{1+z^2}{1-z} \quad& \mbox{if the gluon is the, }3\mbox{ end of the dipole,}\\
                                  \end{cases}    \\
    P_{g\to \tilde gg}^{\text{IS}}(z) &= \frac{1}{2} P_{g\to gg}(z)= C_A \left[ \frac{z}{1-z}+\frac{1-z}{z}+z(1-z)\right],\\
    P_{g\to \tilde q\bar{q}}^{\text{IS}}(z) &=\frac{1}{2}P_{g\to
                                       q\bar{q}}(z)= \frac{1}{2}T_R
                                       \left[(1-z)^2+z^2\right].
  \end{align}
\end{subequations}
The choice in Eqs.~(\ref{eq:PISqtoqg}) and (\ref{eq:PISqbartoqbarg})
to split only the $\bar 3$ or the $3$ end of the dipole is natural in
terms of how the shower organises colour flows.
For possible future extension beyond NLL accuracy (i.e.\ beyond LO
splitting functions) one should keep in mind, however, that such a
choice may introduce effective NLO splitting terms with a spurious
difference between $q\to q'+X$ and $q\to \bar q' + X$.
We leave the study of this question to future work.


\section{Shower mapping coefficients and scale choices}
\label{app:shower-coefficients}
Here we provide the kinematic mappings associated with an
$\itilde \jtilde \to ijk$ dipole splitting.
We denote by $i$ and $j$ the partons that descend from the parent
dipole (and maintain their flavour in the case of gluon emission),
while $k$ is the newly emitted  
parton, whose transverse component, $k_\perp^\mu$, will always be given
by Eq.~(\ref{eq:kperp-decomposition}).

All of our shower implementations will share the factorisation scale
choice 
\begin{equation}
  \label{eq:muF-choice}
  \ln \mu_F = \ln Q + \frac{1}{1+\betaps} \ln \frac{v}{Q}\,,
\end{equation}
where $\betaps$ is taken to be zero for the Dipole-$k_t$ shower
family.
In this work and in our companion paper~\cite{vanBeekveld:2022ukn},
we restrict $v < Q$.
The expression to be used for $\muR$ in the shower branching
probability, Eq.~(\ref{eq:dipole-prob}), will be given separately for
each shower.

\subsection{Dipole-$k_t$}
\label{app:shower-coefficients-dipole-kt}

The kinematic maps that we use for the Dipole-$k_t$ shower follow
from the Catani-Seymour dipole subtraction
formalism~\cite{Catani:1996vz} and, unless otherwise specified, are as
used in a number of dipole
showers~\cite{Schumann:2007mg,Dinsdale:2007mf,Platzer:2009jq,Hoche:2015sya}. 
The mapping coefficients and boosts expressions for each of the 
dipole types are as follows.

\paragraph{{Initial-Initial:}} The dipole map is given by
\begin{subequations}
  \begin{align}
    & p^{\mu}_i = a_i \tilde{p}_i^{\mu}\,,   \\
    & p^{\mu}_j = \tilde{p}_j^{\mu}\,,  \\
    & p^{\mu}_k = a_k \tilde{p}_i^{\mu} + b_k \tilde{p}_j^{\mu} - 
      k_{\perp}^{\mu}\,, 
  \end{align}
\end{subequations}
and the coefficients read 
\begin{align}
& a_k = \frac{1-\zeta}{x_{jab}}, \,\, b_k = \frac{\kappa^2}{1-\zeta}\,\, , a_i = 
\frac{1}{x_{jab}},\,\, |k_\perp|^2=2 a_k b_k \tilde{p}_i \cdot \tilde{p}_j,
\end{align}
where $\zeta = 1-z$ with $z$ as defined in
Eq.~(\ref{eq:eta-dipolekt}), and
\begin{equation}
  \label{eq:dipole-kt-z-kappa-IS}
  \zeta = \frac{1}{1 + \kappa e^{\bar\eta_{\rm dip}}}\,,
  \quad
  \kappa^2 =  \frac{v^2}{\tilde{s}_{ij}}\,,
  \quad
  x_{jab} = \zeta -\frac{\kappa^2}{1-\zeta}\,.
\end{equation}
The boost acts on all of the final-state particles except the newly
created one ($k$), and takes their 
pre-branching total momentum $\tilde{F}^{\mu} = \tilde{p}^{\mu}_i +
\tilde{p}^{\mu}_j$ to a new total momentum  
$F^{\mu} = p^{\mu}_i + p^{\mu}_j - p^{\mu}_k$ where $\widetilde F^2 = F^2$. The boost reads
\begin{eqnarray}
	\label{eq:boost-wo-rotation}
\Lambda^{\mu \nu}({F},\widetilde F) = g^{\mu \nu} + \frac{2  F^{\mu}\widetilde F^{\nu}}{\widetilde F^2} - 
\frac{2 (\widetilde F+ F)^{\mu}(\widetilde F +  F)^{\nu}}{(\widetilde F+ F)^2} 
\,.
\end{eqnarray}
To facilitate more precise numerical evaluation, the boost can be reformulated to be linear in terms of 
\begin{equation}
  F_{\text{diff}}^{\mu} = F^{\mu} - \widetilde F^{\mu} =  \left(\frac{\zeta}{x_{jab}} - 1 \right) \tilde{p}_i^{\mu} - \frac{\kappa^2}{1-\zeta} \tilde{p}_j^{\mu} + k_{\perp}^{\mu},
\end{equation}
which can be evaluated accurately. We find 
\begin{equation}
  \Lambda^{\mu \nu}({F},\widetilde F) = g^{\mu \nu} - 4 \frac{F^{\mu}}{F^2 F_{\text{sum}}^2} \left( F_{\text{diff}}{\cdot}F \, \widetilde F^{\nu} + F^2 \, F_{\text{diff}}^{\nu}\right) + \frac{2}{F_{\text{sum}}^2} F_{\text{diff}}^{\mu} F^{\nu}_{\text{sum}}
\end{equation}
where $F^{\mu}_{\text{sum}} = \tilde F^{\mu}+ F^{\mu}$.

\paragraph{\bf{Initial-Final (local-recoil variant):}} The map is given by 
\begin{subequations}
\begin{align}
  & p^{\mu}_i = a_i \tilde{p}_i^{\mu},   \\
  & p^{\mu}_j = a_j \tilde{p}_i^{\mu} + b_j \tilde{p}_j^{\mu} + k_{\perp}^{\mu},  \\
  & p^{\mu}_k = a_k \tilde{p}_i^{\mu} + b_k \tilde{p}_j^{\mu} - 
  k_{\perp}^{\mu}\,,
\end{align}
\end{subequations}
with
\begin{subequations}
  \begin{align}
    &a_k = \frac{1-\zeta- \kappa^2}{\zeta}\,,\quad  b_k 
      =\frac{\kappa^2}{1-\zeta}\,, \\
    &a_i = \frac{1}{\zeta}\,, 
    \\
    &a_j = \frac{\kappa^2}{\zeta}\,, \quad b_j = 1 - b_k \,,
  \end{align}
\end{subequations}
and $\zeta$, $\kappa^2$ defined as in Eq.~\eqref{eq:dipole-kt-z-kappa-IS}. 

\paragraph{Initial-Final (global-recoil variant):}
We have~\cite{Hoeche:2009xc,Carli:2010cg} 
\begin{subequations}
	\label{eq:IFglobaldipolekt}
\begin{align}
  & \bar p^{\mu}_i = a_i \tilde{p}_i^{\mu} + b_i   \tilde{p}_j^{\mu} +  
  k_{\perp}^{\mu}\,, \\
  & \bar p^{\mu}_j = b_j \tilde{p}_j^{\mu}\,, \\
  & \bar p^{\mu}_k = a_k \tilde{p}_i^{\mu} + b_k \tilde{p}_j^{\mu} + 
  k_{\perp}^{\mu}\, ,
\end{align}
\end{subequations}
with
\begin{subequations}
  \begin{align}
    a_k &= \frac{1-\zeta}{\zeta-u_j} \,,&   b_k &= 
    \frac{u_j}{\zeta}\frac{1-u_j}{\zeta-u_j} 
    \,, \\
    a_i &= \frac{1-u_j}{\zeta-u_j}\,,&  b_i &= 
    \frac{u_j}{\zeta}\frac{1-\zeta}{\zeta-u_j}\,,  
    \\
    b_j &=  1-\frac{u_j}{\zeta}\,, && 
  \end{align}
\end{subequations}
where $u_j=\kappa^2/(1-\zeta)$ and $\zeta$, $\kappa^2$ are defined as in Eq.~\eqref{eq:dipole-kt-z-kappa-IS}. 
As one may observe, this map leaves the initial-state parton
misaligned with the original beam direction.
This is resolved, while ensuring momentum conservation, by performing
a boost and rotation on the initial-state particle that partakes in
the map and on all final-state particles, keeping the other
initial-state particle untouched. Denoting
$p_A^{\mu} \equiv \tilde{p}_i$ (the initial-state parton of the
dipole), $p_B^{\mu}$ the other initial-state parton, and
$p_a^{\mu} \equiv p_i$, the boost and rotation can be written as
\begin{equation} \label{eq:dipole-kt-IF-boost-rotation}
    B^{\mu \nu} = g^{\mu \nu} + \frac{p_B^{\mu} p_a^{\nu} - p_a^{\mu} 
    p_B^{\nu}}{p_a {\cdot} p_B} + \frac{p_A^{\mu} p_B^{\nu} - 
    p_B^{\mu}p_A^{\nu}}{p_A {\cdot} p_B} + \frac{p_A {\cdot} p_a}{(p_A {\cdot} 
    p_B) (p_a {\cdot} p_B)} p_B^{\mu} p_B^{\mu}.
\end{equation}
To achieve a numerically stable version of Eq.~(\ref{eq:dipole-kt-IF-boost-rotation}) we define 
\begin{align}
d_{AB} = p_A{\cdot}p_B, \quad d_{aB} = p_a{\cdot}p_B, \quad d_{aA} = 
p_a{\cdot}p_A\,,
\end{align} 
and decompose $p^\mu_a$ as
\begin{equation}
p^{\mu}_a = \frac{d_{aB}}{d_{AB}} p^\mu_A + \frac{d_{aA}}{d_{AB}} p^\mu_B + 
p_{t}^\mu,
\end{equation} 
where $p_{t}^\mu$ is not the transverse momentum with respect to the
dipole, but that with respect to the beams. 
In this form, the boost reduces to the remarkably simple form 
\begin{align}
    B^{\mu \nu} = g^{\mu \nu} + \frac{1}{d_{aB}} \left( p_B^{\mu} 
    p_{t}^{\nu} - p_{t}^{\mu} p_{B}^{\nu} \right) - \frac{1}{2} 
    \frac{p_{t}^2}{d_{aB}^2} p_B^{\mu} p_B^{\nu}\,.
\end{align} 

\paragraph{\bf{Final-Initial:}} The map is given by  
\begin{subequations}
\begin{align}
  & p^{\mu}_i = a_i \tilde{p}_i^{\mu} + b_i \tilde{p}_j^{\mu} + 
  k_{\perp}^{\mu}\,, \\
  & p^{\mu}_j = b_j \tilde{p}_j^{\mu}\,,  \\
  & p^{\mu}_k = a_k \tilde{p}_i^{\mu} + b_k \tilde{p}_j^{\mu} + 
  k_{\perp}^{\mu}\,,
\end{align}
\end{subequations}
and the coefficients read
\begin{subequations}
\begin{align}
    &a_k = z\,, \quad b_k =(1-z)y\,, \\
    &a_i =  1-z\,,\quad  b_i = zy\,, \\
    &b_j = \frac{1}{x_{ija}}\,,
\end{align}
\end{subequations}
with 
\begin{align}
\label{eq:coeffs-fi-dipole}
    & z = \kappa e^{\bar\eta_{\rm dip}}\,, \quad \kappa^2 = \frac{v^2}{\tilde{s}_{ij}^2}\,, \\
    &  x_{ija} = \frac{\kappa^2}{z}\,, \quad y = \frac{1-x_{ija}}{x_{ija}}\,.
\end{align}

\paragraph{\bf{Final-Final:}} The map is given by 
\begin{subequations}
\begin{align}
  & p^{\mu}_i = a_i \tilde{p}_i^{\mu} + b_i \tilde{p}_j^{\mu} + k_{\perp}^{\mu}\,, \\
  & p^{\mu}_j = b_j \tilde{p}_j^{\mu}\,, \\
  & p^{\mu}_k = a_k \tilde{p}_i^{\mu} + b_k \tilde{p}_j^{\mu} - k_{\perp}^{\mu}\,,
\end{align}
\end{subequations}
where
\begin{subequations}
  \begin{align}
    &a_k = 1 - \tilde{z}\,,\quad b_k = y_{ijk}\tilde{z}\,, \\
    &a_i = \tilde{z}\,,\quad b_i = y_{ijk}(1-\tilde{z})\,, \\
    &b_j = 1 - y_{ijk}\, ,
  \end{align}
\end{subequations}
with $z$ and $\kappa^2$ as in Eqs.~\eqref{eq:coeffs-fi-dipole} and 
\begin{align}
    & \tilde{z} = \frac{\kappa^2 - z(1-z)}{\kappa^2 - z}\,, \quad y_{ijk} = \frac{\kappa^2}{z}\,.
\end{align}
 
For all dipole kinds, the renormalisation scale in
Eq.~(\ref{eq:dipole-prob}) is taken to be
\begin{equation}
  \label{eq:muR-dipole-kt}
  \muR = v\,.
\end{equation}
%

\subsection{PanGlobal}
\label{app:panglobal-map}


The kinematic map is given by Eq.~\eqref{eq:general-global-map}. The
rescaling factors for the radiated particles are linked to the shower
variables via Eq.~\eqref{eq:definition-alphak-betak} where we take
\begin{equation}
  a_k = \alpha_k\,,\qquad
  b_k = \beta_k\,.
  \label{eq:panglobal-ak-bk}
\end{equation}
As usual, the magnitude of the transverse momentum is obtained by imposing $p_k^2 = 0$.
For convenience, in the mapping that we use below, we make use of a Sudakov decomposition of the
momentum of the (pre-branching) hard system with the light-cone
directions given by $\tilde{p}_a$ and $\tilde{p}_b$, i.e., the
incoming partons, 
\begin{equation}
\label{eq:pHtilde-sudakov-decomposition}
    \tilde{p}^{\mu}_H = a_H \tilde{p}^{\mu}_a + b_H \tilde{p}^{\mu}_b
    + \widetilde H^{\mu}_{t}\,.
\end{equation}
with the coefficients given by
\begin{equation}
    a_H = \frac{\tilde{p}_H {\cdot} \tilde{p}_b}{\tilde{p}_a {\cdot} \tilde{p}_b}\,, \quad 
    b_H = \frac{\tilde{p}_H {\cdot} \tilde{p}_a}{\tilde{p}_a {\cdot} \tilde{p}_b}\,, \quad 
    \widetilde H^{\mu}_{t}
    = \tilde{p}^{\mu}_H - a_H \tilde{p}^{\mu}_a - b_H \tilde{p}^{\mu}_b\,.
\end{equation}

A common feature for all dipole types is that the hard system absorbs the
part of the recoil that is transverse to the beam.
For a hard system composed of more than one particle, this is to be
achieved through a boost applied to all particles in the hard
system.
The boost takes the form $\Lambda(p_H,
\tilde p_H)$, cf.\ Eq.~\eqref{eq:boost-wo-rotation},
with $\tilde p_H^\mu$ the pre-branching momentum of the hard system,
and $p_H^\mu$ the momentum that the hard system should have after
absorbing the transverse recoil. 

Our renormalisation scale choice, common to all dipole
kinds, is
\begin{equation}
	\label{eq:muR-panglobal}
	\muR = \kappa_\perp \equiv \rho v e^{\betaps|\bar \eta_Q|}\,.
\end{equation}

\paragraph{\bf{Initial-Initial}:} The mapping (and rescaling) takes the form
\begin{subequations}
  \begin{align}
    p^{\mu}_k & = a_k \tilde p^{\mu}_a + b_k \tilde p^{\mu}_b + k^{\mu}_\perp \,,\\
    p^{\mu}_a & = r_a \tilde p^{\mu}_a\,,  \\
    p^{\mu}_b & = r_b \tilde p^{\mu}_b.
  \end{align}
\end{subequations}
The momentum of the hard system after the rescaling plus boost can be
computed (through momentum conservation), as the momentum of the
`rest of the event', i.e.
\begin{equation}
  p^{\mu}_H = r_a \tilde p^{\mu}_a + r_b \tilde p^{\mu}_b - p^{\mu}_k 
  = (r_a-a_k-1) \tilde p^{\mu}_a + (r_b-b_k-1)\tilde p^{\mu}_b -
  k^{\mu}_\perp + \tilde p^{\mu}_H\,.
\end{equation}
To obtain the right-hand side we have used the map as well as momentum conservation for the pre-branching event.
The coefficients $r_a$ and $r_b$ are obtained by requiring that $p_H$
and $\tilde p_H$ have the same rapidity and invariant mass. Using the
Sudakov decomposition given by Eq.~\eqref{eq:pHtilde-sudakov-decomposition} we get
\begin{subequations}
  \begin{align}
    &0 = (r_a-1-a_k)(r_b-1-b_k) + a_H(r_b-1-b_k) + b_H(r_a-1-a_k) -
        \frac{2\widetilde H_t \cdot k_\perp+k_\perp^2}{\tilde s_{ab}}, \\
    &\frac{a_H}{b_H} = \frac{r_a-1-a_k+a_H}{r_b-1-b_k+b_H}.
  \end{align}
\end{subequations}
Solving this system of equations leads to
\begin{equation}
  r_a = 1+a_k+\omega a_H\,,\qquad
  r_a = 1+b_k+\omega b_H\,,
\end{equation}
with
\begin{equation}
\label{eq:omega-ii-panglobal}
  \omega = \sqrt{1+\frac{2\widetilde H_t\cdot k_\perp+k_\perp^2}{a_H b_H\tilde s_{ab}}}-1\,.
\end{equation}
\paragraph{\bf{Initial-Final}:}
In this case we have
\begin{subequations}
  \begin{align}
    p^{\mu}_k & = a_k \tilde p^{\mu}_a + b_k \tilde p^{\mu}_j + k^{\mu}_\perp\,,\\
    p^{\mu}_a & = r_a \tilde p^{\mu}_a\,, \\
    p^{\mu}_j & = (1-b_k) \tilde p^{\mu}_j\,,  \\
    p^{\mu}_b & = r_b \tilde p^{\mu}_b\,,
  \end{align}
\end{subequations}
where $a$ denotes the incoming leg participating in the splitting, $b$
is the other incoming leg, and $j$ is the final-state colour partner
of $a$.
The rescaling factors $r_a$ and $r_b$ are determined imposing that the mass and the rapidity of the hard system is preserved
\begin{equation}
  p^{\mu}_H = r_a \tilde p^{\mu}_a + r_b \tilde p^{\mu}_b - p^{\mu}_k - p^{\mu}_j - \sum_{l\neq j} \tilde p^{\mu}_l
  = (r_a-1-a_k) \tilde p^{\mu}_a + (r_b-1)\tilde p^{\mu}_b - k^{\mu}_\perp + \tilde p^{\mu}_H \, ,
\end{equation}
where the sum runs over all final-state particles except the new emission.
The virtuality and rapidity constraints can then be written as
\begin{subequations}
  \begin{align}
    &0 = (r_a-1-a_k)(r_b-1) + (a_H-c_b)(r_b-1) + b_H(r_a-1-a_k) -
        (c+b_Hc_b), \\
    &\frac{a_H}{b_H} = \frac{r_a-1-a_k+a_H-c_b}{r_b-1+b_H},
  \end{align}
\end{subequations}
with
\begin{equation}
  c_b = \frac{2\tilde p_b\cdot k_\perp}{\tilde s_{ab}}\,,\qquad
  c = \frac{2\widetilde H_t\cdot k_\perp+k_\perp^2}{\tilde s_{ab}}\,.
\end{equation}
This system admits the following solution (with $\omega$ as in 
Eq.~\eqref{eq:omega-ii-panglobal})
\begin{equation}
  r_a = 1+a_k+\omega a_H+\frac{2\tilde p_b\cdot k_\perp}{\tilde s_{ab}}\,,\qquad
  r_b = 1+\omega b_H\,.
\end{equation}
Since we are considering an antenna shower here, there is no need to
have a separate treatment for an FI dipole.

\paragraph{\bf{Final-Final}:}
  We now have two final-state particles $i$ and $j$ that emit a parton $k$, for 
  which the mapping is
\begin{subequations}
  \begin{align}
    p^{\mu}_k & = a_k \tilde p^{\mu}_i + b_k \tilde p^{\mu}_j + k^{\mu}_\perp\,, \\
    p^{\mu}_i & = (1-a_k) \tilde p^{\mu}_{i} \, , \\
    p^{\mu}_j & = (1-b_k) \tilde p^{\mu}_{j} \, , \\
    p^{\mu}_a & = r_a \tilde p^{\mu}_a  \, ,\\
    p^{\mu}_b & = r_b \tilde p^{\mu}_b \, .
  \end{align}
\end{subequations}
We therefore have
\begin{equation}
  p^{\mu}_H = r_a \tilde p^{\mu}_a + r_b \tilde p^{\mu}_b - p^{\mu}_k - p^{\mu}_j -p^{\mu}_i - \sum_{l\neq i,j,k;l} 
  \tilde p^{\mu}_l
  = (r_a-1) \tilde p^{\mu}_a + (r_b-1)\tilde p^{\mu}_b - k^{\mu}_\perp + \tilde p^{\mu}_H.
\end{equation}
The virtuality and rapidity constraints can then be written as
\begin{subequations}
  \begin{align}
    &0 = (r_a-1)(r_b-1) + (a_H-c_b)(r_b-1) + (b_H-c_a)(r_a-1) -
        (a_Hc_a+b_Hc_b+c), \\
    &\frac{a_H}{b_H} = \frac{r_a-1+a_H-c_b}{r_b-1+b_H-c_a},
  \end{align}
\end{subequations}
with
\begin{equation}
  c_{a,b} = \frac{2\tilde p_{a,b}\cdot k_\perp}{\tilde s_{ab}}\,,\qquad
  c = \frac{2\widetilde H_t \cdot k_\perp+k_\perp^2}{\tilde s_{ab}}
\end{equation}
which admit the following solution (again with $\omega$ as in 
Eq.~\eqref{eq:omega-ii-panglobal})
\begin{equation}
  r_a = 1+\omega a_H+\frac{2\tilde p_b \cdot k_\perp}{\tilde s_{ab}}\,,\qquad
  r_b = 1+\omega b_H+\frac{2\tilde p_a \cdot k_\perp}{\tilde s_{ab}}.
\end{equation}
%

\subsection{PanLocal}
\label{app:panlocal-map}

Here we discuss both the dipole and antenna variants of the PanLocal
shower.
We start with the aspects that are common to both, in particular the
boost and rotation that are to be applied to the event after any
splitting that assigns transverse momentum to an initial-state
particle, i.e.\ initial-initial and initial-final dipoles.
As discussed in the main text, the boost
and rotation realign the incoming partons with the beam axes.
This transformation contains a longitudinal degree of freedom, which
we exploit so as to conserve the rapidity $y_H$ of the hard system
$H$.
We use $\tilde{p}_H$ for the pre-branching momentum of the hard
system, $\bar p_H$ for the momentum after the emission occurred, and
$p_H$ is the final value after the boost has been performed.
For colour-singlet production, we have $\bar p_H=\tilde{p}_H$, but we
maintain the distinction in our notation to keep the discussion general
(for example as concerns future extension to situations with coloured
particles that are part of the hard system).

The first step to obtain the precise form of the boost is to determine the new 
momenta of the incoming partons, which we label with indices $a$ and $b$. 
Using $p_{a,b}$ to refer to the momenta after the boost and rotation, 
we can write
\begin{subequations}
  \begin{align}
    \tilde{p}_{a,b} &= \tilde{x}_{a,b} P_{a,b}\,, \\
    p_{a,b} &= x_{a,b} P_{a,b}\,,
  \end{align}
\end{subequations}
where $P_{a,b} \equiv \widetilde{P}_{a,b}$ are the incoming hadron
momenta.
The hadron momenta must remain unchanged after the combination of kinematic map,
boost and rotation.
Thus to determine the boost we need to find the new values $x_{a,b}$
such that the rapidity of the hard system is unchanged (as defined with
respect to the incoming hadron momenta).
Before the splitting, we have
\begin{align}
	\tilde{y}_H = \frac{1}{2} \ln \frac{\tilde{p}_H \cdot P_b}{\tilde{p}_H \cdot 
		P_a} = \frac{1}{2} \ln \frac{\tilde{p}_H \cdot \tilde{p}_b}{\tilde{p}_H \cdot 
		\tilde{p}_a}+ \frac{1}{2}\ln \frac{\tilde{x}_a}{\tilde{x}_b},
	\label{eq:ytilde}
\end{align}
while after the emission
\begin{equation}
	\label{eq:yprime}
	\begin{split}
		y_H \equiv \frac{1}{2}\ln \frac{p_H \cdot P_b}{p_H \cdot 
			P_a} &= \frac{1}{2}\ln \frac{p_H \cdot p_b}{p_H \cdot 
			p_a} +\frac{1}{2} \ln \frac{x_a}{x_b}
                      \\
		&= \frac{1}{2}\ln \frac{\bar p_H \cdot \bar p_b}{\bar
                  p_H \cdot \bar p_a} +\frac{1}{2} \ln 
		\frac{x_a
		}{x_b},
	\end{split}
\end{equation}
where in the last step we used the Lorentz invariance of the dot product. Equating 
Eq.~\eqref{eq:ytilde} and \eqref{eq:yprime} we obtain
\begin{align}
	\label{eq:xa-one}
  \frac{\tilde{x}_a}{x_a}\frac{x_b}{\tilde{x}_b} =
  \frac{\tilde{p}_H \cdot \tilde{p}_b}{\bar p_H \cdot \bar p_b}\,
  \frac{\bar p_H \cdot \bar p_a}{\tilde{p}_H \cdot \tilde{p}_a}.
\end{align}
Since the Lorentz transformation will preserve the invariant mass of
the final-state system, we also have
\begin{align}
	\label{eq:xa-two}
  2 \bar p_a \cdot \bar p_b = 2 p_a \cdot p_b =
  2 \tilde{p}_a \cdot \tilde{p}_b \frac{x_a x_b}{\tilde{x}_a \tilde{x}_b} .
\end{align}
Solving Eqs.~\eqref{eq:xa-one} and \eqref{eq:xa-two} for the new momentum fractions 
$x_{a,b}$ yields
\begin{align}
  x_a &= \tilde{x}_a  \sqrt{\frac{\bar p_a \cdot \bar p_b}{\tilde{p}_a \cdot \tilde{p}_b}
        \frac{\tilde{p}_H \cdot \tilde{p}_b}{\bar p_H \cdot \bar p_b}
        \frac{\bar p_H \cdot \bar p_a}{\tilde{p}_H \cdot \tilde{p}_a}}
        \,,\qquad
  x_b  = \tilde{x}_b \sqrt{
        \frac{\bar p_a \cdot \bar p_b}{\tilde{p}_a \cdot \tilde{p}_b}
        \frac{\bar p_H \cdot \bar p_b}{\tilde{p}_H \cdot \tilde{p}_b} 
        \frac{\tilde{p}_H \cdot \tilde{p}_a}{\bar p_H \cdot \bar p_a}
        }.
\end{align}
We are now in a position to carry out the boost.
It is useful to introduce a notation for the total hadron-hadron four
momentum before and after the boost as 
\begin{align}
  \bar P^\mu = \frac{\bar p_a^\mu}{x_a}+\frac{\bar p_b^\mu}{x_b}\,,
  \qquad
  P^\mu \equiv \widetilde{P}^\mu = 
  \widetilde{P}_a^\mu + \widetilde{P}_b^\mu\,.
\end{align}
Then every particle in the event is boosted as 
\begin{equation}
  \label{eq:panlocal-boost}
  p^\mu = \Lambda^{\mu \nu}(P, \bar P) \bar p_{\nu}
\end{equation}
with $\Lambda^{\mu\nu}$ as defined in Eq.~(\ref{eq:boost-wo-rotation}).
This boost ensures that the new incoming momenta are back-to-back and
have the correct energy, and we then perform a rotation so as to align
them with the $z$ axis.\footnote{For this last step, we assume that
  the hadrons collide in their centre-of-mass frame, i.e.\ that
  $P^\mu$'s 3-momentum components are all zero. }

In practice, one also has the possibility to perform the boost (and
the final rotation) only at the very end of the parton showering,
provided that the values of the incoming partons energy fractions $x$ are
stored at each step of the shower evolution. These values are used to
compute the beams' momenta in the unboosted frame
\begin{equation}
	\bar P_{a,b}^\mu=\frac{\bar p_{a,b}^\mu}{x_{a,b}}.
	\label{eq:beamsReset}
\end{equation}
Since to define $\bar{\eta} = \bar{\eta}_Q$ and $\rho$ we use the
scalar product with a reference vector $Q^\mu$, we should apply the
inverse rotation and the inverse boost to $Q^\mu$.
For colour-singlet production $Q^\mu$ is chosen to be
\begin{equation}
	Q^\mu = e^y \frac{m}{S} \bar P_a^\mu + e^{-y}\frac{m}{S} \bar P_b^\mu,
	\label{eq:QDY}
\end{equation}
where $m$ and $y$ are the mass and the rapidity of the colour singlet
(in the original frame), while $S=(\bar P_a+\bar P_b)^2$ is the
squared total centre of mass energy.
Thus, in this case we can simply 
evaluate Eq.~\eqref{eq:QDY} after we have recomputed the beam momenta according to 
Eq.~\eqref{eq:beamsReset}.

The renormalisation scale choice for the PanLocal showers is the same
as for the PanGlobal shower, i.e.\ Eq.~(\ref{eq:muR-panglobal}).

\subsubsection{PanLocal dipole}
\paragraph{{Initial-Initial:}} The map is given by
\begin{subequations}
	\begin{align}
		& \bar p^{\mu}_i = a_i \tilde{p}_i^{\mu} + b_i \tilde{p}_j^{\mu} + 
		k_{\perp}^{\mu}\,,   \\
		& \bar p^{\mu}_j = b_j\tilde{p}_j^{\mu}\,,  \\
		& \bar p^{\mu}_k = a_k \tilde{p}_i^{\mu} + b_k \tilde{p}_j^{\mu} +
		k_{\perp}^{\mu}\,, 
	\end{align}
\end{subequations}
and the coefficients read 
\begin{subequations}
\begin{align}
	&a_k = \alpha_k\,, \qquad 
	b_k = \beta_k(1+\alpha_k)^{\frac{2}{1+\betaps}}\,,  \\
	&a_i  = 1 + a_k\,,\qquad b_i = \frac{a_k b_k}{a_i}\,, \\
	&b_j  = 1 - \frac{b_k}{a_i}\,.
\end{align}
\end{subequations}

\paragraph{\bf{Initial-Final:}} The map is given by 
\begin{subequations}
	\begin{align}
		& \bar p^{\mu}_i = a_i \tilde{p}_i^{\mu} + b_i \tilde{p}_j^{\mu} + 
		k_{\perp}^{\mu}\,,   \\
		& \bar p^{\mu}_j = b_j\tilde{p}_j^{\mu}\,,  \\
		& \bar p^{\mu}_k = a_k \tilde{p}_i^{\mu} + b_k \tilde{p}_j^{\mu} +
		k_{\perp}^{\mu}\,, 
	\end{align}
\end{subequations}
with
\begin{subequations}
  \label{eq:PanLocal-dipole-IF-coeffs}
	\begin{align}
		&a_k = \alpha_k\,, \qquad 
		b_k = \beta_k(1+\alpha_k)^{\frac{2}{1+\betaps}}\,,  \\
		&a_i  = 1 + a_k\,,\qquad b_i = \frac{a_k b_k}{a_i}\,, \\
		&b_j  = 1 + \frac{b_k}{a_i}\,.
           \label{eq:PanLocal-dipole-IF-bj}
	\end{align}
\end{subequations}

\paragraph{\bf{Final-Initial:}} The map is given by  
\begin{subequations}
	\begin{align}
		&  p^{\mu}_i = a_i \tilde{p}_i^{\mu} + b_i \tilde{p}_j^{\mu} - 
		k_{\perp}^{\mu}\,,   \\
		& p^{\mu}_j = b_j\tilde{p}_j^{\mu}\,,  \\
		&  p^{\mu}_k = a_k \tilde{p}_i^{\mu} + b_k \tilde{p}_j^{\mu} +
		k_{\perp}^{\mu}\,, 
	\end{align}
\end{subequations}
and the coefficients read
\begin{subequations}
	\begin{align}
		&a_k = \alpha_k\,, \qquad 
		b_k = \beta_k\,,  \\
		&a_i  = 1 - a_k\,, \qquad b_i = \frac{a_k b_k}{a_i}\,, \\
		&b_j  = 1 + \frac{b_k}{a_i}\,.
	\end{align}
\end{subequations}

\paragraph{\bf{Final-Final:}} The map is given by 
\begin{subequations}
	\begin{align}
		& p^{\mu}_i = a_i \tilde{p}_i^{\mu} + b_i \tilde{p}_j^{\mu} - 
		k_{\perp}^{\mu}\,,   \\
		& p^{\mu}_j = b_j\tilde{p}_j^{\mu}\,,  \\
		& p^{\mu}_k = a_k \tilde{p}_i^{\mu} + b_k \tilde{p}_j^{\mu} +
		k_{\perp}^{\mu}\,, 
	\end{align}
\end{subequations}
and the coefficients read
\begin{subequations}
	\begin{align}
		&a_k = \alpha_k\,, \qquad 
		b_k = \beta_k\,,  \\
		&a_i  = 1 - a_k\,, \qquad b_i = \frac{a_k b_k}{a_i}\,, \\
		&b_j  = 1 - \frac{b_k}{a_i}\,.
	\end{align}
\end{subequations}

\subsubsection{PanLocal antenna}

\label{sec:app-panlocal-antenna}
  
\paragraph{{Initial-Initial:}} The map is given by
\begin{subequations}
	\begin{align}
		& \bar p^{\mu}_i = a_i \tilde{p}_i^{\mu} + b_i \tilde{p}_j^{\mu} + 
		f k_{\perp}^{\mu}\,,   \\
		& \bar p^{\mu}_j = a_j \tilde{p}_i^{\mu} +  b_j\tilde{p}_j^{\mu} + (1-f)k_\perp^\mu \,,  \\
		& \bar p^{\mu}_k = a_k \tilde{p}_i^{\mu} + b_k \tilde{p}_j^{\mu} +
		k_{\perp}^{\mu}\,, 
	\end{align}
\end{subequations}
and the coefficients read 
\begin{subequations}
	\begin{align}
		&a_k =  \alpha_k(1+\beta_k)^{\frac{2}{1+\betaps}}\,, \qquad 
		b_k = \beta_k(1+\alpha_k)^{\frac{2}{1+\betaps}}\,,  \\
		& a_i =  \frac{(\sqrt{\lambda_1}+\sqrt{\lambda_2})^2 + 4f^2 a_k b_k}{4(1+b_k)}\,, 
		\qquad b_i =  \frac{(\sqrt{\lambda_1}-\sqrt{\lambda_2})^2 + 4f^2 a_k b_k}{4(1+a_k)}\,, \\ 
		& a_j = \frac{(\sqrt{\lambda_1}-\sqrt{\lambda_2})^2 + 4(1-f)^2 a_k 
			b_k}{4(1+b_k)}\,, 
		\qquad b_j = \frac{(\sqrt{\lambda_1}+\sqrt{\lambda_2})^2 + 4(1-f)^2 a_k 
			b_k}{4(1+a_k)}\,,
	\end{align}
\end{subequations}
with $\lambda_1 = 1+a_k + b_k$ and $\lambda_2 = \lambda_1+4f (1-f) a_k b_k $.

\paragraph{\bf{Initial-Final:}} The map is given by 
\begin{subequations}
	\begin{align}
		& \bar p^{\mu}_i = a_i \tilde{p}_i^{\mu} + b_i \tilde{p}_j^{\mu} + 
		f k_{\perp}^{\mu}\,,   \\
		& \bar p^{\mu}_j = a_j \tilde{p}_i^{\mu} +  b_j\tilde{p}_j^{\mu} - (1-f)k_\perp^\mu \,,  \\
		& \bar p^{\mu}_k = a_k \tilde{p}_i^{\mu} + b_k \tilde{p}_j^{\mu} +
		k_{\perp}^{\mu}\,, 
	\end{align}
\end{subequations}
with
\begin{subequations}
	\begin{align}
		&a_k = \alpha_k\,, \qquad 
		b_k = \beta_k(1+\alpha_k)^{\frac{2}{1+\betaps}}\,,  \\
		& a_i =  \frac{(\sqrt{\lambda_1}+\sqrt{\lambda_2})^2 - 4f^2 a_k b_k}{4(1-b_k)}\,, 
		\qquad  b_i =  \frac{-(\sqrt{\lambda_1}-\sqrt{\lambda_2})^2 + 4f^2 a_k b_k}{4(1+a_k)}\,,\\
		&a_j = \frac{-(\sqrt{\lambda_1}-\sqrt{\lambda_2})^2 + 4(1-f)^2 a_k 
			b_k}{4(1-b_k)}\,, 
		\qquad b_j = \frac{(\sqrt{\lambda_1}+\sqrt{\lambda_2})^2 - 4(1-f)^2 a_k 
			b_k}{4(1+a_k)}\,,
	\end{align}
\end{subequations}
and $\lambda_1 = 1+a_k - b_k$ and $\lambda_2 = \lambda_1-4f (1-f) a_k
b_k $.
Since we are considering an antenna shower here, there is no need to
have a separate treatment of a ``final-initial'' dipole.

\paragraph{\bf{Final-Final:}} The map is given by 
\begin{subequations}
	\begin{align}
		&  p^{\mu}_i = a_i \tilde{p}_i^{\mu} + b_i \tilde{p}_j^{\mu} - 
		f k_{\perp}^{\mu}\,,   \\
		& p^{\mu}_j = a_j \tilde{p}_i^{\mu} +  b_j\tilde{p}_j^{\mu} - (1-f)k_\perp^\mu \,,  \\
		& p^{\mu}_k = a_k \tilde{p}_i^{\mu} + b_k \tilde{p}_j^{\mu} +
		k_{\perp}^{\mu}\,, 
	\end{align}
\end{subequations}
and the coefficients read
\begin{subequations}
	\begin{align}
		&a_k = \alpha_k\,, \qquad 
		b_k = \beta_k\,,  \\
	& a_i =  \frac{(\sqrt{\lambda_1}+\sqrt{\lambda_2})^2 + 4f^2 a_k b_k}{4(1-b_k)}\,, 
	\qquad b_i =  \frac{(\sqrt{\lambda_1}-\sqrt{\lambda_2})^2 + 4f^2 a_k b_k}{4(1-a_k)}\,, \\
	& a_j = \frac{(\sqrt{\lambda_1}-\sqrt{\lambda_2})^2 + 4(1-f)^2 a_k 
		b_k}{4(1-b_k)}\,, 
	\qquad b_j = \frac{(\sqrt{\lambda_1}+\sqrt{\lambda_2})^2 + 4(1-f)^2 a_k 
		b_k}{4(1-a_k)} \,,
	\end{align}
\end{subequations}
with $\lambda_1 = 1-a_k - b_k$ and $\lambda_2 = \lambda_1+4f (1-f) a_k b_k $.


\section{Analytics for sub-leading colour matrix element tests}
\label{app:colour}
Here, we provide all of the analytic ingredients needed to make the
comparisons presented in Section~\ref{sec:colour}. The differential
cross section for the emission of one additional soft gluon  with momentum $k$ and energy fraction $z$ off an amplitude $\dd \sigma_n$ can be written as 
\begin{eqnarray}
\label{eq:colour:general}
\dd \sigma_{n+1} = \dd \sigma_n \frac{{\rm d} z}{z}\frac{{\rm d} \Omega}{2\pi} \frac{\as}{2\pi} \sum_{i,j}C_{ij}(k|ij),
\end{eqnarray}
where ${\rm d} \Omega = {\rm d} \cos \theta {\rm d}\phi $ is the element of solid angle for the emitted gluon, $C_{ij}$ the colour factor, and the eikonal factor $(k|ij)$ is given by
\begin{equation}
    (k|ij)=\frac{1-\cos\theta_{ij}}{(1-\cos\theta_{ki})(1-\cos\theta_{kj})}.
\end{equation}
Starting from Eq.~\eqref{eq:colour:general} we now compute the differential matrix element and integrated rate for two and three emissions. 
\subsection{$q\bar q\to Z$}
Let us begin with the $q\bar q\to Z$ case. 

\paragraph{Two emissions}
One of the two initial configurations is a $q\bar q g_1$ system from which we emit a second soft gluon. The $q\bar q \rightarrow Z g_1 g_2$ amplitude is
\begin{equation}
    \label{eq:colour:qqbarg1g2}
    \frac{|\mathcal{M}_{q\bar q g_1 g_2}|^2}{|\mathcal{M}_{q\bar q g_1}|^2} = \frac{\alpha_s}{2\pi} \left[\frac{C_A}{2}(g_2|g_1q)+\frac{C_A}{2}(g_2|g_1\bar q)+(C_F-\frac{C_A}{2})(g_2|q\bar q)\right]\, ,
\end{equation} 
The integrated emission rate, that we denote  $I^{Zg_1}_{\rm FC}$, is obtained after integrating over the solid angle $\int \dd \Omega = \int \dd \cos\theta \dd \phi$, defined with respect to the direction of the emitting line $i$ or $j$. The azimuthal integration results in the well known property of angular ordering, i.e.
\begin{eqnarray}
    \label{eq:phi-trick}
    \int \frac{\dd \Omega}{2\pi}(k|ij) &=& \frac{1}{2}\left[\int_{\cos \theta_{ij}}^{\cos\theta_{{\rm cut},ik}} \frac{{\rm d}\cos \theta_{ik}}{1-\cos \theta_{ik}} +  \int_{\cos \theta_{ij}}^{\cos\theta_{{\rm cut},jk}} \frac{{\rm d}\cos \theta_{jk}}{1-\cos \theta_{jk}}\right] \\
    &=& \frac{1}{2}\left[-\ln\left(1-\cos\theta_{{\rm cut},ik}\right) -\ln\left(1-\cos\theta_{{\rm cut},jk}\right) + 2\ln\left(1-\cos\theta_{ij}\right)\right],\nonumber
\end{eqnarray}
with ${\theta_{\rm cut}}$ a collinear regulator that can be different for each of the terms. In what follows, we use $\theta_{\rm cut}$ ($\theta_{{\rm cut},s}$) to denote the regulator for primary (secondary) emissions.
\begin{align}
    \label{eq:colour:qqbarg1g2-int1}
    I^{Zg_1}_{\rm FC}&\equiv \int \frac{{\rm d}\Omega}{2\pi}\frac{|\mathcal{M}_{q\bar q g_1 g_2}|^2}{|\mathcal{M}_{q\bar q g_1}|^2} \nonumber \\
    &=  \frac{\alpha_s}{2\pi}\Big\lbrace\frac{C_A}{2}\left[\ln\left(1-\cos\theta_{1q}\right) + \ln(1-\cos\theta_{1\bar{q}}) - \ln(1-\cos\theta_{q\bar{q}}) -\ln(1-\cos\theta_{{\rm cut},s})\right] \nonumber \\
    &\hspace{2cm}  + C_F \left[\ln(1-\cos\theta_{q\bar{q}}) - \ln(1-\cos\theta_{\rm cut}) \right] \Big\rbrace \, ,
\end{align}
Since the initial-state quarks are back-to-back we can replace $\cos \theta_{q\bar{q}} = -1$ and rewrite the angles in terms of rapidities using 
\begin{eqnarray}
\eta = \pm|\ln \tan\theta/2|\to 1-\cos\theta = \frac{2}{1+{\rm e}^{2\eta}}\,.
\end{eqnarray}
We obtain 
\begin{align}
    \label{eq:colour:qqbarg1g2-int2}
    I^{Zg_1}_{\rm FC}    &=  \frac{\alpha_s}{2\pi}\Big\lbrace\frac{C_A}{2}\left[\ln(1+{\rm e}^{2\eta_{{\rm cut},s}}) -\ln(1+{\rm e}^{2\Delta\eta_{1q}})-\ln(1+{\rm e}^{2\Delta\eta_{1\bar{q}}})\right] + C_F \left[\ln(1+{\rm e}^{2\eta_{\rm cut}}) \right] \Big\rbrace \, .
\end{align}
To arrive at our final result, we first use that $\eta_{\rm cut}$ and $\eta_{{\rm cut},s}$ correspond to small angles, such that $\ln(1+{\rm e}^{2\eta_{{\rm cut}(,s)}}) \simeq 2\eta_{{\rm cut}(,s)}$. Secondly, we write $\Delta \eta_{1q} = \eta_1$ and $\Delta \eta_{1\bar{q}} =  -\eta_1 $ (which holds as long as we do only one emission, and is not sensitive to the rapidity of the colour-singlet system). Consequently,
\begin{subequations}
  \begin{align}
    \ln(1+{\rm e}^{2\Delta\eta_{1q}})+\ln(1+{\rm e}^{2\Delta\eta_{1\bar{q}}}) &= 2\eta_1+2\ln(1+{\rm e}^{-2\eta_1})\,, \\
    \ln(1+{\rm e}^{2\Delta\eta_{1q}})-\ln(1+{\rm e}^{2\Delta\eta_{1\bar{q}}}) & = 2\eta_1\,.
  \end{align}
\end{subequations}
Inserting this result into Eq.~\eqref{eq:colour:qqbarg1g2-int2}, we obtain the final expression:
\begin{align}
    \label{eq:colour:qqbarg1g2-int-final}
    I^{Zg_1}_{\rm FC}    &=  \frac{\alpha_s}{2\pi}\left[C_A\left(\eta_{{\rm cut},f} - \eta_1 -\ln(1+{\rm e}^{-2\eta_1})\right) + 2 C_F \eta_{\rm cut} \right] .
\end{align}
Starting from a $q\bar{q}\rightarrow Z$ configuration, we may also
backward-evolve the quark into an initial-state gluon after emitting
an anti-quark. We obtain 
\begin{equation}
    \label{eq:colour:qbargqbar1g2}
    \frac{|\mathcal{M}_{g\bar q \bar{q}_1 g_2}|^2}{|\mathcal{M}_{g\bar q \bar{q}_1}|^2} = \frac{\alpha_s}{2\pi} \left[\frac{C_A}{2}(g_2|\bar{q}g)+\frac{C_A}{2}(g_2|\bar{q}_1g)+(C_F-\frac{C_A}{2})(g_2|\bar{q}\bar{q}_1)\right] ,
\end{equation} 
and 
\begin{subequations}
  \begin{align}
    \label{eq:colour:qbargqbar1g2-int}
    I^{Zq_1}_{\rm FC}    &=  \frac{\alpha_s}{2\pi}\left[C_F\left(\eta_{\rm cut} + \eta_{{\rm cut},f} -\ln\left(1+{\rm e}^{2\Delta \eta_{\bar{q}\bar{q}_1}}\right)\right) \phantom{\frac{C_A}{2}}\right. \nonumber\\
                         &\phantom{= \frac{2\as}{\pi}}\left.+ \frac{C_A}{2}\left(\eta_{\rm cut} + \ln\left(1+{\rm e}^{2\Delta \eta_{\bar{q}\bar{q}_1}}\right) -\ln\left(1+{\rm e}^{2\Delta \eta_{g\bar{q}_1}}\right) \right) \right],  \\
                         &=  \frac{\alpha_s}{2\pi}\left[C_F\left(\eta_{\rm cut} + \eta_{{\rm cut},f} -\ln\left(1+{\rm e}^{-2\eta_1}\right)\right) + C_A \left(\eta_{\rm cut} - \eta_1\right) \right] . 
  \end{align}
\end{subequations}
\paragraph{Three emissions}
Following the same steps as in the previous calculation, we can now consider the emission of a third parton either from a $Zg_1g_2$, $Zg_1q_2$, $Zq_1g_2$ or a $Zq_1q_2$ system. For simplicity, we take the soft and collinear limit. The integrated emissions rates are then given by
\begin{subequations}
  \begin{align}
    I_{\rm FC}^{Zg_1g_2} &= \frac{2\as}{\pi}\left[2C_F \eta_{\rm cut} + 2C_A \eta_{\rm cut, s}\right], \\
    I_{\rm FC}^{Zg_1q_2} &= \frac{2\as}{\pi}\left[C_F \left(\eta_{\rm cut}-\eta_2 +\eta_{\rm cut, s}\right)  + C_A \left(\eta_{\rm cut} + \eta_{\rm cut, s} + \eta_2\right)\right],\\
    I_{\rm FC}^{Zq_1g_2} &= \frac{2\as}{\pi}\left[C_F \left(\eta_{\rm cut}+\eta_1 +\eta_{\rm cut, s}\right) + C_A \left(\eta_{\rm cut} + \eta_{\rm cut, s} - \eta_1\right)\right], \\
    I_{\rm FC}^{Zq_1q_2} &= \frac{2\as}{\pi}\left[C_F \left(2\eta_{\rm cut}+\eta_1 - \eta_2\right) + C_A \left(2\eta_{\rm cut, s} - \eta_1 + \eta_2\right)\right]\,.
  \end{align}
\end{subequations}
\subsection{$gg\to H$}
\paragraph{Two emissions} For the $gg\to H$ cases we first consider the backwards splitting of a gluon into a gluon. Denoting the initial-state gluons with $g_a$ and $g_b$ we obtain 
\begin{equation}
    \label{eq:colour:gagbg1g2}
    \frac{|\mathcal{M}_{g_ag_bg_1g_2}|^2}{|\mathcal{M}_{g_a g_b g_1}|^2} = \frac{\alpha_s}{2\pi} \left[\frac{C_A}{2}(g_2|g_1 g_a)+\frac{C_A}{2}(g_2|g_1 g_b)+\frac{C_A}{2}(g_2|g_a g_b)\right] ,
\end{equation} 
and 
\begin{align}
    \label{eq:colour:gagbg1g2-int}
    I^{Hg_1}_{\rm FC}    &=  \frac{\alpha_s}{2\pi}C_A\left[2\eta_{\rm cut} + \eta_{{\rm cut},s} - \eta_1  -\ln\left(1+{\rm e}^{-2\eta_1}\right) \right]\, . 
\end{align}
The other possibility is a gluon backwards splitting to a quark, for which we obtain
\begin{equation}
    \label{eq:colour:gaqq1g2}
    \frac{|\mathcal{M}_{gq q_1g_2}|^2}{|\mathcal{M}_{gqq_1}|^2} = \frac{\alpha_s}{2\pi} \left[\frac{C_A}{2}(g_2|gq)+\frac{C_A}{2}(g_2|g q_1)+\left(C_F -\frac{C_A}{2}\right)(g_2|q q_1)\right] ,
\end{equation} 
and
\begin{align}
    \label{eq:colour:gaqq1g2-int}
    I^{Hq_1}_{\rm FC}    &=  \frac{\alpha_s}{2\pi}\left[C_F\left(\eta_{\rm cut} + \eta_{{\rm cut},s} -  2\eta_1  -\ln\left(1+{\rm e}^{-2\eta_1}\right)\right) + C_A\left(\eta_{\rm cut} + \eta_1\right)\right] . 
\end{align}
\paragraph{Three emissions}
Finally, we consider three emissions starting from a $gg \to H$ process. The integrated colour rates read
\begin{subequations}
  \begin{align}
    I_{\rm FC}^{Hg_1g_2} &= \frac{2\as}{\pi}\left[2C_A\left( \eta_{\rm cut} + \eta_{\rm cut, s}\right)\right], \\
    I_{\rm FC}^{Hg_1q_2} &= \frac{2\as}{\pi}\left[C_F \left(\eta_{\rm cut}+\eta_2 +\eta_{\rm cut, s}\right) + C_A \left(\eta_{\rm cut} + \eta_{\rm cut, s} - \eta_2\right)\right], \\
    I_{\rm FC}^{Hq_1g_2} &= \frac{2\as}{\pi}\left[C_F \left(\eta_{\rm cut}-\eta_1 +\eta_{\rm cut, s}\right) + C_A \left(\eta_{\rm cut} + \eta_{\rm cut, s} + \eta_1\right)\right], \\
    I_{\rm FC}^{Hq_1q_2} &= \frac{2\as}{\pi}\left[C_F \left(2\eta_{\rm cut}+2\eta_{\rm cut,s}-\eta_1 + \eta_2\right) + C_A \left(\eta_1 - \eta_2\right)\right].
  \end{align}
\end{subequations}
%
\section{Deriving the branching amplitudes for spin correlations}
\label{app:spin}
Here we collect the branching amplitudes in terms for spinor products for
initial-state splittings. The final-state expressions can be found in Appendix~A
of Ref.~\cite{Karlberg:2021kwr}. We first define the spinor product for two
light-like momenta $p_a$ and $p_b$
\begin{eqnarray}
    S_{\lambda}(p_a, p_b) = \bar{u}_{\lambda}(p_a)u_{-\lambda}(p_b)\,,
\end{eqnarray}
where $\lambda = \pm 1$ is the Dirac spinor helicity. This spinor product satisfies
$S_{\lambda}(p_a, p_b) = -S_{\lambda}(p_b, p_a)$. The polarisation vector of a
gluon with momentum $p$ can be written in terms of spinors by using a light-like
reference vector $r$ as
\begin{equation} \label{eq:polarization}
    \epsilon^*_{\lambda}(p) = \frac{1}{\sqrt{2}} \frac{1}{S_{-\lambda}(r,p)} \bar{u}_{\lambda}(p) \gamma^{\mu} u_{\lambda}(r),
\end{equation} 
which obeys $\epsilon^*_{\lambda}(p) = -\epsilon_{-\lambda}(p)$. To perform the
necessary calculations, the Chisholm identity is useful
\begin{eqnarray}
    \slashed{\epsilon}^*_{\lambda}(p) = \frac{\sqrt{2}}{S_{-\lambda}(r,p)}\left[u_{\lambda}(r)\bar{u}_{\lambda}(p) + u_{-\lambda}(p)\bar{u}_{-\lambda}(r)\right].
\end{eqnarray}
We consider a collinear initial-state splitting with $p_i \rightarrow
\tilde{p}_i + p_k$, such that $\tilde{p}_i = (1-z)p_i$ and $p_k = z p_i$
(i.e.~we use the same convention as in Eq.~\eqref{eq:p-initial-state}).
In this limit, any dependence of the branching amplitudes on the gauge vector $r$ vanishes.
The result can be written in terms of the single spinor product $S_{\lambda}(p_i,p_k)$ by using the identities
\begin{align}
S_{\lambda}(p_i, \tilde{p}_i) &= -\sqrt{\frac{z}{1-z}}S_{\lambda}(p_i,p_k), \quad 
S_{\lambda}(p_k, \tilde{p}_i) = -\sqrt{\frac{1}{1-z}}S_{\lambda}(p_i,p_k)\,,
\end{align}
which are valid in the collinear limit. We now compute the relevant collinear
branching amplitudes, stripped from any overall factors as they are not relevant
in the spin correlation algorithm. The results are summarised in
Table~\ref{tab:hel-dep-splittings} in Section~\ref{sec:spin-correl-alg}.
\subsection*{$q_{I} \rightarrow \tilde q_I g_F$}
The branching amplitude is 
\begin{align}
    \mathcal{M}^{\lambda_{\itilde}\, \lambda_i\, \lambda_k}_{q_{I}(p_i) \rightarrow \tilde q_I(\tilde p_{ i}) g_F(p_k)} &= \bar{u}_{\lambda_{\itilde}}(\tilde{p}_{i}) \slashed{\epsilon}^*_{\lambda_k} u_{\lambda_i} (p_i) \\
    &= \frac{\sqrt{2}}{S_{-\lambda_k}(r, p_k)}\bar{u}_{\lambda_{\itilde}}(\tilde{p}_{i})  \bigg[ u_{\lambda_k}(r) \bar{u}_{\lambda_k}(p_k) + u_{-\lambda_k}(p_k)\bar{u}_{-\lambda_k}(r) \bigg] u_{\lambda_i}(p_i).\nonumber 
\end{align} 
Recall that the order of the spin indices in the superscript differs
from the order of the momenta in the subscript.
The amplitude vanishes for $\lambda_{\itilde} = - \lambda_i$, so we set $\lambda = \lambda_{\itilde} = \lambda_i$. 
We find 
\begin{subequations}
  \begin{align}
    &  \mathcal{M}^{\lambda\,\lambda \,\lambda}_{q_{I}(p_i) \rightarrow \tilde q_I(\tilde p_{ i}) g_F(p_k)} =  \frac{\sqrt{2}}{\sqrt{z(1-z)}} S_{\lambda}(p_i, p_k) \,, \\
    & \mathcal{M}^{\lambda \,\lambda\,-\lambda}_{q_{I}(p_i) \rightarrow \tilde q_I(\tilde p_{ i}) g_F(p_k)} = \sqrt{2} \sqrt{\frac{1-z}{z}} S_{-\lambda}(p_i, p_k).
  \end{align} 
\end{subequations}
\subsection*{$g_{I} \rightarrow \tilde g_I g_F$}

The branching amplitude reads
\begin{align}
  \mathcal{M}^{\lambda_{\itilde} \lambda_i, \lambda_k}_{g_{I}(p_i) \rightarrow \tilde g_I(\tilde p_{i}) g_F(p_k)}=& + \epsilon^*_{-\lambda_i}(p_i)\cdot \epsilon^*_{\lambda_k}(p_k) \, p_k \cdot \epsilon^*_{\lambda_{\itilde}}(\tilde{p}_i) - \epsilon^*_{-\lambda_i}(p_i)\cdot \epsilon^*_{\lambda_{\itilde}}(\tilde{p}_i) \, \tilde{p}_i \cdot \epsilon^*_{\lambda_k}(p_k)  \nonumber \\
  &- \epsilon^*_{\lambda_{\itilde}}(\tilde{p}_i)\cdot \epsilon^*_{\lambda_k}(p_k) \, p_k \cdot \epsilon^*_{-\lambda_i}(p_i)\,.
\end{align} 
We take all gluons to have the same gauge vector $r$, in which case we find
\begin{subequations}
  \begin{align}
    & \epsilon^*_{\lambda}(p_a)\cdot \epsilon_{\lambda}^*(p_b) = 0\,,\\
    & \epsilon^*_{\lambda}(p_a)\cdot \epsilon_{-\lambda}^*(p_b) = 1\,,\\
    & \epsilon^*_{\lambda}(p_a)\cdot p_b = \frac{1}{\sqrt{2}}\frac{S_{-\lambda}(p_b,r)}{S_{-\lambda}(r,p_a)}S_{\lambda}(p_a,p_b)\,.
  \end{align}
\end{subequations}
We find that the only non-zero amplitudes are 
\begin{subequations}
  \begin{align}
    & \mathcal{M}^{\lambda\,\lambda\,\lambda}_{g_{I}(p_i) \rightarrow \tilde g_I(\tilde p_{i}) g_F(p_k)} = - \frac{\sqrt{2}}{\sqrt{z}(1-z)} S_{\lambda}(p_i, p_k)\,, \\
    & \mathcal{M}^{\lambda\,\lambda\,-\lambda}_{g_{I}(p_i) \rightarrow \tilde g_I(\tilde p_{i}) g_F(p_k)} = -\sqrt{2} \frac{1-z}{\sqrt{z}} S_{-\lambda}(p_i, p_k)\,, \\
    & \mathcal{M}^{\lambda\,-\lambda\,-\lambda}_{g_{I}(p_i) \rightarrow \tilde g_I(\tilde p_{i}) g_F(p_k)} = -\sqrt{2} \frac{z^{3/2}}{1-z} S_{\lambda}(p_i, p_k)\,.
  \end{align} 
\end{subequations}
%
\subsection*{$q_I \rightarrow \tilde g_I q_F$}
The branching amplitude is 
\begin{align}
    \mathcal{M}^{\lambda_{\itilde}\lambda_i \lambda_k}_{q_I(p_i) \rightarrow \tilde g_I(\tilde p_i) q_F(p_k)} &= \bar{u}_{\lambda_k}(p_k) \slashed{\epsilon}^*_{\lambda_{\itilde}} u_{\lambda_i}(p_i) \\
    &= \frac{\sqrt{2}}{S_{-\lambda_{\itilde}}(r, \tilde{p}_i)} \bar{u}_{\lambda_k}(p_k)  \bigg[ u_{\lambda_{\itilde}}(r) \bar{u}_{\lambda_{\itilde}}(\tilde{p}_i) + u_{-\lambda_{\itilde}}(\tilde{p}_i)\bar{u}_{-\lambda_{\itilde}}(r) \bigg] u_{\lambda_i}(p_i).\nonumber 
\end{align} 
Note that the polarisation vector is complex conjugated, since  the
gluon is in the final state in terms of the $1 \rightarrow 2$
splitting. The amplitude vanishes for $\lambda_k = -\lambda_i$.
Setting to $\lambda_{\itilde} = \lambda$,
we find the non-vanishing contributions
\begin{subequations}
  \begin{align}
    & \mathcal{M}^{\lambda\,\lambda\,\lambda}_{q_I(p_i) \rightarrow \tilde g_I(\tilde p_i) q_F(p_k)}  = - \frac{\sqrt{2}}{1-z} S_{\lambda}(p_i, p_k)\,,  \\
    & \mathcal{M}^{\lambda\,-\lambda\,-\lambda}_{q_I(p_i) \rightarrow \tilde g_I(\tilde p_i) q_F(p_k)}  = -\sqrt{2} \frac{z}{1-z} S_{\lambda}(p_i, p_k)\,.
  \end{align} 
\end{subequations}
\subsection*{$g_I \rightarrow \tilde q_I \bar{q}_F$}
The branching amplitude reads
\begin{align}
    \mathcal{M}^{\lambda_{\itilde} \lambda_i \lambda_k}_{g_I(p_i) \rightarrow \tilde q_I(\tilde p_i) \bar{q}_F(p_k)} &= \bar{u}_{\lambda_{\itilde}} (\tilde{p}_i) \slashed{\epsilon}_{\lambda_i} u_{-\lambda_k}(p_k)  \\
    &= -\bar{u}_{\lambda_{\itilde}} (\tilde{p}_i) \slashed{\epsilon}_{-\lambda_i}^* u_{-\lambda_k}(p_k) \nonumber \\
    &= -\frac{\sqrt{2}}{S_{\lambda_i}(r, p_i)} \bar{u}_{\lambda_{\itilde}}(\tilde{p}_i)  \bigg[ u_{-\lambda_i}(r) \bar{u}_{-\lambda_i}(p_i) + u_{\lambda_i}(p_i)\bar{u}_{\lambda_i}(r) \bigg] u_{-\lambda_k}(p_k) \nonumber.
\end{align} 
Note the $-\lambda_k$, because the final state is an antiquark, and the absence of a complex conjugate on the polarisation, because it is now truly in the initial state of the splitting. The amplitude vanishes for $\lambda_{\itilde} = \lambda_k$. We set $\lambda =\lambda_{\itilde} = -\lambda_k$ and find the following non-vanishing amplitudes
\begin{subequations}
  \begin{align}
    & \mathcal{M}^{\lambda\,\lambda\,-\lambda}_{g_I(p_i) \rightarrow \tilde q_I(\tilde p_i) \bar{q}_F(p_k)} = \sqrt{2} \sqrt{1-z} S_{-\lambda}(p_i,p_k), \\
    & \mathcal{M}^{\lambda\,-\lambda\,-\lambda}_{g_I(p_i) \rightarrow \tilde q_I(\tilde p_i) \bar{q}_F(p_k)}
      = -\sqrt{2} \frac{z}{\sqrt{1-z}}S_{\lambda}(p_i, p_k)\,.
  \end{align}
\end{subequations}

\bibliographystyle{JHEP}
\bibliography{MC}

\providecommand{\href}[2]{#2}\begingroup\raggedright\begin{thebibliography}{100}

\bibitem{vanBeekveld:2022ukn}
M.~van Beekveld, S.~F. Ravasio, K.~Hamilton, G.~P. Salam, A.~Soto-Ontoso,
  G.~Soyez et~al., \emph{{PanScales showers for hadron collisions: all-order
  validation}},  \href{http://arxiv.org/abs/2207.09467}{{\tt 2207.09467}}.

\bibitem{Heinrich:2020ybq}
G.~Heinrich, \emph{{Collider Physics at the Precision Frontier}},
  \href{http://dx.doi.org/10.1016/j.physrep.2021.03.006}{\emph{Phys. Rept.}
  {\bf 922} (2021) 1--69}, [\href{http://arxiv.org/abs/2009.00516}{{\tt
  2009.00516}}].

\bibitem{Frixione:2002ik}
S.~Frixione and B.~R. Webber, \emph{{Matching NLO QCD computations and parton
  shower simulations}},
  \href{http://dx.doi.org/10.1088/1126-6708/2002/06/029}{\emph{JHEP} {\bf 06}
  (2002) 029}, [\href{http://arxiv.org/abs/hep-ph/0204244}{{\tt
  hep-ph/0204244}}].

\bibitem{Nason:2004rx}
P.~Nason, \emph{{A New method for combining NLO QCD with shower Monte Carlo
  algorithms}},
  \href{http://dx.doi.org/10.1088/1126-6708/2004/11/040}{\emph{JHEP} {\bf 11}
  (2004) 040}, [\href{http://arxiv.org/abs/hep-ph/0409146}{{\tt
  hep-ph/0409146}}].

\bibitem{Jadach:2015mza}
S.~Jadach, W.~Płaczek, S.~Sapeta, A.~Siódmok and M.~Skrzypek, \emph{{Matching
  NLO QCD with parton shower in Monte Carlo scheme — the KrkNLO method}},
  \href{http://dx.doi.org/10.1007/JHEP10(2015)052}{\emph{JHEP} {\bf 10} (2015)
  052}, [\href{http://arxiv.org/abs/1503.06849}{{\tt 1503.06849}}].

\bibitem{Hamilton:2012rf}
K.~Hamilton, P.~Nason, C.~Oleari and G.~Zanderighi, \emph{{Merging H/W/Z + 0
  and 1 jet at NLO with no merging scale: a path to parton shower + NNLO
  matching}}, \href{http://dx.doi.org/10.1007/JHEP05(2013)082}{\emph{JHEP} {\bf
  05} (2013) 082}, [\href{http://arxiv.org/abs/1212.4504}{{\tt 1212.4504}}].

\bibitem{Alioli:2013hqa}
S.~Alioli, C.~W. Bauer, C.~Berggren, F.~J. Tackmann, J.~R. Walsh and S.~Zuberi,
  \emph{{Matching Fully Differential NNLO Calculations and Parton Showers}},
  \href{http://dx.doi.org/10.1007/JHEP06(2014)089}{\emph{JHEP} {\bf 06} (2014)
  089}, [\href{http://arxiv.org/abs/1311.0286}{{\tt 1311.0286}}].

\bibitem{Hoche:2014dla}
S.~Hoeche, Y.~Li and S.~Prestel, \emph{{Higgs-boson production through gluon
  fusion at NNLO QCD with parton showers}},
  \href{http://dx.doi.org/10.1103/PhysRevD.90.054011}{\emph{Phys. Rev.} {\bf
  D90} (2014) 054011}, [\href{http://arxiv.org/abs/1407.3773}{{\tt
  1407.3773}}].

\bibitem{Monni:2019whf}
P.~F. Monni, P.~Nason, E.~Re, M.~Wiesemann and G.~Zanderighi,
  \emph{{MiNNLO$_{\text{PS}}$: A new method to match NNLO QCD to parton
  showers}},  \href{http://arxiv.org/abs/1908.06987}{{\tt 1908.06987}}.

\bibitem{Campbell:2021svd}
J.~M. Campbell, S.~H\"oche, H.~T. Li, C.~T. Preuss and P.~Skands,
  \emph{{Towards NNLO+PS Matching with Sector Showers}},
  \href{http://arxiv.org/abs/2108.07133}{{\tt 2108.07133}}.

\bibitem{Prestel:2021vww}
S.~Prestel, \emph{{Matching N3LO QCD calculations to parton showers}},
  \href{http://dx.doi.org/10.1007/JHEP11(2021)041}{\emph{JHEP} {\bf 11} (2021)
  041}, [\href{http://arxiv.org/abs/2106.03206}{{\tt 2106.03206}}].

\bibitem{Catani:2001cc}
S.~Catani, F.~Krauss, R.~Kuhn and B.~R. Webber, \emph{{QCD matrix elements +
  parton showers}},
  \href{http://dx.doi.org/10.1088/1126-6708/2001/11/063}{\emph{JHEP} {\bf 11}
  (2001) 063}, [\href{http://arxiv.org/abs/hep-ph/0109231}{{\tt
  hep-ph/0109231}}].

\bibitem{Mangano:2001xp}
M.~L. Mangano, M.~Moretti and R.~Pittau, \emph{{Multijet matrix elements and
  shower evolution in hadronic collisions: $W b \bar{b}$ + $n$ jets as a case
  study}}, \href{http://dx.doi.org/10.1016/S0550-3213(02)00249-3}{\emph{Nucl.
  Phys.} {\bf B632} (2002) 343--362},
  [\href{http://arxiv.org/abs/hep-ph/0108069}{{\tt hep-ph/0108069}}].

\bibitem{Krauss:2002up}
F.~Krauss, \emph{{Matrix elements and parton showers in hadronic
  interactions}},
  \href{http://dx.doi.org/10.1088/1126-6708/2002/08/015}{\emph{JHEP} {\bf 08}
  (2002) 015}, [\href{http://arxiv.org/abs/hep-ph/0205283}{{\tt
  hep-ph/0205283}}].

\bibitem{Lavesson:2008ah}
N.~Lavesson and L.~Lonnblad, \emph{{Extending CKKW-merging to One-Loop Matrix
  Elements}},
  \href{http://dx.doi.org/10.1088/1126-6708/2008/12/070}{\emph{JHEP} {\bf 12}
  (2008) 070}, [\href{http://arxiv.org/abs/0811.2912}{{\tt 0811.2912}}].

\bibitem{Hoeche:2009rj}
S.~Hoeche, F.~Krauss, S.~Schumann and F.~Siegert, \emph{{QCD matrix elements
  and truncated showers}},
  \href{http://dx.doi.org/10.1088/1126-6708/2009/05/053}{\emph{JHEP} {\bf 05}
  (2009) 053}, [\href{http://arxiv.org/abs/0903.1219}{{\tt 0903.1219}}].

\bibitem{Hamilton:2009ne}
K.~Hamilton, P.~Richardson and J.~Tully, \emph{{A Modified CKKW matrix element
  merging approach to angular-ordered parton showers}},
  \href{http://dx.doi.org/10.1088/1126-6708/2009/11/038}{\emph{JHEP} {\bf 11}
  (2009) 038}, [\href{http://arxiv.org/abs/0905.3072}{{\tt 0905.3072}}].

\bibitem{Giele:2011cb}
W.~T. Giele, D.~A. Kosower and P.~Z. Skands, \emph{{Higher-Order Corrections to
  Timelike Jets}},
  \href{http://dx.doi.org/10.1103/PhysRevD.84.054003}{\emph{Phys. Rev. D} {\bf
  84} (2011) 054003}, [\href{http://arxiv.org/abs/1102.2126}{{\tt 1102.2126}}].

\bibitem{Platzer:2012bs}
S.~Pl\"atzer, \emph{{Controlling inclusive cross sections in parton shower +
  matrix element merging}},
  \href{http://dx.doi.org/10.1007/JHEP08(2013)114}{\emph{JHEP} {\bf 08} (2013)
  114}, [\href{http://arxiv.org/abs/1211.5467}{{\tt 1211.5467}}].

\bibitem{Lonnblad:2012ix}
L.~L\"onnblad and S.~Prestel, \emph{{Merging Multi-leg NLO Matrix Elements with
  Parton Showers}},
  \href{http://dx.doi.org/10.1007/JHEP03(2013)166}{\emph{JHEP} {\bf 03} (2013)
  166}, [\href{http://arxiv.org/abs/1211.7278}{{\tt 1211.7278}}].

\bibitem{Frederix:2012ps}
R.~Frederix and S.~Frixione, \emph{{Merging meets matching in MC@NLO}},
  \href{http://dx.doi.org/10.1007/JHEP12(2012)061}{\emph{JHEP} {\bf 12} (2012)
  061}, [\href{http://arxiv.org/abs/1209.6215}{{\tt 1209.6215}}].

\bibitem{Lonnblad:2012ng}
L.~Lonnblad and S.~Prestel, \emph{{Unitarising Matrix Element + Parton Shower
  merging}}, \href{http://dx.doi.org/10.1007/JHEP02(2013)094}{\emph{JHEP} {\bf
  02} (2013) 094}, [\href{http://arxiv.org/abs/1211.4827}{{\tt 1211.4827}}].

\bibitem{Bellm:2017ktr}
J.~Bellm, S.~Gieseke and S.~Pl\"atzer, \emph{{Merging NLO Multi-jet
  Calculations with Improved Unitarization}},
  \href{http://dx.doi.org/10.1140/epjc/s10052-018-5723-2}{\emph{Eur. Phys. J.
  C} {\bf 78} (2018) 244}, [\href{http://arxiv.org/abs/1705.06700}{{\tt
  1705.06700}}].

\bibitem{Brooks:2020mab}
H.~Brooks and C.~T. Preuss, \emph{{Efficient multi-jet merging with the Vincia
  sector shower}},
  \href{http://dx.doi.org/10.1016/j.cpc.2021.107985}{\emph{Comput. Phys.
  Commun.} {\bf 264} (2021) 107985},
  [\href{http://arxiv.org/abs/2008.09468}{{\tt 2008.09468}}].

\bibitem{Martinez:2021chk}
A.~B. Martinez, F.~Hautmann and M.~L. Mangano, \emph{{TMD evolution and
  multi-jet merging}},
  \href{http://dx.doi.org/10.1016/j.physletb.2021.136700}{\emph{Phys. Lett. B}
  {\bf 822} (2021) 136700}, [\href{http://arxiv.org/abs/2107.01224}{{\tt
  2107.01224}}].

\bibitem{Sjostrand:2006za}
T.~Sjostrand, S.~Mrenna and P.~Z. Skands, \emph{{PYTHIA 6.4 Physics and
  Manual}}, \href{http://dx.doi.org/10.1088/1126-6708/2006/05/026}{\emph{JHEP}
  {\bf 05} (2006) 026}, [\href{http://arxiv.org/abs/hep-ph/0603175}{{\tt
  hep-ph/0603175}}].

\bibitem{Giele:2007di}
W.~T. Giele, D.~A. Kosower and P.~Z. Skands, \emph{{A simple shower and
  matching algorithm}},
  \href{http://dx.doi.org/10.1103/PhysRevD.78.014026}{\emph{Phys. Rev.} {\bf
  D78} (2008) 014026}, [\href{http://arxiv.org/abs/0707.3652}{{\tt
  0707.3652}}].

\bibitem{Schumann:2007mg}
S.~Schumann and F.~Krauss, \emph{{A Parton shower algorithm based on
  Catani-Seymour dipole factorisation}},
  \href{http://dx.doi.org/10.1088/1126-6708/2008/03/038}{\emph{JHEP} {\bf 03}
  (2008) 038}, [\href{http://arxiv.org/abs/0709.1027}{{\tt 0709.1027}}].

\bibitem{Platzer:2009jq}
S.~Platzer and S.~Gieseke, \emph{{Coherent Parton Showers with Local Recoils}},
  \href{http://dx.doi.org/10.1007/JHEP01(2011)024}{\emph{JHEP} {\bf 01} (2011)
  024}, [\href{http://arxiv.org/abs/0909.5593}{{\tt 0909.5593}}].

\bibitem{Hoche:2015sya}
S.~Hoeche and S.~Prestel, \emph{{The midpoint between dipole and parton
  showers}}, \href{http://dx.doi.org/10.1140/epjc/s10052-015-3684-2}{\emph{Eur.
  Phys. J.} {\bf C75} (2015) 461}, [\href{http://arxiv.org/abs/1506.05057}{{\tt
  1506.05057}}].

\bibitem{Cabouat:2017rzi}
B.~Cabouat and T.~Sjöstrand, \emph{{Some Dipole Shower Studies}},
  \href{http://dx.doi.org/10.1140/epjc/s10052-018-5645-z,
  10.1140/s10052-018-5645-z}{\emph{Eur. Phys. J.} {\bf C78} (2018) 226},
  [\href{http://arxiv.org/abs/1710.00391}{{\tt 1710.00391}}].

\bibitem{Sjostrand:2014zea}
T.~Sjöstrand, S.~Ask, J.~R. Christiansen, R.~Corke, N.~Desai, P.~Ilten et~al.,
  \emph{{An Introduction to PYTHIA 8.2}},
  \href{http://dx.doi.org/10.1016/j.cpc.2015.01.024}{\emph{Comput. Phys.
  Commun.} {\bf 191} (2015) 159--177},
  [\href{http://arxiv.org/abs/1410.3012}{{\tt 1410.3012}}].

\bibitem{Bahr:2008pv}
M.~Bahr et~al., \emph{{Herwig++ Physics and Manual}},
  \href{http://dx.doi.org/10.1140/epjc/s10052-008-0798-9}{\emph{Eur. Phys. J.}
  {\bf C58} (2008) 639--707}, [\href{http://arxiv.org/abs/0803.0883}{{\tt
  0803.0883}}].

\bibitem{Bellm:2019zci}
J.~Bellm et~al., \emph{{Herwig 7.2 release note}},
  \href{http://dx.doi.org/10.1140/epjc/s10052-020-8011-x}{\emph{Eur. Phys. J.
  C} {\bf 80} (2020) 452}, [\href{http://arxiv.org/abs/1912.06509}{{\tt
  1912.06509}}].

\bibitem{Gleisberg:2008ta}
T.~Gleisberg, S.~Hoeche, F.~Krauss, M.~Schonherr, S.~Schumann, F.~Siegert
  et~al., \emph{{Event generation with SHERPA 1.1}},
  \href{http://dx.doi.org/10.1088/1126-6708/2009/02/007}{\emph{JHEP} {\bf 02}
  (2009) 007}, [\href{http://arxiv.org/abs/0811.4622}{{\tt 0811.4622}}].

\bibitem{Sherpa:2019gpd}
{\scshape Sherpa} collaboration, E.~Bothmann et~al., \emph{{Event Generation
  with Sherpa 2.2}},
  \href{http://dx.doi.org/10.21468/SciPostPhys.7.3.034}{\emph{SciPost Phys.}
  {\bf 7} (2019) 034}, [\href{http://arxiv.org/abs/1905.09127}{{\tt
  1905.09127}}].

\bibitem{Gieseke:2003rz}
S.~Gieseke, P.~Stephens and B.~Webber, \emph{{New formalism for QCD parton
  showers}}, \href{http://dx.doi.org/10.1088/1126-6708/2003/12/045}{\emph{JHEP}
  {\bf 12} (2003) 045}, [\href{http://arxiv.org/abs/hep-ph/0310083}{{\tt
  hep-ph/0310083}}].

\bibitem{Catani:1990rr}
S.~Catani, B.~R. Webber and G.~Marchesini, \emph{{QCD coherent branching and
  semiinclusive processes at large x}},
  \href{http://dx.doi.org/10.1016/0550-3213(91)90390-J}{\emph{Nucl. Phys.} {\bf
  B349} (1991) 635--654}.

\bibitem{Catani:1992ua}
S.~Catani, L.~Trentadue, G.~Turnock and B.~R. Webber, \emph{{Resummation of
  large logarithms in $e^+ e^-$ event shape distributions}},
  \href{http://dx.doi.org/10.1016/0550-3213(93)90271-P}{\emph{Nucl. Phys.} {\bf
  B407} (1993) 3--42}.

\bibitem{Banfi:2006gy}
A.~Banfi, G.~Corcella and M.~Dasgupta, \emph{{Angular ordering and parton
  showers for non-global QCD observables}},
  \href{http://dx.doi.org/10.1088/1126-6708/2007/03/050}{\emph{JHEP} {\bf 03}
  (2007) 050}, [\href{http://arxiv.org/abs/hep-ph/0612282}{{\tt
  hep-ph/0612282}}].

\bibitem{Dasgupta:2001sh}
M.~Dasgupta and G.~Salam, \emph{{Resummation of nonglobal QCD observables}},
  \href{http://dx.doi.org/10.1016/S0370-2693(01)00725-0}{\emph{Phys. Lett. B}
  {\bf 512} (2001) 323--330}, [\href{http://arxiv.org/abs/hep-ph/0104277}{{\tt
  hep-ph/0104277}}].

\bibitem{Dasgupta:2018nvj}
M.~Dasgupta, F.~A. Dreyer, K.~Hamilton, P.~F. Monni and G.~P. Salam,
  \emph{{Logarithmic accuracy of parton showers: a fixed-order study}},
  \href{http://dx.doi.org/10.1007/JHEP09(2018)033}{\emph{JHEP} {\bf 09} (2018)
  033}, [\href{http://arxiv.org/abs/1805.09327}{{\tt 1805.09327}}].

\bibitem{Dasgupta:2020fwr}
M.~Dasgupta, F.~A. Dreyer, K.~Hamilton, P.~F. Monni, G.~P. Salam and G.~Soyez,
  \emph{{Parton showers beyond leading logarithmic accuracy}},
  \href{http://dx.doi.org/10.1103/PhysRevLett.125.052002}{\emph{Phys. Rev.
  Lett.} {\bf 125} (2020) 052002}, [\href{http://arxiv.org/abs/2002.11114}{{\tt
  2002.11114}}].

\bibitem{Hamilton:2020rcu}
K.~Hamilton, R.~Medves, G.~P. Salam, L.~Scyboz and G.~Soyez, \emph{{Colour and
  logarithmic accuracy in final-state parton showers}},
  \href{http://dx.doi.org/10.1007/JHEP03(2021)041}{\emph{JHEP} {\bf 03} (2021)
  041}, [\href{http://arxiv.org/abs/2011.10054}{{\tt 2011.10054}}].

\bibitem{Karlberg:2021kwr}
A.~Karlberg, G.~P. Salam, L.~Scyboz and R.~Verheyen, \emph{{Spin correlations
  in final-state parton showers and jet observables}},
  \href{http://dx.doi.org/10.1140/epjc/s10052-021-09378-0}{\emph{Eur. Phys. J.
  C} {\bf 81} (2021) 681}, [\href{http://arxiv.org/abs/2103.16526}{{\tt
  2103.16526}}].

\bibitem{Hamilton:2021dyz}
K.~Hamilton, A.~Karlberg, G.~P. Salam, L.~Scyboz and R.~Verheyen, \emph{{Soft
  spin correlations in final-state parton showers}},
  \href{http://arxiv.org/abs/2111.01161}{{\tt 2111.01161}}.

\bibitem{Bewick:2019rbu}
G.~Bewick, S.~Ferrario~Ravasio, P.~Richardson and M.~H. Seymour,
  \emph{{Logarithmic Accuracy of Angular-Ordered Parton Showers}},
  \href{http://arxiv.org/abs/1904.11866}{{\tt 1904.11866}}.

\bibitem{Bewick:2021nhc}
G.~Bewick, S.~Ferrario~Ravasio, P.~Richardson and M.~H. Seymour, \emph{{Initial
  State Radiation in the Herwig 7 Angular-Ordered Parton Shower}},
  \href{http://arxiv.org/abs/2107.04051}{{\tt 2107.04051}}.

\bibitem{Forshaw:2020wrq}
J.~R. Forshaw, J.~Holguin and S.~Pl\"atzer, \emph{{Building a consistent parton
  shower}}, \href{http://dx.doi.org/10.1007/JHEP09(2020)014}{\emph{JHEP} {\bf
  09} (2020) 014}, [\href{http://arxiv.org/abs/2003.06400}{{\tt 2003.06400}}].

\bibitem{Holguin:2020joq}
J.~Holguin, J.~R. Forshaw and S.~Pl\"atzer, \emph{{Improvements on dipole
  shower colour}},
  \href{http://dx.doi.org/10.1140/epjc/s10052-021-09145-1}{\emph{Eur. Phys. J.
  C} {\bf 81} (2021) 364}, [\href{http://arxiv.org/abs/2011.15087}{{\tt
  2011.15087}}].

\bibitem{Nagy:2020dvz}
Z.~Nagy and D.~E. Soper, \emph{{Summations by parton showers of large
  logarithms in electron-positron annihilation}},
  \href{http://arxiv.org/abs/2011.04777}{{\tt 2011.04777}}.

\bibitem{Parisi:1979se}
G.~Parisi and R.~Petronzio, \emph{{Small Transverse Momentum Distributions in
  Hard Processes}},
  \href{http://dx.doi.org/10.1016/0550-3213(79)90040-3}{\emph{Nucl. Phys. B}
  {\bf 154} (1979) 427--440}.

\bibitem{Collins:1984kg}
J.~C. Collins, D.~E. Soper and G.~F. Sterman, \emph{{Transverse Momentum
  Distribution in Drell-Yan Pair and W and Z Boson Production}},
  \href{http://dx.doi.org/10.1016/0550-3213(85)90479-1}{\emph{Nucl. Phys. B}
  {\bf 250} (1985) 199--224}.

\bibitem{Nagy:2009vg}
Z.~Nagy and D.~E. Soper, \emph{{On the transverse momentum in Z-boson
  production in a virtuality ordered parton shower}},
  \href{http://dx.doi.org/10.1007/JHEP03(2010)097}{\emph{JHEP} {\bf 03} (2010)
  097}, [\href{http://arxiv.org/abs/0912.4534}{{\tt 0912.4534}}].

\bibitem{Sjostrand:1985xi}
T.~Sjostrand, \emph{{A Model for Initial State Parton Showers}},
  \href{http://dx.doi.org/10.1016/0370-2693(85)90674-4}{\emph{Phys. Lett.} {\bf
  157B} (1985) 321--325}.

\bibitem{Gustafson:1987rq}
G.~Gustafson and U.~Pettersson, \emph{{Dipole Formulation of QCD Cascades}},
  \href{http://dx.doi.org/10.1016/0550-3213(88)90441-5}{\emph{Nucl. Phys. B}
  {\bf 306} (1988) 746--758}.

\bibitem{Catani:1996vz}
S.~Catani and M.~H. Seymour, \emph{{A General algorithm for calculating jet
  cross-sections in NLO QCD}},
  \href{http://dx.doi.org/10.1016/S0550-3213(96)00589-5,
  10.1016/S0550-3213(98)81022-5}{\emph{Nucl. Phys.} {\bf B485} (1997)
  291--419}, [\href{http://arxiv.org/abs/hep-ph/9605323}{{\tt
  hep-ph/9605323}}].

\bibitem{Sjostrand:2004ef}
T.~Sjostrand and P.~Z. Skands, \emph{{Transverse-momentum-ordered showers and
  interleaved multiple interactions}},
  \href{http://dx.doi.org/10.1140/epjc/s2004-02084-y}{\emph{Eur. Phys. J.} {\bf
  C39} (2005) 129--154}, [\href{http://arxiv.org/abs/hep-ph/0408302}{{\tt
  hep-ph/0408302}}].

\bibitem{Jager:2020hkz}
B.~J\"ager, A.~Karlberg, S.~Pl\"atzer, J.~Scheller and M.~Zaro,
  \emph{{Parton-shower effects in Higgs production via Vector-Boson Fusion}},
  \href{http://dx.doi.org/10.1140/epjc/s10052-020-8326-7}{\emph{Eur. Phys. J.
  C} {\bf 80} (2020) 756}, [\href{http://arxiv.org/abs/2003.12435}{{\tt
  2003.12435}}].

\bibitem{Hoche:2021mkv}
S.~H\"oche, S.~Mrenna, S.~Payne, C.~T. Preuss and P.~Skands, \emph{{A Study of
  QCD Radiation in VBF Higgs Production with Vincia and Pythia}},
  \href{http://dx.doi.org/10.21468/SciPostPhys.12.1.010}{\emph{SciPost Phys.}
  {\bf 12} (2022) 010}, [\href{http://arxiv.org/abs/2106.10987}{{\tt
  2106.10987}}].

\bibitem{Fischer:2016vfv}
N.~Fischer, S.~Prestel, M.~Ritzmann and P.~Skands, \emph{{Vincia for Hadron
  Colliders}},
  \href{http://dx.doi.org/10.1140/epjc/s10052-016-4429-6}{\emph{Eur. Phys. J.}
  {\bf C76} (2016) 589}, [\href{http://arxiv.org/abs/1605.06142}{{\tt
  1605.06142}}].

\bibitem{Brooks:2020upa}
H.~Brooks, C.~T. Preuss and P.~Skands, \emph{{Sector Showers for Hadron
  Collisions}}, \href{http://dx.doi.org/10.1007/JHEP07(2020)032}{\emph{JHEP}
  {\bf 07} (2020) 032}, [\href{http://arxiv.org/abs/2003.00702}{{\tt
  2003.00702}}].

\bibitem{Andersson:1988gp}
B.~Andersson, G.~Gustafson, L.~Lonnblad and U.~Pettersson, \emph{{Coherence
  Effects in Deep Inelastic Scattering}},
  \href{http://dx.doi.org/10.1007/BF01550942}{\emph{Z. Phys.} {\bf C43} (1989)
  625}.

\bibitem{Kuraev:1977fs}
E.~A. Kuraev, L.~N. Lipatov and V.~S. Fadin, \emph{{The Pomeranchuk Singularity
  in Nonabelian Gauge Theories}}, {\emph{Sov. Phys. JETP} {\bf 45} (1977)
  199--204}.

\bibitem{Balitsky:1978ic}
I.~I. Balitsky and L.~N. Lipatov, \emph{{The Pomeranchuk Singularity in Quantum
  Chromodynamics}}, {\emph{Sov. J. Nucl. Phys.} {\bf 28} (1978) 822--829}.

\bibitem{Kirschner:1983di}
R.~Kirschner and L.~N. Lipatov, \emph{{Double Logarithmic Asymptotics and Regge
  Singularities of Quark Amplitudes with Flavor Exchange}},
  \href{http://dx.doi.org/10.1016/0550-3213(83)90178-5}{\emph{Nucl. Phys. B}
  {\bf 213} (1983) 122--148}.

\bibitem{Jung:2010si}
H.~Jung et~al., \emph{{The CCFM Monte Carlo generator CASCADE version 2.2.03}},
  \href{http://dx.doi.org/10.1140/epjc/s10052-010-1507-z}{\emph{Eur. Phys. J.
  C} {\bf 70} (2010) 1237--1249}, [\href{http://arxiv.org/abs/1008.0152}{{\tt
  1008.0152}}].

\bibitem{Baranov:2021uol}
S.~Baranov et~al., \emph{{CASCADE3 A Monte Carlo event generator based on
  TMDs}}, \href{http://dx.doi.org/10.1140/epjc/s10052-021-09203-8}{\emph{Eur.
  Phys. J. C} {\bf 81} (2021) 425},
  [\href{http://arxiv.org/abs/2101.10221}{{\tt 2101.10221}}].

\bibitem{Andersen:2009nu}
J.~R. Andersen and J.~M. Smillie, \emph{{Constructing All-Order Corrections to
  Multi-Jet Rates}},
  \href{http://dx.doi.org/10.1007/JHEP01(2010)039}{\emph{JHEP} {\bf 01} (2010)
  039}, [\href{http://arxiv.org/abs/0908.2786}{{\tt 0908.2786}}].

\bibitem{Andersen:2009he}
J.~R. Andersen and J.~M. Smillie, \emph{{The Factorisation of the t-channel
  Pole in Quark-Gluon Scattering}},
  \href{http://dx.doi.org/10.1103/PhysRevD.81.114021}{\emph{Phys. Rev. D} {\bf
  81} (2010) 114021}, [\href{http://arxiv.org/abs/0910.5113}{{\tt 0910.5113}}].

\bibitem{Andersen:2011hs}
J.~R. Andersen and J.~M. Smillie, \emph{{Multiple Jets at the LHC with High
  Energy Jets}}, \href{http://dx.doi.org/10.1007/JHEP06(2011)010}{\emph{JHEP}
  {\bf 06} (2011) 010}, [\href{http://arxiv.org/abs/1101.5394}{{\tt
  1101.5394}}].

\bibitem{Hartgring:2013jma}
L.~Hartgring, E.~Laenen and P.~Skands, \emph{{Antenna Showers with One-Loop
  Matrix Elements}},
  \href{http://dx.doi.org/10.1007/JHEP10(2013)127}{\emph{JHEP} {\bf 10} (2013)
  127}, [\href{http://arxiv.org/abs/1303.4974}{{\tt 1303.4974}}].

\bibitem{Nagy:2014nqa}
Z.~Nagy and D.~E. Soper, \emph{{Ordering variable for parton showers}},
  \href{http://dx.doi.org/10.1007/JHEP06(2014)178}{\emph{JHEP} {\bf 06} (2014)
  178}, [\href{http://arxiv.org/abs/1401.6366}{{\tt 1401.6366}}].

\bibitem{Nagy:2017dxh}
Z.~Nagy and D.~E. Soper, \emph{{Jets and threshold summation in Deductor}},
  \href{http://dx.doi.org/10.1103/PhysRevD.98.014035}{\emph{Phys. Rev. D} {\bf
  98} (2018) 014035}, [\href{http://arxiv.org/abs/1711.02369}{{\tt
  1711.02369}}].

\bibitem{Gustafson:1992uh}
G.~Gustafson, \emph{{Multiplicity distributions in QCD cascades}},
  \href{http://dx.doi.org/10.1016/0550-3213(93)90203-2}{\emph{Nucl. Phys.} {\bf
  B392} (1993) 251--280}.

\bibitem{Banfi:2004yd}
A.~Banfi, G.~P. Salam and G.~Zanderighi, \emph{{Principles of general
  final-state resummation and automated implementation}},
  \href{http://dx.doi.org/10.1088/1126-6708/2005/03/073}{\emph{JHEP} {\bf 03}
  (2005) 073}, [\href{http://arxiv.org/abs/hep-ph/0407286}{{\tt
  hep-ph/0407286}}].

\bibitem{Friberg:1996xc}
C.~Friberg, G.~Gustafson and J.~Hakkinen, \emph{{Color connections in $e^+ e^-$
  annihilation}},
  \href{http://dx.doi.org/10.1016/S0550-3213(97)00064-3}{\emph{Nucl. Phys.}
  {\bf B490} (1997) 289--305}, [\href{http://arxiv.org/abs/hep-ph/9604347}{{\tt
  hep-ph/9604347}}].

\bibitem{Nagy:2015hwa}
Z.~Nagy and D.~E. Soper, \emph{{Effects of subleading color in a parton
  shower}}, \href{http://dx.doi.org/10.1007/JHEP07(2015)119}{\emph{JHEP} {\bf
  07} (2015) 119}, [\href{http://arxiv.org/abs/1501.00778}{{\tt 1501.00778}}].

\bibitem{Platzer:2018pmd}
S.~Plaetzer, M.~Sjodahl and J.~Thorén, \emph{{Color matrix element corrections
  for parton showers}},
  \href{http://dx.doi.org/10.1007/JHEP11(2018)009}{\emph{JHEP} {\bf 11} (2018)
  009}, [\href{http://arxiv.org/abs/1808.00332}{{\tt 1808.00332}}].

\bibitem{Nagy:2019pjp}
Z.~Nagy and D.~E. Soper, \emph{{Parton showers with more exact color
  evolution}}, \href{http://dx.doi.org/10.1103/PhysRevD.99.054009}{\emph{Phys.
  Rev. D} {\bf 99} (2019) 054009}, [\href{http://arxiv.org/abs/1902.02105}{{\tt
  1902.02105}}].

\bibitem{DeAngelis:2020rvq}
M.~De~Angelis, J.~R. Forshaw and S.~Pl\"atzer, \emph{{Resummation and
  Simulation of Soft Gluon Effects beyond Leading Color}},
  \href{http://dx.doi.org/10.1103/PhysRevLett.126.112001}{\emph{Phys. Rev.
  Lett.} {\bf 126} (2021) 112001}, [\href{http://arxiv.org/abs/2007.09648}{{\tt
  2007.09648}}].

\bibitem{Forshaw:2021mtj}
J.~R. Forshaw, J.~Holguin and S.~Pl\"atzer, \emph{{Rings and strings: a basis
  for understanding subleading colour and QCD coherence beyond the two-jet
  limit}},  \href{http://arxiv.org/abs/2112.13124}{{\tt 2112.13124}}.

\bibitem{Platzer:2012np}
S.~Platzer and M.~Sjodahl, \emph{{Subleading $N_c$ improved Parton Showers}},
  \href{http://dx.doi.org/10.1007/JHEP07(2012)042}{\emph{JHEP} {\bf 07} (2012)
  042}, [\href{http://arxiv.org/abs/1201.0260}{{\tt 1201.0260}}].

\bibitem{Hoche:2020pxj}
S.~H\"oche and D.~Reichelt, \emph{{Numerical resummation at subleading color in
  the strongly ordered soft gluon limit}},
  \href{http://dx.doi.org/10.1103/PhysRevD.104.034006}{\emph{Phys. Rev. D} {\bf
  104} (2021) 034006}, [\href{http://arxiv.org/abs/2001.11492}{{\tt
  2001.11492}}].

\bibitem{Frixione:2021yim}
S.~Frixione and B.~R. Webber, \emph{{The role of colour flows in matrix element
  computations and Monte Carlo simulations}},
  \href{http://dx.doi.org/10.1007/JHEP11(2021)045}{\emph{JHEP} {\bf 11} (2021)
  045}, [\href{http://arxiv.org/abs/2106.13471}{{\tt 2106.13471}}].

\bibitem{Forshaw:2006fk}
J.~R. Forshaw, A.~Kyrieleis and M.~H. Seymour, \emph{{Super-leading logarithms
  in non-global observables in QCD}},
  \href{http://dx.doi.org/10.1088/1126-6708/2006/08/059}{\emph{JHEP} {\bf 08}
  (2006) 059}, [\href{http://arxiv.org/abs/hep-ph/0604094}{{\tt
  hep-ph/0604094}}].

\bibitem{Catani:2011st}
S.~Catani, D.~de~Florian and G.~Rodrigo, \emph{{Space-like (versus time-like)
  collinear limits in QCD: Is factorization violated?}},
  \href{http://dx.doi.org/10.1007/JHEP07(2012)026}{\emph{JHEP} {\bf 07} (2012)
  026}, [\href{http://arxiv.org/abs/1112.4405}{{\tt 1112.4405}}].

\bibitem{Nagy:2019rwb}
Z.~Nagy and D.~E. Soper, \emph{{Exponentiating virtual imaginary contributions
  in a parton shower}},
  \href{http://dx.doi.org/10.1103/PhysRevD.100.074005}{\emph{Phys. Rev.} {\bf
  D100} (2019) 074005}, [\href{http://arxiv.org/abs/1908.11420}{{\tt
  1908.11420}}].

\bibitem{Becher:2021zkk}
T.~Becher, M.~Neubert and D.~Y. Shao, \emph{{Resummation of Super-Leading
  Logarithms}},
  \href{http://dx.doi.org/10.1103/PhysRevLett.127.212002}{\emph{Phys. Rev.
  Lett.} {\bf 127} (2021) 212002}, [\href{http://arxiv.org/abs/2107.01212}{{\tt
  2107.01212}}].

\bibitem{Banfi:2010xy}
A.~Banfi, G.~P. Salam and G.~Zanderighi, \emph{{Phenomenology of event shapes
  at hadron colliders}},
  \href{http://dx.doi.org/10.1007/JHEP06(2010)038}{\emph{JHEP} {\bf 06} (2010)
  038}, [\href{http://arxiv.org/abs/1001.4082}{{\tt 1001.4082}}].

\bibitem{Forshaw:2021fxs}
J.~R. Forshaw and J.~Holguin, \emph{{Coulomb gluons will generally destroy
  coherence}}, \href{http://dx.doi.org/10.1007/JHEP12(2021)084}{\emph{JHEP}
  {\bf 12} (2021) 084}, [\href{http://arxiv.org/abs/2109.03665}{{\tt
  2109.03665}}].

\bibitem{Dokshitzer:1997in}
Y.~L. Dokshitzer, G.~D. Leder, S.~Moretti and B.~R. Webber, \emph{{Better jet
  clustering algorithms}},
  \href{http://dx.doi.org/10.1088/1126-6708/1997/08/001}{\emph{JHEP} {\bf 08}
  (1997) 001}, [\href{http://arxiv.org/abs/hep-ph/9707323}{{\tt
  hep-ph/9707323}}].

\bibitem{Wobisch:1998wt}
M.~Wobisch and T.~Wengler, \emph{{Hadronization corrections to jet
  cross-sections in deep inelastic scattering}},  in \emph{{Workshop on Monte
  Carlo Generators for HERA Physics (Plenary Starting Meeting)}}, pp.~270--279,
  4, 1998.
\newblock \href{http://arxiv.org/abs/hep-ph/9907280}{{\tt hep-ph/9907280}}.

\bibitem{Collins:1987cp}
J.~C. Collins, \emph{{Spin Correlations in Monte Carlo Event Generators}},
  \href{http://dx.doi.org/10.1016/0550-3213(88)90654-2}{\emph{Nucl. Phys.} {\bf
  B304} (1988) 794--804}.

\bibitem{Knowles:1987cu}
I.~Knowles, \emph{{Angular Correlations in QCD}},
  \href{http://dx.doi.org/10.1016/0550-3213(88)90653-0}{\emph{Nucl. Phys. B}
  {\bf 304} (1988) 767--793}.

\bibitem{Knowles:1988vs}
I.~Knowles, \emph{{Spin Correlations in Parton - Parton Scattering}},
  \href{http://dx.doi.org/10.1016/0550-3213(88)90092-2}{\emph{Nucl. Phys. B}
  {\bf 310} (1988) 571--588}.

\bibitem{Knowles:1988hu}
I.~G. Knowles, \emph{{A Linear Algorithm for Calculating Spin Correlations in
  Hadronic Collisions}},
  \href{http://dx.doi.org/10.1016/0010-4655(90)90063-7}{\emph{Comput. Phys.
  Commun.} {\bf 58} (1990) 271--284}.

\bibitem{Kleiss:1985yh}
R.~Kleiss and W.~Stirling, \emph{{Spinor Techniques for Calculating $p\bar p
  \to W^{\pm} / Z^0 +$ Jets}},
  \href{http://dx.doi.org/10.1016/0550-3213(85)90285-8}{\emph{Nucl. Phys. B}
  {\bf 262} (1985) 235--262}.

\bibitem{Dinsdale:2007mf}
M.~Dinsdale, M.~Ternick and S.~Weinzierl, \emph{{Parton showers from the dipole
  formalism}}, \href{http://dx.doi.org/10.1103/PhysRevD.76.094003}{\emph{Phys.
  Rev. D} {\bf 76} (2007) 094003}, [\href{http://arxiv.org/abs/0709.1026}{{\tt
  0709.1026}}].

\bibitem{Hoeche:2009xc}
S.~Hoeche, S.~Schumann and F.~Siegert, \emph{{Hard photon production and
  matrix-element parton-shower merging}},
  \href{http://dx.doi.org/10.1103/PhysRevD.81.034026}{\emph{Phys. Rev. D} {\bf
  81} (2010) 034026}, [\href{http://arxiv.org/abs/0912.3501}{{\tt 0912.3501}}].

\bibitem{Carli:2010cg}
T.~Carli, T.~Gehrmann and S.~Hoeche, \emph{{Hadronic final states in
  deep-inelastic scattering with Sherpa}},
  \href{http://dx.doi.org/10.1140/epjc/s10052-010-1261-2}{\emph{Eur. Phys. J.
  C} {\bf 67} (2010) 73--97}, [\href{http://arxiv.org/abs/0912.3715}{{\tt
  0912.3715}}].

\end{thebibliography}\endgroup

\end{document}